\documentclass[a4paper,11pt]{article}
\pdfoutput=1
\usepackage{jcappub}

\usepackage{amsmath,amssymb,graphicx}
\usepackage{soul}
\usepackage[latin1]{inputenc}
\usepackage{dsfont}

\usepackage{subfig}

\usepackage{array}
\usepackage{mathtools}
\usepackage{bbold}

\usepackage{multirow}

\usepackage{dcolumn} 
\usepackage{bm}
\usepackage{color}
\usepackage{slashed}
\usepackage{cancel}

\usepackage{comment}

\usepackage{booktabs}
\usepackage{dcolumn}
\newcolumntype{d}[1]{D{.}{.}{#1}}
\newcommand{\spaceclebsches}[0]{\addlinespace[0.2cm]}
\newcommand{\spacelines}[0]{\addlinespace[0.1cm]}
\newcommand{\aligncell}[1]{\multicolumn{1}{r}{#1}}

\usepackage{caption}
\usepackage{longtable}

\def\SYM{\ddagger}

\def\YU{\mathbf{Y}_u}
\def\YD{\mathbf{Y}_d}
\def\YE{\mathbf{Y}_e}
\def\YNU{\mathbf{Y}_\nu}

\usepackage[section, below, above]{placeins}

\usepackage{afterpage}

\begin{document}

\begin{titlepage}

\vspace*{-15mm}
\vspace*{0.7cm}

\begin{center}

{\Large {\bf Predicting $\delta^\text{PMNS}$, $\theta_{23}^\text{PMNS}$  and fermion mass ratios \\[1mm] from flavour GUTs with CSD2}}\\[8mm]

Stefan Antusch$^{\star}$\footnote{Email: \texttt{stefan.antusch@unibas.ch}},  Christian Hohl$^\star$\footnote{Email: \texttt{ch.hohl@unibas.ch}}, 
Charanjit K. Khosa$^{\SYM}$\footnote{Email: \texttt{charanjit.kaur@sussex.ac.uk}}, and Vasja Susi\v{c}$^\star$\footnote{Email: \texttt{vasja.susic@unibas.ch}}

\end{center}

\vspace*{0.20cm}

\centerline{$^{\star}$ \it
Department of Physics, University of Basel,}
\centerline{\it
Klingelbergstr.\ 82, CH-4056 Basel, Switzerland}

\vspace*{0.4cm}

\centerline{$^{\SYM}$ \it
Department of Physics and Astronomy, University of Sussex,}
\centerline{\it
Brighton, BN1 9RH, United Kingdom}

\vspace*{1.2cm}

\begin{abstract}
\noindent
Constrained Sequential neutrino Dominance of type 2 (referred to as CSD2) is an attractive building block for flavour Grand Unified Theories (GUTs)  because it predicts a non-zero leptonic mixing angle $\theta_{13}^\text{PMNS}$, a deviation of $\theta_{23}^\text{PMNS}$ from $\pi /4$, as well as a leptonic Dirac CP phase $\delta^\text{PMNS}$ which is directly linked to the CP violation relevant for generating the baryon asymmetry via the leptogenesis mechanism. When embedded into GUT flavour models, these predictions are modified in a specific way, depending on which GUT operators are responsible for generating the entries of fermion Yukawa matrices. In this paper, we systematically investigate and classify the resulting predictions from supersymmetric $\mathrm{SU}(5)$ based flavour models by fitting the known fermion mass and mixing data, in order to provide a roadmap for future model building. Interestingly, the promising models predict the lepton Dirac CP phase $\delta^\mathrm{PMNS}$ between $230^\circ$ and $290^\circ$, and the quark CP phase $\delta^\mathrm{CKM}$ in accordance with a right-angled unitarity triangle ($\alpha_\mathrm{UT}=90^\circ$). Also, our model setup predicts the quantities $\theta_{23}^\mathrm{PMNS}$ and $m_d/m_s$ with less uncertainty than current experimental precision, and allowing with future sensitivity to discriminate between them.
\end{abstract}

\end{titlepage}
\tableofcontents
\section{Introduction \label{sec:introduction}}
Grand Unified Theories (GUTs) offer an attractive framework for model building beyond the Standard Model (SM). Fermion unification in the large GUT representations, on top of gauge coupling unification, makes them a natural environment for addressing the flavour puzzle, i.e.\ the question about the origin of the observed fermion masses, mixings and CP violating phases. Popular GUT models are, e.g., based on the unifying gauge groups $\mathrm{SU}(5)$ \cite{Georgi:1979df} or $\mathrm{SO}(10)$ \cite{Fritzsch:1974nn,Georgi:1974my}. In this work we will focus on the framework of $\mathrm{SU}(5)$ based GUTs.

Depending on the GUT gauge group and on the choice of the GUT-Higgs representations involved in the GUT operators for the Yukawa matrices, the unification of fermions in GUT-matter representations leads to a variety of close connections between the elements of the Yukawa matrices, and thus between the masses and mixings in the quark and lepton sectors (cf.\ \cite{Antusch:2009gu,Antusch:2013rxa}). Furthermore, in particular towards understanding the observed charged fermion mass hierarchies and the large mixing in the lepton sector, family symmetries are often considered in addition to the unifying gauge symmetry. In the literature many options have been considered for family symmetries, including continuous or discrete symmetries, Abelian and/or non-Abelian groups etc., for reviews see e.g.\ \cite{King:2017guk,Meloni:2017cig,King:2013eh}. In such a ``flavour GUT"  scenario, the vacuum expectation values (VEVs) of the family symmetry breaking fields (known as ``flavons'') play a crucial role in generating the Yukawa couplings.

Although the origin of the observed fermion masses, mixings and CP violating phases has been among the most important puzzles of particle physics already for a long time, the discovery of a non-zero leptonic mixing angle 
$\theta_{13}^\text{PMNS}$ a few years back by T2K~\cite{Abe:2011sj}, Double Chooz~\cite{Abe:2011fz}, RENO~\cite{Ahn:2012nd}, and in particular Daya Bay~\cite{An:2012eh} has  triggered new proposals for models towards its solution. In particular, the nowadays very precise result from combining the latest measurements in a global fit yielding $\theta_{13}^\text{PMNS} \approx 8.54^\circ \pm 0.15^\circ$~\cite{Esteban:2016qun} requires a substantial deviation of leptonic mixing (described by the Pontecorvo-Maki-Nakagawa-Sakata (PMNS) matrix) from the tri-bimaximal (TB) mixing pattern \cite{Harrison:2002er,Xing:2002sw}, which for some time was considered a valid mixing scheme for the lepton sector. 

Classifying flavour models as ``direct'' or ``indirect'' as in \cite{King:2009ap}, depending on whether residual symmetries are used (``direct'' models) or whether a family symmetry gets completely broken to generate the flavour structure (``indirect'' models), different routes were followed. In the context of ``direct'' models it was found that more and more complicated groups had to be chosen in order to be in approximate agreement with the $\theta_{13}^\text{PMNS}$ measurement (cf.\ e.g.\ Refs.\ in \cite{King:2017guk,Meloni:2017cig}). In ``indirect'' models, on the other hand, those became appreciated which had the ``corrections'' to a leading order mixing pattern with zero $1$-$3$ mixing (such as the TB mixing pattern)  present \textsl{a priori}.

The latter situation is typical for flavour GUTs, since due to the GUT relation between quarks and leptons the charged lepton Yukawa matrix typically features some non-zero mixing related to the mixing in the down-type Yukawa matrix, which can correct a leading order mixing pattern from the neutrino mass matrix. Ideas in this direction have been studied some time back under the name ``quark-lepton complementarity (QLC)'' \cite{Minakata:2004xt,Smirnov:2004ju,Raidal:2004iw,Frampton:2004vw,Antusch:2005ca,Li:2005yj,Minakata:2005rf,Hochmuth:2007wq,Goswami:2009yy,Qin:2010hn,Patel:2010hr,Qin:2011ub,Ahn:2011yj,Ahn:2011ep,Zheng:2011uz,Ahn:2011if} with a leading order pattern of ``bimaximal mixing'' \cite{Barger:1998ta} in the neutrino sector, which gets a correction from the charged lepton sector (with ``CKM-like'' mixing angles). Such a scenario was able to explain the large lepton mixing (with a large but non-maximal $\theta_{12}^\text{PMNS}$) at that time. 

Furthermore, it was proposed (cf.~e.g.~\cite{King:2012vj,Antusch:2012fb}) and also realised (cf.~\cite{Meroni:2012ty,Antusch:2013kna,Zhao:2014qwa,Shimizu:2014ria}) in the context of flavour GUTs that a leading order TB mixing pattern in the neutrino sector, which then gets modified by a charged lepton mixing correction (via a predicted 1-2 mixing angle from the charged lepton Yukawa matrix related to the Cabibbo angle $\theta_\mathrm{C}$, cf.\ \cite{Antusch:2011qg,Marzocca:2011dh}), could be an interesting scheme for model building. Such a leading order TB mixing pattern in the neutrino sector can, e.g., be realised in ``indirect'' models via a type I seesaw mechanism with so-called Constrained Sequential Dominance (CSD, also referred to as CSD1) \cite{King:2005bj}. In CSD1, the VEVs of the ``flavons'', which break the family symmetry, point in the specific directions $(0,1,-1)$ and $(1,1,1)$ in flavour space, corresponding to two of the columns of the TB mixing matrix.   

Interestingly, it was found that the general scenario that $\theta_{13}^\text{PMNS}$ emerges entirely from a charged lepton $1$-$2$ mixing contribution leads not only to the relation $\theta_{13}^\text{PMNS} = s_{23}^\text{PMNS} \theta_{12}^\text{e}$ (in leading order), with $\theta_{12}^\text{e}$ potentially related to the Cabibbo angle $\theta_\mathrm{C}$ in the context of GUTs, but also to a so-called lepton mixing sum rule which allows to predict $\delta^\mathrm{PMNS}$ once the mixing pattern in the neutrino mass matrix is fixed \cite{King:2005bj,Masina:2005hf,Antusch:2005kw,Antusch:2007rk,Antusch:2012fb,Girardi:2014faa,Ballett:2014dua}. Taking the $2$-$3$ mixing in the neutrino sector maximal and $\theta_{12}^\text{e} \approx \theta_\mathrm{C}$, we arrive at the prediction $\theta_{13}^\text{PMNS} \approx \theta_\mathrm{C}/{\sqrt{2}} \approx 9.2^\circ $ \cite{King:2012vj,Antusch:2012fb}. This value was originally close to the observed one, but is now disfavoured with the more precise measurement of $\theta_{13}^\text{PMNS}=8.54^\circ \pm 0.15^\circ$~\cite{Esteban:2016qun}. Furthermore, the experimentally preferred region for $s_{23}^\text{PMNS}$ is now larger than $1/\sqrt{2}$, making the prediction for $\theta_{13}^\text{PMNS}$ under the assumption of $\theta_{12}^\text{e} \approx \theta_\mathrm{C}$ even worse.

Already before the very precise measurements of $\theta_{13}^\text{PMNS}$, the pattern of CSD2 \cite{Antusch:2011ic} was proposed for the neutrino sector, as an alternative to CSD1.
Here, the ``flavons'' which break the family symmetry point in the directions $(0,1,-1)$ and $(1,2,0)$ (or ($1,0,2$)) in flavour space. 
CSD2 features the same attractive prediction for the neutrino sector $1$-$2$ mixing $\theta_{12}^\nu$, close to the measured PMNS value of about $35^\circ$, but it predicts a non-zero mixing angle $\theta_{13}^{\nu}$ already in the neutrino sector, a deviation of $\theta_{23}^\text{PMNS}$ from $45^\circ$, as well as a leptonic Dirac CP phase $\delta^\text{PMNS}$ which has been shown in \cite{Antusch:2011ic} to be directly linked to the CP violation relevant for generating the baryon asymmetry via the leptogenesis mechanism \cite{Fukugita:1986hr}. When CSD2 is realised in the context of GUTs, then the combined mixing from the charged lepton sector (predicted by GUT relations) and the neutrino sector leads to an attractive class of models for explaining the observed PMNS parameters. Specific $\mathrm{SU}(5)$ GUT models realising this idea have been constructed in \cite{Antusch:2013wn,Antusch:2017ano}. 

The purpose of this paper is to perform a systematic analysis of the above-described novel class of models. After defining the model class and identifying the possible choices of GUT operators and the free parameters, we will systematically investigate and classify the resulting predictions by fitting the known experimental results for fermion masses and mixings, in order to select the most promising routes for future model building. It will turn out that the promising models predict the lepton and quark Dirac CP phases $\delta^\mathrm{PMNS}, \delta^\mathrm{CKM}$, with $\delta^\mathrm{PMNS}$ between $230^\circ$ and $290^\circ$ and $\delta^\mathrm{CKM}$ in accordance with a right-angle unitarity triangle ($\alpha_\mathrm{UT}=90^\circ$). They also $\theta_{23}^\mathrm{PMNS}$ and $m_d/m_s$ with much less uncertainty than the experimentally allowed ranges. Such predictions of the considered class of models will be probed by future experiments. The DUNE experiment, for instance, can measure $\theta_{23}^\mathrm{PMNS}$ with a precision of less than  $1^\circ$, and $\delta^\mathrm{PMNS}$ with a precision of ${\cal O}(10^\circ)$ \cite{Abi:2018dnh,Abi:2018alz,Abi:2018rgm}.

The paper is organized as follows. In Section~\ref{sec:model} we describe the class of models we consider in our study, including the specification of all fermion sectors and an extensive discussion on the texture and various predictive mechanisms used. In Section~\ref{sec:implementationmodel} we analyse the predictive power of the models and determine the best approach to a numerical analysis. In Section~\ref{sec:numericalresults} we present the results. In Section~\ref{sec:conclusions} we conclude with a summary of our work, as well as discuss the future outlook and application of our results.

\section{A new class of  models: CSD2 in a simple and predictive GUT setup \label{sec:model}}

\subsection{General $\mathrm{SU}(5)$ GUT setup}\label{sec:GUTsetup}

In this section we define the setup for the class of models we consider in this paper. The general idea is to take supersymmetric (SUSY) $\mathrm{SU}(5)$ GUT models and assume a texture in the Yukawa sector which is as predictive as possible. We shall not be concerned with how these textures are dynamically achieved, i.e.~we shall not specify a flavour theory, as we want to do an analysis which is as model independent as possible.

We assume that the fermion sector consists of the usual three families of $\mathbf{\bar{5}}\oplus\mathbf{10}$, which decompose under the SM group $\mathrm{SU}(3)_C\times\mathrm{SU}(2)_L\times\mathrm{U}(1)_Y$ as
\begin{align}
\mathbf{\bar{5}}_i&=(\mathbf{\bar{3}},\mathbf{1},+\tfrac{1}{3})_i\;\oplus\;(\mathbf{1},\mathbf{2},-\tfrac{1}{2})_i\equiv d^c_i\oplus L_i,\\
\mathbf{10}_i&=(\mathbf{3},\mathbf{2},+\tfrac{1}{6})_i\;\oplus\;(\mathbf{\bar{3}},\mathbf{1},-\tfrac{2}{3})_i\;\oplus\; (\mathbf{1},\mathbf{1},+1)_i\equiv Q_i\oplus u^c_i\oplus e^c_i,
\end{align}
\noindent
where the family index $i$ goes from $1$ to $3$. In addition, the implementation of CSD2 via seesaw type~I in the neutrino sector would require additional right-handed neutrinos in the representation $\mathbf{1}$ of $\mathrm{SU}(5)$.

We make, however, no explicit assumptions on the Higgs sector or any top-down flavour theory, although these do have implicit requirements by the choice of our Yukawa texture. Since no Higgs sector is given, we remain agnostic about the the exact superpotential terms of the Yukawa sector in the $\mathrm{SU}(5)$ theory. The most appropriate level at which such a superpotential is to be written is that of the MSSM with right-handed neutrinos:
\begin{align}
\begin{split}
W_{\text{Yuk}}&=\sum_{i,j}\;\; (\YU)_{ij}\;Q_i\cdot H_u\,u_j^c-
(\YD)_{ij}\;Q_i\cdot H_d\,d^c_j-(\YE)_{ij}\;L_i\cdot H_d\,e^c_j\\
&\quad +\sum_{ik}(\YNU)_{ik}\; L_i\cdot H_u \,\nu_k^c+\sum_{kl}(\mathbf{M}_R)_{kl}\;\nu^c_k\,\nu^c_l,
\end{split}\label{eq:MSSM-superpotential}
\end{align}
\noindent
where the $H_u\sim (\mathbf{1},\mathbf{2},+\tfrac{1}{2})$ and $H_d\sim (\mathbf{1},\mathbf{2},-\tfrac{1}{2})$ are the two Higgs fields of the MSSM. The dot $\cdot$ represents a contraction of $\mathrm{SU}(2)$ fundamental indices of the form 
\begin{align}
X\cdot Y\equiv \varepsilon_{ab}\;X^a\,Y^b,
\end{align} 
\noindent
where $\varepsilon_{ab}$ is the completely anti-symmetric tensor with two indices and $\varepsilon_{12}=1$. The indices $i$ and $j$ run from $1$ to $3$, while we do not assume necessarily the same for $k$ and $l$. We have suppressed in this notation the $\mathrm{SU}(3)$ indices. The Yukawa matrices are written in the left-right convention, and the signs in front of the terms are chosen, so that we get positive terms for the fermion mass terms when the electrically neutral components of $H_u$ and $H_d$ acquire a VEV; this convention is equivalent to that of \cite{Martin:1997ns}.

The free parameters in Eq.~\eqref{eq:MSSM-superpotential} are the $3\times 3$ Yukawa matrices $\YU$, $\YD$ and $\YE$, the $3\times n$ neutrino Yukawa matrix $\YNU$ and the $n\times n$ Majorana mass matrix $\mathbf{M}_{R}$, where $n$ is the number of right-handed neutrinos. At the $\mathrm{SU}(5)$ level, the various Yukawa terms are coming from the following type of operators, where each of $X,Y,Z$ stand for a GUT-Higgs field (or a product of GUT-Higgs fields) in $\mathrm{SU}(5)$ representations, such that the terms form an $\mathrm{SU}(5)$ invariant:
\begin{align}
(\YU)_{ij}:&\qquad \mathbf{10}_i\;\mathbf{10}_j\;X,\label{eq:operator-u}\\
(\YD)_{ij}:&\qquad \mathbf{10}_i\;\mathbf{\bar{5}}_j\;Y,\label{eq:operator-d}\\
(\YE^\textsf{T})_{ij}:&\qquad \mathbf{10}_i\;\mathbf{\bar{5}}_j\;Y,\label{eq:operator-e}\\
(\YNU)_{ik}:&\qquad \mathbf{\bar{5}}_i\;\mathbf{1}_k\;Z.\label{eq:operator-nu}
\end{align}
This is a list of $3$ different types of operators, and therefore gauge unification imprints itself only in the form of relations between $\YD$ and $\YE$, while all other Yukawa parameters are completely independent. We therefore use the GUT concept to relate the parameters in the down-type quark and charged lepton sector in a particular manner, which we discuss later. Note though that the unknown parts $X,Y,Z$ are in general different for different choices of indices $i$ and $j$. In order to be as predictive as possible, we assume that the matrices $\YU$, $\YD$, $\YE$, $\YNU$ and $\mathbf{M}_R$ have special textures at the GUT scale, which we discuss and motivate below. 
 
The GUT setup will be studied in the framework of supersymmetry, however we will mainly be concerned with the predictions for the ``SM part'', i.e.\ for the prediction for the fermion masses, mixings and CP phases. SUSY enters mainly via the RGEs when we run the parameters from the GUT scale to the SUSY scale, and via the one-loop SUSY threshold corrections \cite{Hempfling:1993kv,Hall:1993gn,Carena:1994bv,Blazek:1995nv,Antusch:2008tf} for which we will use a general parameterisation as discussed later in section \ref{sec:GUT-operators} following \cite{Antusch:2013jca}. We will include the SUSY threshold correction parameters as free parameters in our analysis. Since they will be determined by the fit to the experimental data for the fermion flavour structure, they can give interesting constraints on the SUSY sparticle spectrum, and, together with the measured mass of the SM Higgs particle and the GUT constraints on the soft SUSY breaking terms, can even fully determine the sparticle spectrum (cf.\ \cite{Antusch:2017ano,Antusch:2016nak,Antusch:2015nwi}). We will leave the investigation of the consequences of the considered models for the sparticle spectrum for a future study.

\subsection{Choice of Yukawa sector\label{sec:Yukawa-sector}}

We now focus on our choice of textures in the Yukawa sector. We shall choose a specific texture based on previous analyses and model building ideas, as we discuss below, but remain agnostic regarding the underlying flavour theory.

The choice of our Yukawa textures will be guided by the principles of simplicity and predictivity, and we shall choose the explanation of the CKM CP violating phase as an important starting point. A summary of the train of thought determining the textures of both the quark and lepton sectors is the following:
\begin{enumerate}
\item \textbf{Phase sum rule}: as a guide to obtaining a viable CKM CP violating phase, we choose the phase sum rule \cite{Antusch:2009hq} that a unitarity triangle angle of $\alpha_{UT}=\delta^{dL}_{12}-\delta^{uL}_{12}\approx 90^\circ$ gives a good prediction; a necessary condition for the implementation is that $\theta^{uL}_{13}=\theta^{dL}_{13}=0$.
\item \textbf{Simplicity in $\YD$}: we choose the down-sector to have no mixing between the first two families and the third family at all, i.e.~$\theta^{dL}_{23}=\theta^{dL}_{13}=0$, and further simplify the texture by taking $(\YD)_{11}=0$ and use phase redefinitions of the fields $\mathbf{10}_i$ to eliminate unphysical phases in the entries of $\YD$.
\item \textbf{CP violation in $\YU$}: in the up-type quark  sector the matrix $\YU$ is symmetric; we achieve $\theta^{uL}_{13}=0$ by $(\YU)_{13}=0$ and implement the quark sector CP violating phase via a realisation of the ``phase sum rule mechanism'' \cite{Antusch:2009hq} by taking all entries real, except the $(\YU)_{12}$ entry to be imaginary.
\footnote{We would like to emphasize that any choice that realizes the phase sum rule leads to the same predictions. We therefore do not lose generality by our particular implementation.} 

\item \textbf{Single operator dominance}: we assume that each non-vanishing entry in $\YD$ comes dominantly from only one operator of the type of Eq.~\eqref{eq:operator-d}, consequently relating $\YE$ to $\YD$ in the simplest and most predictive way.
\item \textbf{CSD2 for $\YNU$}: we choose the form of ``Constrained Sequential Dominance 2'' (CSD2) for the neutrino sector. 
\end{enumerate}

\noindent
Each of these arguments will be thoroughly explored later in this section; we shall discuss each one separately and also flesh-out the connections on how  each one then leads to the next. For now, we simply specify the form that Yukawa matrices take considering the points mentioned above: the quark and charged lepton sector Yukawa matrices at the scale $M_{\text{GUT}}$ take the form

\begin{align}
\YU = \begin{pmatrix}   u_1 & iu_2 & 0 \cr iu_2 & u_3 & u_4 \cr 0 & u_4 & u_5 \end{pmatrix}\,\quad 
\YD = \begin{pmatrix}   0 & z & 0 \cr y e^{i \gamma} & x & 0 \cr 0 & 0 & y_b  \end{pmatrix} \,, \quad
\YE = \begin{pmatrix}   0 & c_y ye^{i \gamma} & 0 \cr c_z z & c_x x & 0 \cr 0 & 0 & y_{\tau}  \end{pmatrix}\,,\label{eq:Yukawa-texture-ude}
\end{align}
while the left-handed neutrino mass matrix at $M_{\text{GUT}}$ takes one of the following two forms:
\begin{align}
\mathbf{M}_{\nu}^{(102)} = m_a \begin{pmatrix} \epsilon e^{i\alpha} &0& 2 \epsilon e^{i\alpha} \cr 0&1&-1\cr 2\epsilon e^{i\alpha} & -1&1+4\epsilon e^{i\alpha} \end{pmatrix}\,,\quad
\mathbf{M}_{\nu}^{(120)} =  m_a \begin{pmatrix} \epsilon e^{i\alpha} & 2 \epsilon e^{i\alpha} & 0 \cr 2\epsilon e^{i\alpha} & 1+4\epsilon e^{i\alpha} & -1 \cr 0 & -1 & 1  \end{pmatrix}\,.\label{eq:CSD2-2variants}
\end{align}
The reason behind two possible forms of $\mathbf{M}_\nu$ will be explained in Section~\ref{sec:reasoning-Ynu}.

Above, the Yukawa sector is parametrized by $14$ real parameters in total; this includes the $12$ parameters
\begin{align}
u_1,\quad u_2,\quad u_3,\quad u_4,\quad u_5, \quad x,\quad y,\quad z, \quad m_a, \quad \epsilon,\quad y_b, \quad y_\tau, \label{eq:list_parameters_1}
\end{align}
\noindent
and the $2$ phases 
\begin{align}
\alpha,\quad \gamma. \label{eq:list_parameters_2}
\end{align}

In addition, the factors $c_x,c_y,c_z$ in $\YE$ are Clebsch-Gordan (CG) coefficients, which are fixed by the choice of particular GUT operators in Eqs.~\eqref{eq:operator-d} and \eqref{eq:operator-e}. We postpone a more in-depth discussion on the possible values of these coefficients to Section~\ref{sec:GUT-operators}.

The above texture considerably reduces the number of free parameters, allowing it to make a number of predictions. The way this texture works is the following:
\begin{itemize}
\item Since $\YD$ is block diagonal, the angles $\theta_{23}^{\text{CKM}}$ and $\theta^{\text{CKM}}_{13}$ are coming only from $\YU$. Since $(\YU)_{13}=0$ and $(\YD)_{13}=0$, the angle $\theta_{13}^\text{CKM}$ is generated indirectly: 
	\begin{align}
	\theta_{13}^\text{CKM}\approx \theta_{12}^{uL} \theta_{23}^{uL}.\label{eq:theta13-generated-indirectly}
	\end{align}
	All in all, this means that the parameters $u_i$ in $\YU$, where $i=1\ldots 5$, are fitted to accommodate the three mass eigenvalues $m_u$, $m_c$, $m_t$, and two mixing angles $\theta_{23}^{\text{CKM}}$ and $\theta_{13}^{\text{CKM}}$ (the latter generated indirectly). Since $\theta_{23}^{uL}$ is fixed by the CKM angle $\theta_{23}^\text{CKM}$, the angle $\theta_{12}^{uL}$ is determined by Eq.~\eqref{eq:theta13-generated-indirectly}. Since there is no left mixing phase $\delta^{dL}_{12}$ in $\YD$, the relative factor $i$ in $\YU$ predicts the $\alpha_{\text{UT}}$ angle in the unitarity triangle $\alpha_{UT}\approx-\delta^{uL}_{12}\approx \pi/2$, and thus $\delta^{\text{CKM}}$.

\item In $\YE$ and $\YD$, the parameters $x$, $y$, $z$ are used to correctly fit the two well measured charged lepton masses $m_{e}$ and $m_{\mu}$, and produce a suitable $\theta_{12}^{dL}$ in order to produce the correct remaining CKM angle $\theta_{12}^{\text{CKM}}$. Given fixed Clebsch coefficients $c_x$, $c_y$ and $c_z$ (they are not free parameters in a chosen model), this automatically predicts the down-type masses $m_d$ and $m_s$ and  the charged lepton mixing angle $\theta_{12}^{eL}$. The prediction of the masses $m_d$ and $m_s$ can be alternatively thought of as the prediction of the ratio $m_d/m_s$, which turns out to be fixed almost solely by the three Clebsch coefficients, while the SUSY threshold correction parameter $\eta_q$ (to be discussed later in Section~\ref{sec:GUT-operators}) is fit so as to give a correct overall scale for the $m_d$ and $m_s$ masses. Finally, the parameters $y_b$ and $y_\tau$ can be set independently, thus determining the correct $m_b$ and $m_\tau$ mass, respectively.
\item The form of CSD2 in the neutrino sector predicts one neutrino mass to be zero. The parameters $m_a$ and $\epsilon$ in $\mathbf{M}_{\nu}$ can be used to fit the two non-zero masses $m_{\nu_2}$ and $m_{\nu_3}$. The two remaining parameters are the phases $\alpha$ and $\gamma$ in the matrices $\mathbf{M}_\nu$ and $\YE$, respectively. These 2 parameters have to be used to fit the three PMNS mixing angles, and they also determine the PMNS CP violating phase. From a simplified perspective, successfully fitting $3$ angles with 2 parameters implies that $1$ angle is determined by the other $2$; we choose the least well measured of the angles, the angle $\theta_{23}^{\text{PMNS}}$, to be the predicted one. All in all, that means that the CSD2 form of the neutrino sector and the given texture in $\YE$ make 2 predictions: $\delta^{\text{PMNS}}$ and $\theta_{23}^{\text{PMNS}}$. 
\end{itemize}
Given the considerations above, we see that the chosen textures make $4$ predictions, which we summarize in a table given below:
\begin{center}
	\begin{tabular}{ll}
	\toprule
	predicted quantity&root cause\\
	\midrule
	$\delta^{\text{CKM}}$& phase sum rule\\\addlinespace[4pt]
	$m_d/m_s$& GUT connection\\\addlinespace[4pt]
	$\theta_{23}^{\text{PMNS}}$& $\YE$ texture and CSD2\\\addlinespace[4pt]
	$\delta^{\text{PMNS}}$& $\YE$ texture and CSD2\\
	\bottomrule	
	\end{tabular}
\end{center} 
Additionally, two more interesting quantities are fit: the charged lepton mixing angle $\theta_{12}^{eL}$ and the SUSY threshold parameter for the first two down-type families $\eta_{q}$. The quantity $\theta_{12}^{eL}$ may be of interest for more general model building approaches, e.g.\ when the charged fermion GUT setup may be combined with a different scheme for the neutrino sector. The value for $\eta_{q}$ would have to be realized by a realistic model of SUSY breaking, which can lead to interesting constraints on the sparticle spectrum as discussed at the end of Section~\ref{sec:GUTsetup}.

We note that complete analysis has to take into account the RGE running of the Yukawa matrices to low energies, as well as SUSY threshold corrections. Such a complete analysis of all input parameters and observables of the model with careful consideration of the involved energy scales is performed later in Section~\ref{sec:implementationmodel}. The discussion in this section was intended only for demonstrative purposes of what the chosen textures can achieve.

It has thus been established that the chosen Yukawa textures are both simple and predictive. Based on our $5$-point step-by-step reasoning, we also claim that the choice of the texture is far from arbitrary, and that the various motivational points lead from one to the other. 
For an example how such an $\mathrm{SU}(5)$ GUT texture for the charged fermions can be realised in an explicit model, we refer the interested reader to Ref.\ \cite{Antusch:2013tta}. 
We now return to this step-by-step motivation of the textures, discussing each of the $5$ considerations in greater detail. 

\subsubsection{Phase sum rule \label{sec:reasoning-phase}}
The starting point for the considerations to determine our textures was the ``phase sum rule mechanism'' in the quark sector, proposed in~\cite{Antusch:2009hq}, which leads to a predictive scheme for CP violation in the quark sector featuring a right-angled unitarity triangle with $\alpha_{\mathrm{UT}}= 90^\circ$ (corresponding to a prediction $\delta^{\text{CKM}} = 1.188\pm 0.016$, well within the current experimental range of $1.208 \pm 0.054$ )\footnote{This is a prediction for $\alpha_\mathrm{UT}=90^\circ$, with the CKM angles $\theta_{12}^\text{CKM}$, $\theta_{23}^\text{CKM}$ and $\theta_{13}^\text{CKM}$ taken in their $1\sigma$ experimental ranges, with data at $M_{Z}$ taken from~\cite{Antusch:2013jca}. The experimental range for $\delta^{\text{CKM}}$ at $M_Z$ is also from~\cite{Antusch:2013jca}.}. 

In~\cite{Antusch:2009hq} it was shown that a number of ``quark mixing sum rules'' arise under the condition that the 1-3 mixings from both $\mathbf{U}_{u}^{L}$ and $\mathbf{U}_{d}^{L}$ are zero (see Appendix~\ref{app:ckm_pmns} for notation). Assuming $\theta_{13}^{uL}=\theta_{13}^{dL}=0$ and the small angle approximation, 
the mixing sum rules can be written as 
\begin{align}
 \delta^{dL}_{12}-\delta^{uL}_{12}&\approx \alpha_{\text{UT}}\approx \arg \left(1-\frac{\theta_{12}^{\text{CKM}}\theta_{23}^{\text{CKM}}}{\theta_{13}^{\text{CKM}}}e^{-i\delta^{\text{CKM}}}\right),\\
\theta_{12}^{uL}&\approx \frac{\theta_{13}^{\text{CKM}}}{\theta_{23}^{\text{CKM}}}, \label{eq:theta12uL}\\
\theta_{12}^{dL}&\approx \bigg|\theta_{12}^{\text{CKM}}-\frac{\theta_{13}^{\text{CKM}}}{\theta_{23}^{\text{CKM}}}\;e^{-i\delta^{\text{CKM}}}\bigg|,
\end{align}
where $\alpha_{\text{UT}}$ is the upper angle in the unitarity triangle (labelled $\alpha$ in PDG~\cite{Patrignani:2016xqp}). Taking the central values and $1\sigma$ errors at the scale $M_{Z}$ from \cite{Antusch:2013jca}, we thus arrive to the conclusion that the numerical values for the left-hand side quantities are
\begin{align}
\delta^{dL}_{12}-\delta^{uL}_{12}&\approx 88.5^\circ\pm 3.2^\circ\,, \\
\theta_{12}^{uL}&\approx 4.96^\circ \pm 0.19^\circ\,,\\
\theta_{12}^{dL}&\approx 12.18^\circ \pm 0.27^\circ\,.\label{eq:sum-rule-d}
\end{align} 

Experimental data is thus consistent with the intriguing possibility that $\alpha_{\mathrm{UT}}= 90^\circ$. It has been proposed in \cite{Antusch:2009hq} that simple textures realising $\delta_{12}^{dL} - \delta_{12}^{uL} = \pi/2$ could thus be used for building predictive models for CP violation in the quark sector. This idea has been applied, e.g., in the GUT flavour models in Refs.~\cite{Antusch:2011sx,Meroni:2012ty,Antusch:2013wn,Antusch:2013kna,Antusch:2013tta,Antusch:2013rla,Antusch:2017ano}. Future more precise measurements of the CKM phase have the potential to verify or exclude this $90^\circ$ prediction.

As a final comment on the phase sum rule, we would like to point out that the generation of a CP violating phase from a phase $\pi/2$ in one of the entries is attractive from a model building point of view, since it can arise from an underlying discrete symmetry or spontaneous breaking thereof. 
In models of flavour, where the structure of the Yukawa matrices arises from the vacuum expectation values of so-called ``flavons'', which break a certain family symmetry, phase differences of $\pi/2$ between different flavons, or between the different components of one flavon, can emerge in various ways, e.g.\ via ``discrete vacuum alignment' with $\mathbb{Z}_4$ symmetry combined with spontaneous CP violation \cite{Antusch:2011sx} or from a flavon potential as discussed in \cite{Antusch:2013kna}. 

\subsubsection{$\YD$: simplicity and predictivity \label{sec:reasoning-Yd}}
From among the fermion sectors we first turn to the down-type quark sector and discussion the form that the matrix $\YD$ takes. Here we rely on the principles of simplicity and predictivity. When we apply these principles to the down-sector, there are added benefits also in the charged-lepton sector, since the two sector are related due to gauge unification in the underlying $\mathrm{SU}(5)$ setup.

An important prerequisite for the phase mixing sum rule of Section~\ref{sec:reasoning-phase} to work was to have vanishing $1$-$3$ mixing angles; for $\YD$ this means that $\theta_{13}^{dL}=0$, and we can approximately achieve this by taking a texture zero by $(\YD)_{13}=0$. We can further simplify $\YD$ by assuming that the $\theta_{23}^{dL}$ angle is also zero, so that all CKM mixing between the first two and the last family is coming from the up sector. With this assumption only the largest CKM mixing angle $\theta_{12}^{\text{CKM}}$ in the hierarchy
\begin{align}
1&\gg \theta_{12}^{\text{CKM}}\gg\theta_{23}^{\text{CKM}}\gg \theta_{13}^{\text{CKM}}
\end{align} 
is generated from the down sector. A simple way with a minimal number of free parameters is to choose a $2+1$ block diagonal structure. This structure then needs to generate the $4$ relevant quantities (excluding any phases): $3$ down-type masses, as well as the $1$-$2$ contribution via the angle $\theta_{12}^{dL}$.
A $2+1$ block structure has $5$ non-zero entries, so we can still explain the $4$ relevant quantities if we eliminate one of the entries in the $2\times 2$ block. The dominant entry in this block will be generating the mass $m_s$, so we require $(\YD)_{22}\neq 0$, while the non-zero $\theta_{12}^{dL}$ angle contribution will benefit from $(\YD)_{12}\neq 0$. Since the right mixing will dominantly come from $(\YD)_{21}$, and this mixing can be of use later in the lepton sector, we would like to keep that as well. We thus eliminate the parameter in the $1$-$1$ entry: $(\YD)_{11}=0$.\footnote{We like to note that another possibility here would be to take the $2$-$1$ entry to be zero. This, however, would decouple the mixing in the quark and lepton sectors since then $\theta_{12}^{dL}$ would vanish. While this may be of interest for different model building ideas, we prefer to stick to $(\YD)_{11}=0$ in the following since CSD2 will make use of a non-zero $\theta_{12}^{dL}$.}

The non-zero entries in such a texture are in general complex. We have the freedom, however, to absorb phases into redefinitions of the fields. Since we are using the left-right convention for the matrix $\YD$, the basis of the rows comes from the  $\mathrm{SU}(5)$ representations $\mathbf{10}_{i}$ (where the left-handed down-quarks live), while the basis for columns comes from $\mathbf{\bar{5}}_{i}$. We use only the freedom of the phase in $\mathbf{10}_i$, with which we can make $3$ entries, one in each row, to be real. A redefinition of $\mathbf{\bar{5}}_{i}$, on the other hand, would influence the neutrino mass matrix; we prefer not to absorb the one remaining phase in $\YD$ into $\mathbf{\bar{5}}_{i}$ due to greater clarity later when considering the neutrino sector. The choice of which phases to absorb has now fixed the basis of the $\mathbf{10}_i$. As will be discussed below, to eliminate phases in the neutrino Yukawa matrix we will globally redefine the $\mathbf{\bar{5}}_i$, such that the bases of all the Yukawa matrices are fixed and only the physical phases remain; since in a flavon setup the three families of $\mathbf{\bar{5}}_i$ form a triplet, there is only the freedom to absorb one phase.

Given the considerations above, we have thus arrived to the following form of $\YD$:
\begin{align}
	\YD&\sim\begin{pmatrix}
		0&\ast&0\\
		\ast&\ast&0\\
		0&0&\ast\\
	\end{pmatrix} \quad \to
	\begin{pmatrix}
		0&\star&0\\
		\ast&\star&0\\
		0&0&\star\\
	\end{pmatrix}	.
\end{align}
The symbols $\ast$ denote non-zero complex entries, while the $\star$ represents positive real entries. The arrow ``$\to$'' represents the absorbing of phases in to redefinitions of $\mathbf{10}_i$, which shows our choice of entries from which the phase is eliminated, arriving to the final form of $\YD$ given in Eq.~\eqref{eq:Yukawa-texture-ude}. We note that in this parametrisation the remaining complex entry $(\YD)_{21}$ does not affect the CKM CP phase, but will have an influence on CP violation in the lepton sector since in the considered $\mathrm{SU}(5)$ framework, $\YD$ is related to $\YE^\textsf{T}$. We will parametrise the complex $2$-$1$ element of $\YD$ as $ye^{i \gamma}$ with real parameter $y$ (cf.\ Eq.~\eqref{eq:Yukawa-texture-ude}). 

We finish the motivation of $\YD$ with a remark comparing our texture to the one, which gives rise to the Gatto Sartori Tonin (GST) relations \cite{Gatto:1968ss}: 
The vanishing $1$-$1$ entry in the $2\times 2$ block connects the two mixing angles (left and right) and the two singular values of the block in a relation. Adapting the notation to the concrete case of $\YD$, the $2$-$2$ entry is roughly equal to the bigger singular value and thus to the strange quark mass $m_s$. We can then write the block using the small angle approximation $\theta_{12}^{dL},\theta_{12}^{dR}\ll 1$ as
\begin{align}
(\YD)_{2\times 2}&\approx\begin{pmatrix}
0&\theta_{12}^{dL}m_s\\
\theta_{12}^{dR}m_s&\phantom{\theta_{12}^{dL}}m_s
\end{pmatrix},
\end{align}
from which we can derive the relation
\begin{align}
\theta_{12}^{dL}\theta_{12}^{dR}&\approx \frac{m_d}{m_s},
\end{align}
where $m_d$ is the down quark mass, which is the smaller of the two singular values; since $m_d\ll m_s$ experimentally, the small angle approximation is justified. 

The aforementioned GST relation has a texture zero in the same $1$-$1$ location, but it is valid only when the matrix is symmetric, and the $1$-$2$ mixing angle is taken to be the Cabibbo angle $\theta_\mathrm{C}$: if $\theta_{12}^{dL}=\theta_{12}^{dR}\approx \theta_\mathrm{C}$, then we get the GST relation
\begin{align}
\sqrt{\frac{m_d}{m_s}}\approx \theta_\mathrm{C}.
\end{align}

We stress, however, that in our case the GST relation is not valid; beside our texture not being symmetric, it is also important that not all the $\theta_{12}^{\text{CKM}}$ mixing is generated from $\YD$, such that we do not obtain a prediction for $m_d/m_s$ in terms of the Cabbibo angle $\theta_\mathrm{C}$. In our texture, the parameters $x,y$ and $z$ are determined by the very accurately measured $m_e$ and $m_\mu$, and by 
$\theta_{12}^{dL}$ which in turn is fixed by the quark mixing sum rule in Eq.~\eqref{eq:sum-rule-d}. The masses $m_d$ and $m_s$, or more precisely the ratio $m_d/m_s$ and the SUSY threshold correction parameter $\eta_q$, are then obtained as predictions once $x,y$ and $z$ are fitted. The model predictions for $m_d/m_s$ thus in general differ from the one of the GST relation.

\subsubsection{$\YU$: generating CP violation in CKM \label{sec:reasoning-Yu}}
The up sector Yukawa matrix $\YU$ is taken to be symmetric. The masses in the up-quark sector are hierarchical and the mixing angles small. We will take a general $\YU$ under the two conditions that (i) the 1-3 mixing in the up-type quark sector is vanishing, which we achieve in a very good approximation by $(\YU)_{13} = 0$ and which is a condition for the phase sum rule to hold, and (ii) the phase of the 1-2 mixing is equal to $- \pi/2$, which, applying the ``phase sum rule'' relation $\delta_{12}^{dL} - \delta_{12}^{uL} = \pi/2$  \cite{Antusch:2009hq} gives $\alpha_{\mathrm{UT}}= 90^\circ$. 

Due to the texture zero in the entries $(\YU)_{13}$ and $(\YD)_{13}$ the relation
\begin{align}
\theta_{13}^\text{CKM}\approx \theta_{12}^{uL}\,\theta_{23}^{uL}\approx \frac{(\YU)_{12}\,(\YU)_{23}}{(\YU)_{22}\,(\YU)_{33}},
\end{align} 
holds, with $\theta_{12}^{uL}$ given by Eq.~\eqref{eq:theta12uL} and where $\theta_{23}^{uL} = \theta_{23}^{\text{CKM}}$. 
Since the down sector is block diagonal, it contributes only $\theta_{12}^{dL}$, and the CKM angles $\theta^{\text{CKM}}_{13}$ and $\theta_{23}^{\text{CKM}}$ are thus generated exclusively from the up sector, while $\theta_{12}^\text{CKM}$ gets contributions from the up-sector and the down-sector.

For the following analysis it is only relevant that $\delta_{12}^{uL} = -\pi/2$, however in order to be specific we will choose a special representative of the possible $\YU$ with this property, namely the case where most entries are real, except for the $(\YU)_{12}$ (and $(\YU)_{21}$) entry which has a complex phase of $\pi/2$ (cf.\ \cite{Antusch:2009hq}). The placement of the $i$ in $\YU$ is the sole generator of the quark CP phase  $\delta^{\text{CKM}}$. Note that the freedom for phase redefinitions of $\mathbf{10}_i $ was already used for $\YD$, so basis for $\YU$ is thus already fixed and there is no phase freedom remaining. In summary, the texture we consider for $\YU$ and $\YD$ is
	\begin{align}
	\YU&\sim \begin{pmatrix}
		\star&i\star& 0\\
		i\star&\star&\star\\
		0&\star&\star\\
		\end{pmatrix},&
	\YD&\sim \begin{pmatrix}
		0&\star& 0\\
		\ast&\star&0\\
		0&0&\star\\
		\end{pmatrix},
		\label{eq:general_texture_yu_yd}
\end{align}
where $\star$ denote any positive real values, and the zero entries $(\YU)_{13}=(\YD)_{13}=0$ ensure approximately zero $1$-$3$ mixing, ensuring the validity of the phase sum rule to a good approximation. The asterisk $\ast$ denotes arbitrary complex entries, and a complex phase in the 2-1 entry of $\YD$ only contributes to right-mixing and not to the left mixing matrix relevant for the construction of the CKM mixing matrix. 
Since $\YD$ is related to $\YE^T$ in our $\mathrm{SU}(5)$ setup, however, this phase appears in the left-side mixing matrix of the charged lepton sector, helping to generate $\delta^{\text{PMNS}}$.
We shall make use of this form of the matrices $\YU$ and $\YD$ in the next steps.

\subsubsection{$\YE$: single operator dominance \label{sec:reasoning-Ye}}
In the following we will furthermore assume that the entries in Yukawa matrices are each dominantly generated by a single GUT operator of the type given in Eqs.~\eqref{eq:operator-u}--\eqref{eq:operator-nu}, which could be a tree-level operator (e.g.\ for the case of the $3$-$3$ element of $\YU$ to generate the comparatively large top quark mass) or an effective operator (which helps to explain the hierarchy of the quark and charged lepton masses). We refer to this principle as  \textbf{single operator dominance}. The assumption is that possible effects of subdominant operators can be neglected.\footnote{It has been checked in explicit GUT flavour models, e.g.\ in \cite{Antusch:2013kna}, that this principle works very well, unless two operators are engineered to both contribute with similar strength. With the single operator dominance principle, one arrives at more predictive models, while engineering two operators to contribute with similar strength would introduce a new parameter to soften correlations which are otherwise induced by the GUT operators. To be as predictive as possible, we choose not to rely on such assumptions.} 

This assumption enables to establish a direct relation between $\YD$ and $\YE$ due to the same operator contributing to one entry of each of these matrices, cf.~Eq.~\eqref{eq:operator-d} and \eqref{eq:operator-e}; the entries are related by a group-theoretic $\mathrm{SU}(5)$ Clebsch-Gordan coefficient depending on which product of representations $Y$ represents in the stated equations. Each entry in $\YD$ can be coming from a different type of operator, so each matrix entry $(\YE^\mathsf{T})_{ij}$ can have a different CG coefficient relative to the entry $(\YD)_{ij}$. The possibilities of which values the Clebsch coefficients $c_x,c_y,c_z$ can take will be discussed later. 
	
As we have already mentioned briefly in Section~\ref{sec:GUTsetup}, and as we will discuss in more detail in Section~\ref{sec:GUT-operators}, the relation between $\YD$ and $\YE$ is affected by RG running between the GUT scale and low energies, and also by the SUSY threshold correction when matching the MSSM to the SM at loop level. The latter effects can be particularly large since there are contributions that are loop suppressed but $\tan \beta$ enhanced  \cite{Hempfling:1993kv,Hall:1993gn,Carena:1994bv,Blazek:1995nv,Antusch:2008tf,Antusch:2009gu}. For the $1$-$2$ blocks of $\YD$ and $\YE$, we will show that to a very good approximation both effects can be subsumed into a single factor that merely rescales one block compared to the other. 

Regarding $(\YD)_{33}$, in an explicit model, there may also be a Clebsch coefficient relating $y_b$ and $y_\tau$ at the GUT scale, and an additional SUSY threshold correction parameter $\eta_b$ (analogously to $\eta_q$ for the $1$-$2$ block) which is fit to match the measured bottom and tau quark masses. This gives an additional constraint on the SUSY particle spectrum (cf.\ discussion in section 2.1), but it will not be discussed any further in this paper. Possible Clebsch factors between $y_b$ and $y_\tau$ in $\mathrm{SU}(5)$ are e.g.\ $y_\tau/y_b = 1$ (i.e.\ $b$-$\tau$ unification \cite{Georgi:1979df}) or $y_\tau/y_b =3/2$ (see for example~\cite{Antusch:2009gu}). However, as already mentioned earlier, we will simply fit $y_b$ and $y_\tau$ to the experimental data, since we want our analysis to be as model independent as possible.

\subsubsection{Neutrino sector: using CSD2 \label{sec:reasoning-Ynu}}
In the neutrino sector, we choose the CSD2 texture for the light neutrinos, which is known to be very predictive~\cite{Antusch:2011ic,Antusch:2013wn}. This section provides a brief summary and motivation for this texture; for understanding the remainder of the paper, however, it is sufficient to simply note the forms of the light neutrino mass matrices of Eq.~\eqref{eq:CSD2-2variants} (description with $3$ real parameters $m_a$, $\epsilon$ and $\alpha$) and the approximate mixing angle predictions in Eqs.~\eqref{eq:12pmns}--\eqref{eq:23pmns}.

The discovery of neutrino oscillations made clear that at least two out of the three observed left-handed neutrinos possess mass, and that there is a mismatch between the flavour eigenbasis $\{\nu_e,\nu_\mu,\nu_\tau\}$ and the mass eigenbasis $\{\nu_1,\nu_2,\nu_3\}$. The two bases are related by the unitary PMNS matrix; see Appendix~\ref{app:ckm_pmns} for details on notation.

\vspace{0.5cm}
{\bf Large lepton mixing in type I seesaw models via Sequential Dominance:}\newline 
To understand the origin of the two large lepton mixing angles in the context of the type I seesaw mechanism, the concept of Sequential Dominance (SD) \cite{King:2002nf,Antusch:2004gf} of the right-handed neutrino contributions to the neutrino mass matrix was proposed. Writing
\begin{align}
\YNU=\begin{pmatrix}   A_1 & B_1 & C_1 \cr A_2 & B_2  & C_2 \cr A_3 & B_3  & C_3 \end{pmatrix}  \,, 
\quad  \mathbf{M}_{R}=\begin{pmatrix} M_A & 0 & 0 \cr 0 & M_B & 0 \cr 0 & 0 & M_C \end{pmatrix} \,,\label{nuyuk}
\end{align}
then according to the type~I seesaw mechanism, the neutrino masses are given by
\begin{align}
\mathbf{M}_{\nu} =v^2 \YNU \mathbf{M}_R^{-1} \YNU^\mathsf{T}=v^2 \left[ \frac{AA^\mathsf{T}}{M_A}+ \frac{BB^\mathsf{T}}{M_B}+ \frac{CC^\mathsf{T}}{M_C}\right]\,,\label{eq:Mnu-general}
\end{align}
where $A$, $B$, $C$ are the column vectors of neutrino Yukawa matrix, e.g.~$A=(A_1,A_2,A_3)^\mathsf{T}$.
SD is the assumption that
\begin{align}
\frac{AA^\mathsf{T}}{M_A}\gg \frac{BB^\mathsf{T}}{M_B}\gg \frac{CC^\mathsf{T}}{M_C} \;,
\end{align}
i.e.\ that the contribution of one of the right-handed neutrinos, the one with mass $M_A$, dominates $\mathbf{M}_{\nu}$, the one with mass $M_B$ is subdominant, and the one with mass $M_C$ can be neglected. Sequential Dominance thus 
 corresponds to strong normal hierarchy, i.e. $m_{3}^{\nu}\gg m_{2}^{\nu}\gg m_{1}^{\nu}$. With this hierarchy and the simplifying assumption $A_1=0$,  the neutrino mixing angles in leading order satisfy \cite{King:2002nf}
\begin{align}
\tan \theta_{12}^{\nu} &\approx \frac{|B_1|}{c^{\nu}_{23}|B_2|\cos(\phi^\prime_{B_2})-s^{\nu}_{23}|B_3|\cos(\phi^\prime_{B_3})}\,, \label{eq:SD_th12}\\
\theta_{13}^{\nu} &\approx \frac{|B_1||A_2^* B_2+A_3^* B_3|}{(|A_2|^2+|A_3|^2)^{3/2}} \frac{M_A}{M_B}\,,\label{eq:SD_th13} \\
\tan \theta_{23}^{\nu} &\approx\frac{|A_2|}{|A_3|}\,. \label{eq:SD_th23}
\end{align}
We used the definitions
\begin{align}
\phi^\prime_{B_2} &= \phi_{B_2} - \phi_{B_1} - \phi^\nu_2 - \chi^\nu\,, \label{eq:def_phase2}\\
\phi^\prime_{B_3} &= \phi_{B_3} - \phi_{B_1} + \phi_{A_2} - \phi_{A_3} - \phi^\nu_2 - \chi^\nu\,, \label{eq:def_phase3}
\end{align}
and the (complex) parameters in $\YNU$ are written in the form $X=|X|e^{i\phi_X}$ $(X\in\{A_i,B_i\})$. With no loss of generality $M_A$, $M_B$, $M_C$ are chosen real and positive. The values of the two auxiliary phases $\phi^\nu_2$ and $\chi^\nu$ (see the convention in Eq.~\eqref{eq:general_matrix_3}) are fixed by the equations
\begin{align}
\phi^\nu_2 - \phi_{A_2} + \phi_{B_1} &\approx \arg{(A_2^* B_2+A_3^* B_3)}\,, \\
c^{\nu}_{23}|B_2|\sin(\phi^\prime_{B_2}) &\approx s^{\nu}_{23}|B_3|\sin(\phi^\prime_{B_3})\,,
\end{align}
such that the angles $\theta_{12}^\nu$ and $\theta_{13}^\nu$ are real (which is already assumed in Eq.~\eqref{eq:SD_th12} and \eqref{eq:SD_th13}). 

\vspace{0.5cm}
{\bf Before the measurement of the $\theta_{13}^\text{PMNS}$: TB mixing via CSD1}\newline 
Before the measurement of the ``reactor angle'' $\theta_{13}^\text{PMNS}$, the values of the mixing angles were consistent with the simple scenario
\begin{align}
\sin^2\theta^{\text{PMNS}}_{12}\approx \tfrac{1}{3},\\
\sin^2\theta^{\text{PMNS}}_{13}\approx 0,\\
\sin^2\theta^{\text{PMNS}}_{23}\approx \tfrac{1}{2},
\end{align}
which can be summarized with a PMNS matrix of the form
\begin{align}
U_\text{PMNS}&=
\begin{pmatrix}
\sqrt{\tfrac{2}{3}} & \sqrt{\tfrac{1}{3}} & 0 \\
-\sqrt{\tfrac{1}{6}} & \sqrt{\tfrac{1}{3}} & \sqrt{\tfrac{1}{2}} \\
\sqrt{\tfrac{1}{6}} & -\sqrt{\tfrac{1}{3}} & \sqrt{\tfrac{1}{2}} \\
\end{pmatrix},
\end{align}
with bases for the rows and columns defined by the standard PDG convention stated in Eq.~\eqref{eq:matrix_explicit}: the matrix is $U_{fi}$, where the indices $f=e,\mu,\tau$ and $i=1,2,3$. This pattern is called tri-bimaximal mixing~\cite{Harrison:2002er,Xing:2002sw}. If the PMNS matrix takes the TB form, the atmospheric angle $\theta^{\text{PMNS}}_{23}$ is maximal, while the reactor angle $\theta^{\text{PMNS}}_{13}$ is predicted to be zero, and there is no complex CP-violating phase.

The TB mixing matrix in the neutrino sector can be realized with SD by imposing the conditions
 $|A_1|=0\,,
 |A_2|=|A_3|\,,  |B_1|=|B_2|=|B_3|\,,  \phi'_{B_2}=0\,, 
 \phi'_{B_3}=\pi$, 
 which corresponds to $\YNU$ and $ \mathbf{M}_{R}$ of the form
\begin{align}
\YNU=\begin{pmatrix}   0 & b \cr a & b \cr -a & b  \end{pmatrix}  \,, \quad  \mathbf{M}_{R}=\begin{pmatrix} M_A & 0 \cr 0 & M_B \end{pmatrix} \,,\label{eq:alignement-CSD}
\end{align}  
where the parameters $a$, $b$ are in general complex. 
We assume that the heaviest right-handed neutrino is completely decoupled, either because it is very heavy or because the corresponding neutrino Yukawa couplings are very small, thus the contribution from the subsubleading term $CC^\mathsf{T}/M_C$ in the light neutrino mass matrix is neglected.\footnote{ Alternatively, in $\mathrm{SU}(5)$ models, we may assume that only two right-handed neutrinos exist.} 
When $\YE$ is diagonal, the PMNS matrix will be completely determined by the mixing in the neutrino sector. In this case the PMNS mixing matrix has the TB form discussed above:

The condition $|A_2|=|A_3|$ gives rise to $\tan{\theta^\nu_{23}}=1$, whereas $|B_1|=|B_2|=|B_3|$ with the phase relations  imply $\tan{\theta^\nu_{12}}=1/2$. Furthermore, substituting $ \phi'_{B_2}=0$ and $  \phi'_{B_3}=\pi$ into the definitions in Eq.~\eqref{eq:def_phase2} and \eqref{eq:def_phase3}, we get the relation $(\phi_{B_3}-\phi_{A_3})-(\phi_{B_2}-\phi_{A_2})=\pi$ and it follows immediately that $\theta_{13}^\nu = 0$. The combination of SD and the above set of relations for the neutrino Yukawa couplings $A_i$ and $B_i$ is known as constrained sequential dominance \cite{King:2005bj}, which we may also refer to as CSD1.

The exact TB mixing pattern in the PMNS matrix, however, has been ruled out ever since the measurement of a non-zero $\theta^{\text{PMNS}}_{13}$ mixing angle at T2K~\cite{Abe:2011sj}, Double Chooz~\cite{Abe:2011fz}, RENO~\cite{Ahn:2012nd} and Daya Bay~\cite{An:2012eh}. This implies that the TB structure of the PMNS matrix needs to be perturbed in some way.

\vspace{0.5cm}
{\bf TB neutrino mixing plus charged lepton mixing contribution:}\newline 
After the $\theta^{\text{PMNS}}_{13}$ measurement, it was realised that in a GUT context, where a $1$-$2$ mixing contribution from the charged lepton sector is typically present due to GUT relations between $\YE$ and $\YD$ (with $\YD$ often being the dominant source for the CKM mixing and thus featuring a sizeable $1$-$2$ mixing), TB mixing could still be an attractive mixing pattern in the neutrino sector. The angle $\theta^{\text{PMNS}}_{13}$ is in this scenario generated via the $1$-$2$ charged lepton mixing contribution.   

In typical flavour GUT models $\theta^{eL}_{12}$ will be dominant, because it is related to the largest (Cabibbo) mixing angle in the quark sector. This motivates the assumption that only $\theta^{eL}_{12}$ is non-zero ($\theta^{eL}_{13}=0$, $\theta^{eL}_{23}=0$). Under the assumption that $\theta^{eL}_{12},\ll 1$ and remembering that TB mixing implies $\theta_{13}^\nu=0$, the general formulas for the lepton mixing angles from Eqs.~\eqref{appeq:eqs12}--\eqref{appeq:eqs23}, including charged lepton contributions, give (cf. \cite{Antusch:2005kw}) 
\begin{align}
s_{12}^\text{PMNS} e^{-i\delta_{12}^\text{PMNS}} &\approx s_{12}^{\nu} e^{-i (\delta_{12}^{\nu}+\theta_{12}^{eL} t_{12}^\nu c_{23}^\nu \sin(\delta_{12}^\nu-\delta_{12}^{eL}))} + \theta_{12}^{eL} c_{12}^{\nu} c_{23}^{\nu} e^{-i\delta_{12}^{eL}}\,, \label{eqs12} \\
s_{13}^\text{PMNS} e^{-i \delta_{13}^\text{PMNS}} &\approx \theta_{12}^{eL} s_{23}^\nu e^{-i (\delta_{23}^{\nu}+\delta_{12}^{eL})}\,,  \label{eqs13} \\
s_{23}^\text{PMNS} e^{-i \delta_{23}^\text{PMNS}} &\approx s_{23}^{\nu} e^{-i \delta_{23}^{\nu}} \,, \label{eqs23}
\end{align}
where $s^\nu_{ij}\equiv\sin\theta^\nu_{ij}$, $c^\nu_{ij}\equiv\cos\theta^\nu_{ij}$ and $t^\nu_{ij}\equiv\tan\theta^\nu_{ij}$. In particular, from Eq.~\eqref{eqs13} we obtain for the PMNS angle $\theta_{13}^\text{PMNS}$: 
\begin{align}
s_{13}^\text{PMNS}&\approx \theta_{13}^\text{PMNS} \approx \theta_{12}^{eL} s_{23}^\nu \;,\label{eq:13pmnsCSD1}
\end{align}
in leading order in $\theta_{12}^{eL}$. With approximate TB mixing realised in the neutrino sector, e.g.\ via CSD1, $s_{23}^\nu = s_{23}^\text{PMNS} =  1/\sqrt{2}$ and we obtain $\theta_{13}^\text{PMNS} \approx \theta_{12}^{eL}/\sqrt{2}$. It has been pointed out that with $\theta_{12}^{eL} \approx \theta_\mathrm{C}$ one would obtain $\theta_{13}^\text{PMNS} \approx 9.2^\circ$, close to the experimental value at that time, and models along this line have been constructed e.g.\ in \cite{Meroni:2012ty,Antusch:2013kna}. However, with the present rather accurate measurement of $\theta_{13}^\text{PMNS}$ it has turned out that the predicted value for $\theta_{13}^\text{PMNS}$ from this consideration is not in agreement with the experimental data. 

\vspace{0.5cm}
{\bf A novel scheme for PMNS mixing: CSD2 plus charged lepton corrections}\newline
In \cite{Antusch:2011ic,Antusch:2013wn} it was proposed to use a novel vacuum alignment of the flavons, such that the form of $\YNU$ is different from the one in Eq.~\eqref{eq:alignement-CSD}. In particular, the alternative flavon vacuum alignment retains the dominant flavon VEV in the first column, but a different subdominant flavon choice in the second column. This new form is called CSD2~\cite{Antusch:2011ic}, and it comes along in two varieties based on two different VEV alignments of the subdominant column.\footnote{We remark that alongside CSD1 and CSD2, there exist still more possible interesting types of vacuum alignment of flavons, such as CSD3~\cite{King:2013iva} and CSD4~\cite{King:2013xba,King:2013hoa}. These alignments though generate a good $\theta^{\text{PMNS}}_{13}$ from the neutrino sector alone, making them less attractive in a GUT setup (where the charged lepton contribution is linked to the down sector). We shall thus not consider CSD3 or CSD4 further in this paper.} They are denoted by $\phi_{102}$ and $\phi_{120}$, and they respectively correspond to the following neutrino Yukawa matrices:
\begin{align}
\YNU^{(102)}=\begin{pmatrix}   0 & b \cr a & 0 \cr -a & 2b  \end{pmatrix}\,, \quad 
\YNU^{(120)}=\begin{pmatrix}   0 & b \cr a & 2b \cr -a & 0  \end{pmatrix}\,.
\label{yukcsd2}
\end{align}
With these alternative vacuum alignments, the seesaw mechanism from Eq.~\eqref{eq:Mnu-general} (and $CC^\mathsf{T}/M_C \to 0$) delivers the following mass matrices for the left handed neutrinos:
\begin{align}
\mathbf{M}_{\nu}^{(102)} &= m_a \begin{pmatrix} 0 & 0 & 0 \cr 0 & 1 & -1 \cr 0 & -1 & 1 \end{pmatrix} + m_b \begin{pmatrix} 1 & 0 & 2 \cr 0 & 0 & 0 \cr 2 & 0 & 4 \end{pmatrix}
= m_a \begin{pmatrix} \epsilon e^{i\alpha} &0& 2 \epsilon e^{i\alpha} \cr 0&1&-1\cr 2\epsilon e^{i\alpha} & -1&1+4\epsilon e^{i\alpha} \end{pmatrix}\,,\label{eq:Mnu-CSD2-b}\\
\mathbf{M}_{\nu}^{(120)} &= m_a \begin{pmatrix} 0 & 0 & 0 \cr 0 & 1 & -1 \cr 0 & -1 & 1 \end{pmatrix} + m_b \begin{pmatrix} 1 & 2 & 0 \cr 2 & 4 & 0 \cr 0 & 0 & 0 \end{pmatrix}
= m_a \begin{pmatrix} \epsilon e^{i\alpha} & 2 \epsilon e^{i\alpha} & 0 \cr 2\epsilon e^{i\alpha} & 1+4\epsilon e^{i\alpha} & -1 \cr 0 & -1 & 1  \end{pmatrix}\,,\label{eq:Mnu-CSD2-a}
\end{align}
where the complex mass parameters $m_a$ and $m_b$ are defined by
\begin{align}
m_a:=\frac{v^2 a^2}{M_A}\,,\quad
m_b:=\frac{v^2 b^2}{M_B}, \label{eq:ma-mb}
\end{align}
while their ratio is parametrized by the modulus $\epsilon$ and phase angle $\alpha$ via
\begin{align}
\frac{m_b}{m_a} &\equiv \epsilon e^{i\alpha}.
\end{align}
By using the overall phase freedom for $\mathbf{\bar{5}}_i$ (which form a flavour triplet in a complete flavour theory, as already mentioned earlier), we can absorb the phase from the parameter $m_a$, making $m_a$ real\footnote{In the effective theory with no right-handed neutrinos, the phase redefinitions of $\mathbf{1}_k$ (in contrast to the phase from $\mathbf{5}_i$) do not appear anywhere; the phases do not change $m_a$ and $m_b$, since they cancel in the fractions of Eq.~\eqref{eq:ma-mb}.}. the light neutrino mass matrix in Eq.~\eqref{eq:Mnu-CSD2-b} or \eqref{eq:Mnu-CSD2-a} will thus be parametrized by $3$ real parameters: $m_a$, $\epsilon$ and $\alpha$.

It is clear from the above equation that the neutrino sector mixing matrix will depend only on the ratio $m_b/m_a = \epsilon e^{i\alpha}$, while the size of the parameter $m_a$ determines the overall scale of the masses. With the assumption $M_A \ll M_B$ we get $|m_b|\ll|m_a|$ and $\epsilon$ can be used as an expansion parameter for the angles in the neutrino rotation matrix. Beside the contribution from the neutrino sector to the lepton mixing parameters in the PMNS matrix, there is also one coming from the charged leptons. Using Eqs.~\eqref{appeq:12pmns_102}--\eqref{appeq:23pmns_102} we obtain the PMNS angles as an expansion in the parameters $\epsilon$ and $\theta_{12}^{eL}$ when both lepton sectors contribute (cf.\ \cite{Antusch:2013wn,King:2005bj})
\begin{align}
\theta_{12}^\text{PMNS} &\approx 35.3^\circ - \frac{\theta_{12}^{eL}}{\sqrt{2}}\cos{\gamma}\,, \label{eq:12pmns} \\
\theta_{13}^\text{PMNS} &\approx \frac{1}{\sqrt{2}} \big( \epsilon^2 + {\theta_{12}^{eL}}^2 \pm 2\epsilon\theta_{12}^{eL} \cos{(\alpha + \gamma)} \big)^{1/2}\,, \label{eq:13pmns}\\
\theta_{23}^\text{PMNS} &\approx 45^\circ \mp \epsilon \cos{\alpha}\,, \label{eq:23pmns}
\end{align}
for the two CSD2 scenarios $\mathbf{M}_{\nu}^{(102)}$ and $\mathbf{M}_{\nu}^{(120)}$. As we can see, CSD1 and CSD2 share the good prediction that in leading order $\theta_{12}^\text{PMNS} \approx 35.3^\circ$ and $\theta_{23}^\text{PMNS} \approx 45^\circ $, as in the TB mixing pattern, however a non-zero $\theta_{13}^\text{PMNS}$ is already predicted from the neutrino sector (even if $\theta_{12}^{eL}$ was zero). Interestingly, in contrast to CSD1 models where the decay asymmetry for leptogenesis is suppressed \cite{Antusch:2006cw}, it has been shown in \cite{Antusch:2011ic} that in CSD2 it is unsuppressed and directly linked to the leptonic Dirac CP phase $\delta^\text{PMNS}$. 

A CSD2 set-up in combination with charged lepton corrections was considered the first time in a renormalizable model based on an $\mathrm{A}_4$ family symmetry and a specific $\mathrm{SU}(5)$ GUT set-up in Ref.~\cite{Antusch:2013wn}, and more recently also in \cite{Antusch:2017ano}. In the models considered in the present paper, we assume that the neutrino mass matrix $\mathbf{M}_{\nu}$ has the CSD2 form of either the $\phi_{102}$ or $\phi_{120}$ flavon vacuum alignment, as written down in Eq.~\eqref{eq:Mnu-CSD2-b} and \eqref{eq:Mnu-CSD2-a}, and we explore the possible sets of GUT operators, which essentially predict $\theta_{12}^{eL}$, to find out which of them are most promising for model building.

\subsection{Candidates for GUT operators in the Yukawa sector\label{sec:GUT-operators}}
We have just set the texture of $\YU$, $\YD$, $\YE$ and $\mathbf{M}_{\nu}$ at the GUT scale in the previous part of this section, but there are still undetermined quantities which are an integral part of a model: the Clebsch-Gordan coefficients $c_x$, $c_y$ and $c_z$. Their values depend on the yet unspecified choices for the unknown parts $Y$ of the $\mathrm{SU}(5)$ GUT operator in Eq.~\eqref{eq:operator-d} and \eqref{eq:operator-e}. 

The possible Clebsch factors between the down and charged lepton sector in these operators have been classified in \cite{Antusch:2009gu,Antusch:2013rxa}. It would  seem that we have potentially a very large number of viable possibilities which Clebsch coefficients to take. The exact relations with Clebsch coefficients are valid only at the GUT scale, while the measured masses and mixing angles of the fermion sector are considered at $M_Z$. The RGE running of these parameters from the GUT scale to low scales, as well as unknown SUSY threshold corrections at the SUSY scale, can to some extent ``repair'' the high-scale relations  so that they are compatible with experiment at low energy.  Therefore it might appear that there are few constraints on the combination of Clebsch factors yielding realistic low energy masses and mixing angles.

It turns out, however, that we can greatly limit the number of Clebsch combinations by considering the following double ratio of the first two generations of Yukawa couplings (which, as we shall show below, is approximately invariant under RGE and SUSY threshold corrections):
\begin{align}
d:= \frac{y_\mu y_d}{ y_e y_s}.
\end{align}
In the model under consideration at the GUT scale, this ratio can be approximately written as a ratio of Clebsch factors
\begin{align}
d\Big|_{M_{\text{GUT}}} = \frac{y_\mu y_d}{ y_e y_s}\Big|_{M_{\text{GUT}}} \approx \Big|\frac{c_x^2}{c_y c_z}\Big |.\label{eq:doubleratio-GUT}
\end{align}
The last approximation comes the following approximate formulas for the Yukawa couplings:
\begin{align}
y_d \approx \Big|\frac{yz}{x}\Big|\,, \quad y_s \approx |x| \,, \quad y_e \approx \Big|\frac{c_y c_z}{c_x}\Big| \Big|\frac{yz}{x}\Big|\,, \quad y_\mu \approx |c_x| |x|\,,
\end{align}
in the case where $x \gg y,z$, using the texture of Eq.~\eqref{eq:Yukawa-texture-ude}. On the other hand, this ratio is experimentally determined at low energies (at the $Z$ scale) to be 
\begin{align}
d\Big|_{M_Z} = \frac{m_\mu m_d}{ m_e m_s} = 10.7\,{}^{+1.6}_{-0.9}\,,\label{eq:doubleratio-Z}
\end{align}
with the errors coming mostly from the quark masses $m_s$ and $m_d$, while the lepton masses $m_e$ and $m_\mu$ are very well measured. The values at $M_Z$ were taken from \cite{Antusch:2013jca}; the asymmetry of the error mostly comes from the measurement of $m_d$. 

The ratio $d$ has the remarkable property that it is stable both under RGE running and under SUSY threshold corrections~\cite{Hempfling:1993kv,Hall:1993gn,Carena:1994bv,Blazek:1995nv,Antusch:2008tf}. This is easy to see by noticing that single ratios $y_d/y_s$ and $y_e/y_\mu$ within the same sector are already stable; the reason for using the double ratio is that its GUT scale expression depends only on group-theoretical Clebsch factors, and not on any of the unknown parameter values.

We now argue that the ratios $y_d/y_s$ and $y_e/y_\mu$ are stable under RGE and SUSY threshold corrections: 
\begin{enumerate}
\item \textbf{RGE running}\par
We consider the $1$-loop RGE equations in the MSSM \cite{Martin:1993zk} for the down  and charged lepton Yukawa (written in the LR convention):\footnote{We neglect the effects of the neutrino Yukawa couplings here (cf.\ e.g.\ \cite{Antusch:2002ek}). We can assume they are small, since they would stem from an effective operator in a model realisation. }
\begin{align}
\tfrac{d}{dt}\YD&=\tfrac{1}{16\pi^2}\left(\mathrm{Tr}(3\YD^\dagger\YD+\YE^\dagger\YE)+3\YD\YD^\dagger+\YU\YU^\dagger-\tfrac{16}{3}g_3^2-3g_2^2-\tfrac{7}{15}g_1^2\right)\,\YD,\\
\tfrac{d}{dt}\YE&=\tfrac{1}{16\pi^2}\left(\mathrm{Tr}(3\YD^\dagger\YD+\YE^\dagger\YE)+3\YE\YE^\dagger-3g_2^2-\tfrac{9}{5}g_1^2\right)\,\YE,
\end{align}
where $t=\log\mu$ is the log of the renormalization scale $\mu$, and the explicit writing of unit matrices next to the scalar terms has been suppressed in the above notation. We use the left and right basis of the matrices $\YD$ and $\YE$, where they are diagonal, which is simplest for our considerations. Due to the strong hierarchy in the down and charged lepton sector masses, in particular
$y_d,y_s\ll y_b$ and $y_e,y_\mu\ll y_\tau$, the 3rd generation Yukawa terms from the trace and the gauge coupling terms dominate the RGE beta functions of the first two families, and the contributions of first two generation Yukawas can be neglected; thus
	\begin{align}
	\tfrac{d}{dt}y_d&\approx \tfrac{1}{16\pi^2}\,y_d\,(3|y_b|^2+|y_\tau|^2-\tfrac{16}{3}g_3^2-3g_2^2-\tfrac{7}{15}g_1^2),\\
	\tfrac{d}{dt}y_s&\approx \tfrac{1}{16\pi^2}\,y_s\,(3|y_b|^2+|y_\tau|^2-\tfrac{16}{3}g_3^2-3g_2^2-\tfrac{7}{15}g_1^2),\\
	\tfrac{d}{dt}y_e&\approx\tfrac{1}{16\pi^2}\,y_e\,(3|y_b|^2+|y_\tau|^2-3g_2^2-\tfrac{9}{5}g_1^2),\\
	\tfrac{d}{dt}y_\mu&\approx\tfrac{1}{16\pi^2}\,y_\mu\,(3|y_b|^2+|y_\tau|^2-3g_2^2-\tfrac{9}{5}g_1^2).
	\end{align}
	Now it is clear that $dy_d/y_d\approx dy_s/y_s$ and $dy_e/y_e\approx dy_\mu/y_\mu$, consequently keeping the ratios $y_d/y_s$ and $y_e/y_\mu$ approximately constant under RG running in the MSSM. Similar arguments hold also for the SM RG running below the SUSY scale. 
\item \textbf{SUSY threshold corrections}\par
At the SUSY scale, where the MSSM is matched to the SM, the SUSY threshold corrections \cite{Hempfling:1993kv,Hall:1993gn,Carena:1994bv,Blazek:1995nv,Antusch:2008tf} of the Yukawa couplings are implemented as the following~\cite{Antusch:2013jca}:
\begin{align}
\label{thresh:yu}
\YU^\text{MSSM} &\approx \frac{\YU^\text{SM}}{\sin{\beta}}\,,\\
\label{thresh:yd}
\YD^\text{MSSM} &\approx  \text{diag}\left(\frac{1}{1+\eta_q},\frac{1}{1+\eta_q},\frac{1}{1+\eta_b}\right) \frac{\YD^\text{SM}}{\cos{\beta}}\,,\\
\label{thresh:yl}
\YE^\text{MSSM} &\approx  \frac{\YE^\text{SM}}{\cos{\beta}}\,.
\end{align}
The above equations are written in the basis where $\YU$ is diagonal and in the left-right convention for the Yukawa matrices. The SUSY threshold corrections are parametrized by $\eta_q$ and $\eta_b$, and they also depend on the $\tan\beta$ parameter defined as the ratio of the VEVs of the Higgs fields $H_u$ and $H_u$ in the MSSM: $\tan{\beta} := v_u/v_d$. Note that in Eqs.~\eqref{thresh:yu}--\eqref{thresh:yl}, we only considered $\tan{\beta}$ enhanced contributions from down type quarks.\footnote{The analysis already covers also the general case. The 3rd family SUSY threshold corrections can be absorbed into $\tan\beta$ and relabelling it into $\tan\bar{\beta}$, cf.~\cite{Antusch:2013jca}. The $\eta_l$ correction to the first two families also has no qualitative effect on predictions of observables: it would change the overall scale of the $1$-$2$ charged lepton block, which can be compensated by a change in $x$,$y$ and $z$ by a common factor, while the consequent change in the overall change of scale in the down sector can then be absorbed by a shift in $\eta_q$. This leads to the same low energy prediction at a shifted parameter point.} From Eq.~\eqref{thresh:yd}, it is clear that the $1/\cos\beta$ and $1/(1+\eta_q)$ factors drop out of the ratio $y_d/y_s$. Similarly, according to Eq.~\eqref{thresh:yl} the $1/\cos\beta$ factor drops out of the ratio $y_e/y_\mu$.
\end{enumerate}
\par
We have thus seen that the double ratio $d$ is a very useful quantity and that it has approximately the same value at all scales. Equating Eq.~\eqref{eq:doubleratio-GUT} and \eqref{eq:doubleratio-Z}, the ratio of Clebsch factors $|c_x^2/(c_y c_z)|$ must thus be around $10.7$. This guideline enables to greatly reduce the number of relevant Clebsch factor combinations that one needs to consider, since we automatically know that large deviations from the ratio will not provide a good fit to the low energy observables.

Taking the list of possible Clebsch factors of operators from \cite{Antusch:2009gu,Antusch:2013rxa}, we compute all the combinations giving a good value for the double ratio. Although the double ratio is only sensitive to the product $c_y c_z$, permutations of $c_y$ and $c_z$ are considered as different cases since the model predictions will be dependent on the individual values. As will be argued in Section~\ref{sec:parameters}, a change of sign in any of the Clebsch-Gordan coefficients returns an equivalent solution. Therefore only the absolute values $|c_x|$, $|c_y|$ and $|c_z|$ are distinguished. In addition, for $c_x$ we restrict the values to $3$, $9/2$ and $6$ (when we run up  the ratio $y_\mu /y_s $ to the GUT scale, its value becomes roughly $4.5$ if there were no threshold corrections, while threshold effects can readily change this to $3$ and $6$, but not much further) and the values for $c_y$ and $c_z$ are then chosen in a way that the double ratio lies between $9$ and $14$, corresponding to a roughly $2\sigma$ region in Eq.~\eqref{eq:doubleratio-Z}. 
 
We identify in this way the potentially good Clebsch combinations, and list them in Table~\ref{tab1}. This is the list of combinations we shall consider further in the numerical analysis of the models in the next sections.

\renewcommand{\arraystretch}{1.3}
\begin{table}
\begin{align*}
\begin{array}{c@{\kern1.0em}c@{\kern1.0em}c}
\toprule
{c_x,c_y,c_z} & {c_x,c_y,c_z} & {c_x,c_y,c_z} \\
\midrule
{3,\frac{1}{6},\frac{9}{2}} & {\frac{9}{2},\frac{1}{6},9} & {6,\frac{1}{2},6} \\
{3,\frac{1}{6},6} & {\frac{9}{2},\frac{1}{2},3} & {6,\frac{2}{3},\frac{9}{2}} \\
{3,\frac{1}{2},\frac{3}{2}} & {\frac{9}{2},\frac{1}{2},\frac{9}{2}} & {6,\frac{2}{3},6} \\
{3,\frac{1}{2},2} & {\frac{9}{2},\frac{2}{3},3} & {6,1,3} \\
{3,\frac{2}{3},1} & {\frac{9}{2},1,\frac{3}{2}} & {6,\frac{3}{2},2} \\
{3,\frac{2}{3},\frac{3}{2}} & {\frac{9}{2},1,2} & {6,2,\frac{3}{2}} \\
{3,1,\frac{2}{3}} & {\frac{9}{2},\frac{3}{2},1} & {6,2,2} \\
{3,1,1} & {\frac{9}{2},\frac{3}{2},\frac{3}{2}} & {6,3,1} \\
{3,\frac{3}{2},\frac{1}{2}} & {\frac{9}{2},2,1} & {6,\frac{9}{2},\frac{2}{3}} \\
{3,\frac{3}{2},\frac{2}{3}} & {\frac{9}{2},3,\frac{1}{2}} & {6, 6,\frac{1}{2}} \\
{3,2,\frac{1}{2}} & {\frac{9}{2},3,\frac{2}{3}} & {6,6,\frac{2}{3}} \\
{3,\frac{9}{2},\frac{1}{6}} & {\frac{9}{2},\frac{9}{2},\frac{1}{2}} & \\
{3,6,\frac{1}{6}} & {\frac{9}{2},9,\frac{1}{6}} & \\
\bottomrule
\end{array}
\end{align*}
\caption{The list of all combinations of $\mathrm{SU}(5)$ Clebsch-Gordan coefficients (only absolute values), which provide the
 Yukawa double ratio $\frac{y_\mu y_d}{y_s y_e} \approx \Big|\frac{c_x^2}{c_y c_z}\Big|$ in the range between $9$ and $14$. The possible values of these coefficients were taken from the classification in \cite{Antusch:2009gu,Antusch:2013rxa}.}
\label{tab1}
\end{table}
\renewcommand{\arraystretch}{1.0}

\section{Model implementation and analysis}
\label{sec:implementationmodel}
The model is implemented at the GUT scale, using the texture described in the previous section for specific combinations of CG coefficients in the charged lepton Yukawa couplings. In order to compare the observables with experimental data and fit the parameters of the model, we use the MSSM and SM RGEs for the running, including SUSY threshold corrections, with boundary conditions at the GUT scale. The fitting is done by calculating the $\chi^2$ of the observables.

\subsection{Model setup}
\label{sec:modelsetup}
\subsubsection{Texture}
\label{sec:texture}
The texture for the down-type quark and charged lepton Yukawa couplings is stated in Eq.~\eqref{eq:Yukawa-texture-ude}. The matrices are given by
\begin{align}
\YD = \begin{pmatrix}   0 & z & 0 \cr y e^{i \gamma} & x & 0 \cr 0 & 0 & y_b  \end{pmatrix} \,, \quad \YE = \begin{pmatrix}   0 & c_y ye^{i \gamma} & 0 \cr c_z z & c_x x & 0 \cr 0 & 0 & y_{\tau}  \end{pmatrix}\,.
\label{yuk:ydye}
\end{align}
According to the texture in Eq.~\eqref{eq:Yukawa-texture-ude}, the (symmetric) up-type Yukawa matrix is implemented in the following way
\begin{align}
\YU= \mathbf{U}_{23}(\theta^\text{CKM}_{23})\, \mathbf{U}_{12}(\theta^{uL}_{12})\, \text{diag}(y_u,y_c,y_t)\,\mathbf{U}^\textsf{T}_{12}(\theta^{uL}_{12})\,\mathbf{U}^\textsf{T}_{23}(\theta^\text{CKM}_{23})\,,
\label{yuk:yu}
\end{align}
with the unitary matrices
\begin{align}
\mathbf{U}_{23}(\theta^\text{CKM}_{23}) = \begin{pmatrix} 1 & 0 & 0 \\ 0 & \cos{\theta^\text{CKM}_{23}} & \sin{\theta^\text{CKM}_{23}} \\ 0 & -\sin{\theta^\text{CKM}_{23}} & \cos{\theta^\text{CKM}_{23}} \end{pmatrix}\,,\quad
\mathbf{U}_{12}(\theta^{uL}_{12}) = \begin{pmatrix} \cos{\theta^{uL}_{12}} & i\sin{\theta^{uL}_{12}} & 0 \\ i\sin{\theta^{uL}_{12}} & \cos{\theta^{uL}_{12}} & 0 \\ 0 & 0 & 1 \end{pmatrix}\,.
\end{align}
The factor $i$ in $\mathbf{U}_{12}$, which corresponds to a phase $\delta_{12}^{uL}=-\pi/2$, is introduced to get an imaginary $1$-$2$ element in $\YU$, and a potential rotation angle $\theta_{13}^{uL}$ in Eq.~\eqref{yuk:yu} (cf. Eq.~\eqref{eq:general_matrix_2_1}) is chosen equal to zero, such that the $1$-$3$ element is negligible, which realizes the texture in Eq.~\eqref{eq:general_texture_yu_yd} to a very good approximation.\footnote{Although according to Eq.~\eqref{yuk:yu} the $1$-$3$ and $3$-$1$ elements of $\YU$ do not vanish exactly, the relative correction of $\theta_{13}^\text{CKM}$ compared to the texture in Eq.~\eqref{eq:general_texture_yu_yd}, where the two entries are zero, is of order $y_c/y_t$, which is much smaller than the experimental uncertainty.} The values for $y_b$, $y_\tau$, $y_u$, $y_c$, $y_t$ and $\theta^\text{CKM}_{23}$ in Eq.~\eqref{yuk:ydye} and \eqref{yuk:yu} are set to the experimental values at the GUT scale provided in \cite{Antusch:2013jca}.

The CSD2 mechanism provides two choices of flavon VEVs which determine the neutrino Yukawa matrices, i.e. $\YNU^{(102)}$ and $\YNU^{(120)}$, as stated in Eq.~\eqref{yukcsd2}. After integrating out the right-handed neutrinos, the corresponding mass matrices of the left-handed neutrinos are given by (as stated in Eq.~\eqref{eq:Mnu-CSD2-b} and \eqref{eq:Mnu-CSD2-a})
\begin{align}
\mathbf{M}_\nu^{(102)} = m_a \begin{pmatrix} \epsilon e^{i\alpha} & 0 & 2\epsilon e^{i\alpha} \\ 0 & 1 & -1 \\ 2\epsilon e^{i\alpha} & -1 & 1+4\epsilon e^{i\alpha} \end{pmatrix}\,,\quad
\mathbf{M}_\nu^{(120)} = m_a \begin{pmatrix} \epsilon e^{i\alpha} & 2\epsilon e^{i\alpha} & 0 \\ 2\epsilon e^{i\alpha} & 1+4\epsilon e^{i\alpha} & -1 \\ 0 & -1 & 1 \end{pmatrix}\,.
\label{mass:neutrino}
\end{align}
At the SUSY scale, where the MSSM is matched to the SM, the threshold corrections of the Yukawa couplings are implemented according to Eqs.~\eqref{thresh:yu}--\eqref{thresh:yl}.

\subsubsection{Observables}
\label{sec:observables}
With the implementation shown above the model provides $12$ experimentally measured observables, namely the Yukawa couplings $y_e$, $y_\mu$, $y_d$, $y_s$, the CKM angles and CKM Dirac phase $\theta_{12}^\text{CKM}$, $\theta_{13}^\text{CKM}$, $\delta^\text{CKM}$, the PMNS angles $\theta_{12}^\text{PMNS}$, $\theta_{13}^\text{PMNS}$, $\theta_{23}^\text{PMNS}$ and the neutrino mass squared differences $\Delta m^2_\text{21}$, $\Delta m^2_\text{31}$. There also exist other observables, which are in one-to-one correspondence with a parameter (such as the up-type Yukawa couplings, $3$rd generation Yukawa couplings in $\YE$ and $\YD$, and $\theta_{23}^\text{CKM}$) and can be fitted independently; these observable-parameter pairs are not counted. As mentioned earlier, the CSD2 scenario implies normal hierarchy for neutrino masses. Furthermore, the model predicts three observables which are not (or not well) measured: the PMNS Dirac phase $\delta^\text{PMNS}$, the ratio of the Yukawa couplings $\frac{y_d}{y_s}$ and the effective mass $\langle m_{\beta\beta} \rangle$ in neutrinoless double-beta ($0\nu\beta\beta$) decay. Although $\theta_{23}^\text{PMNS}$ is measured by experiment, the range of $\theta_{23}^\text{PMNS}$ predicted by the model for the different combinations of CG coefficients is usually much smaller than the uncertainty in the experimental data. The same holds true for the ratio $\frac{y_d}{y_s}$, which is stable under the RGE running and the SUSY threshold corrections. The general formula for the effective mass $\langle m_{\beta\beta} \rangle$ is given by
\begin{align}
\begin{split}
\langle m_{\beta\beta} \rangle &= \Big{|} \sum_i (\mathbf{U}^\text{PMNS}_{1i})^2 m_{\nu_i} \Big{|} \\
&= c_{12}^2 c_{13}^2 e^{-i\varphi^\text{PMNS}_1} m_{\nu_1} + s_{12}^2 c_{13}^2 e^{-i\varphi^\text{PMNS}_2} m_{\nu_2} + s_{13}^2 e^{-2i\delta^\text{PMNS}} m_{\nu_3}\,,
\end{split}
\label{eq:meff}
\end{align}
with the PMNS matrix $\mathbf{U}^\text{PMNS}$ and the abbreviations $c_{ij}=\cos{\theta_{ij}^\text{PMNS}}$, $s_{ij}=\sin{\theta_{ij}^\text{PMNS}}$. The left-handed neutrino masses are labelled by $m_{\nu_i}$ (where \hbox{$m_{\nu_1} < m_{\nu_2} < m_{\nu_3}$}) and $\varphi^\text{PMNS}_1$, $\varphi^\text{PMNS}_2$ are the two PMNS Majorana phases. Since the neutrino sector contains only two right handed neutrinos, we have $m_{\nu_1}=0$ and consequently $\varphi^\text{PMNS}_1$ is unphysical; $\langle m_{\beta\beta} \rangle$ thus acts as a proxy for the Majorana phase $\varphi^\text{PMNS}_2$.

\begin{figure}
\centering
\includegraphics[width=0.8\textwidth]{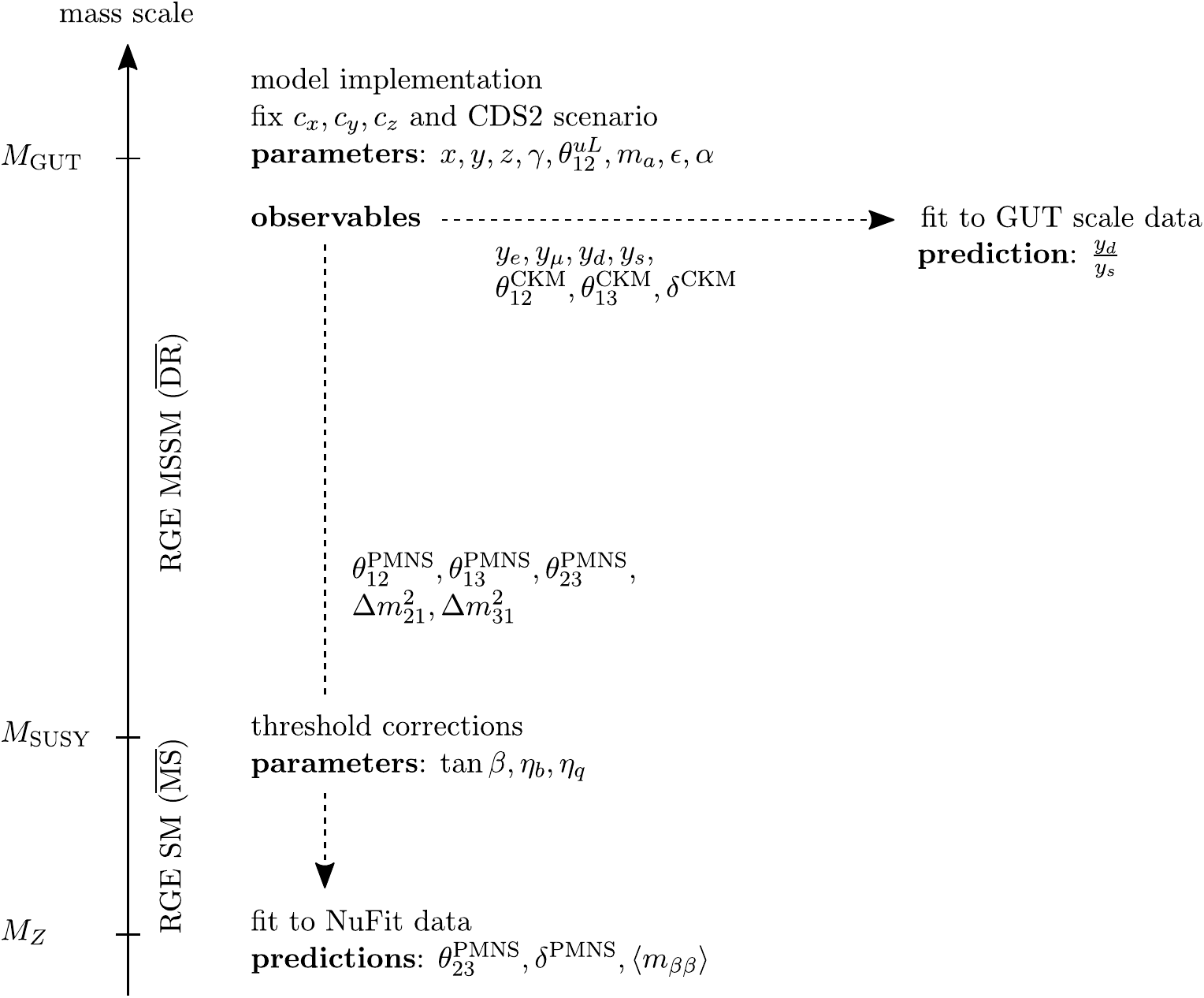}
\caption{Schematic illustration of the model quantities concerning the different mass scales. On the y-axis the three mass scales $M_\text{GUT}$, $M_\text{SUSY}$ and $M_Z$ are indicated, as well as the type of the RGEs needed for the running. At $M_\text{GUT}$ the model is implemented by fixing the CG coefficients, the CSD2 scenario and $8$ of the model parameters. Then $7$ observables corresponding to the Yukawa couplings and the CKM matrix are fitted directly to the experimental values at the GUT scale using the data from~\cite{Antusch:2013jca}. The other $5$ observables, corresponding to the neutrino masses and the PMNS matrix are run down to the $Z$ scale, where they are fitted to the data from NuFIT 3.2 (2018)~\cite{Esteban:2016qun}. At the SUSY scale, the threshold corrections are specified by $3$ parameters and we also switch from the $\overline{\text{DR}}$ to the $\overline{\text{MS}}$ scheme when matching. While the Yukawa ratio of the down and the strange quark is predicted at $M_\text{GUT}$, the other predictions are run down to $M_Z$.}
\label{fig:scales}
\end{figure}

The total $\chi^2$ of the model is given by the sum of the individual $\chi^2$ of each measured observable, which are calculated by using the experimental data. The $\chi^2$ therefore consists of $12$ terms. If the $1\sigma$ experimental range for any of them is asymmetric relative to the central value, we took this into account. An exception is the observable $\theta_{23}^\text{PMNS}$ for which we used the exact $\Delta\chi^2$ function provided by NuFIT 3.2 (2018)~\cite{Esteban:2016qun}. The experimental values for the Yukawa couplings and the CKM parameters are taken at the GUT scale. They are provided in \cite{Antusch:2013jca}, including the corresponding $1\sigma$ errors, as functions of the parameters $\tan{\beta}$, $\eta_b$ and $\eta_q$. The PMNS angles and the neutrino mass squared differences are determined at the $Z$ boson mass scale $M_Z$, where they are fitted to the experimental values from NuFIT 3.2 (2018)~\cite{Esteban:2016qun}. Furthermore, the predictions for the PMNS Dirac phase and the effective mass in $0\nu\beta\beta$ decay are calculated at $M_Z$ too. A schematic illustration of the model quantities at the different scales is shown in Figure~\ref{fig:scales}. For all the observables determined at low scale the change of their values when running them from $M_\text{GUT}$ to $M_Z$ is calculated by using an interpolated data table, whose implementation is discussed in detail in Appendix~\ref{app:rg_running}.  The data table is available under the link stated in \cite{running_data}.

\subsubsection{Parameters}
\label{sec:parameters}
Once the CG coefficients ${c_x,c_y,c_z}$ in the charged lepton Yukawa matrix are fixed, the model contains $11$ parameters according to Eqs.~\eqref{thresh:yu}--\eqref{thresh:yl} and Eqs.~\eqref{yuk:ydye}--\eqref{mass:neutrino}. These parameters are $x$, $y$, $z$, $\gamma$, $\theta_{12}^{uL}$, $m_a$, $\epsilon$, $\alpha$, $\tan{\beta}$, $\eta_b$ and $\eta_q$ (see also Figure~\ref{fig:scales}). In fact, the two parameters $\tan{\beta}$ and $\eta_b$ have only a minor impact via RGE effects on the observables. Thus the $12$ measured observables are basically fitted by $9$ parameters. The parameter and observable counting excludes direct pairs of parameter-observable, where a fit of the pair can be performed independently; thus for the fit of the model less parameters are effectively used than present in Eq.~\eqref{eq:list_parameters_1} and \eqref{eq:list_parameters_2}.

In particular, the parameters $x$, $y$, $z$, $\theta_{12}^{uL}$ and $\eta_q$ are used to fit the four Yukawa couplings in the down-type quark and charged lepton sector, the two CKM angles and the CKM Dirac phase, while $\gamma$, $m_a$, $\epsilon$ and $\alpha$ determine the three PMNS angles and the two neutrino mass squared differences. Furthermore, all parameters are real, as discussed in Section~\ref{sec:model}.

Considering the parametrization in Section~\ref{sec:texture}, it turns out that there is some redundancy in the values of the parameters and CG coefficients, i.e. for certain different parameters and CG coefficients the model retains the same values for the observables. For example, a change of sign $c_y \rightarrow -c_y$ can be compensated by the shift $\gamma \rightarrow \gamma+\pi$. In the same manner $c_x \rightarrow -c_x$ is compensated, and $c_z \rightarrow -c_z$ has no impact on the observables at all. Thus, with no loss of generality all CG coefficients can be chosen positive as assumed in Table~\ref{tab1}. Furthermore, a simultaneous change of sign in $z$ and $\theta_{12}^{uL}$ or in $x$, $y$ and $z$ do not change the observables and a change of sign in $y$ can again be compensated by the shift of $\gamma$ by $\pi$. In order to keep the factor $i$ in the $1$-$2$ element in $\YU$, which predicts a viable CKM Dirac phase, the quantities $\frac{z}{x}$ and $s_{12}^{uL}$ must have the same sign. Therefore $x$, $y$, $z$ and $\theta_{12}^{uL}$ are chosen non-negative in the analysis below.

For the numerical analysis we choose the following parameter ranges:
\begin{align}
\begin{split}
&x,\theta_{12}^{uL} \in [0,0.1]\,,\quad y,z \in [0,0.01]\,,\quad \gamma,\alpha \in [0,2\pi]\,,\quad \epsilon \in [0,1]\,,\quad m_a \in [0,0.1]\,\mathrm{eV}\,, \\
&\tan{\beta} \in [20,50]\,,\quad \eta_b,\eta_q \in [-0.6,0.6]\,,
\end{split}
\end{align}
and the different mass scales are fixed as
\begin{align}
M_Z=91.2\,\mathrm{GeV}\,,\quad M_\text{SUSY}=3\cdot10^3\,\mathrm{GeV}\,,\quad M_\text{GUT}=2\cdot10^{16}\,\mathrm{GeV}\,.
\end{align}

\subsection{Analytical considerations in the lepton sector}
\label{sec:analyticalconsideration}
\begin{figure}
\centering
\includegraphics[scale=0.9]{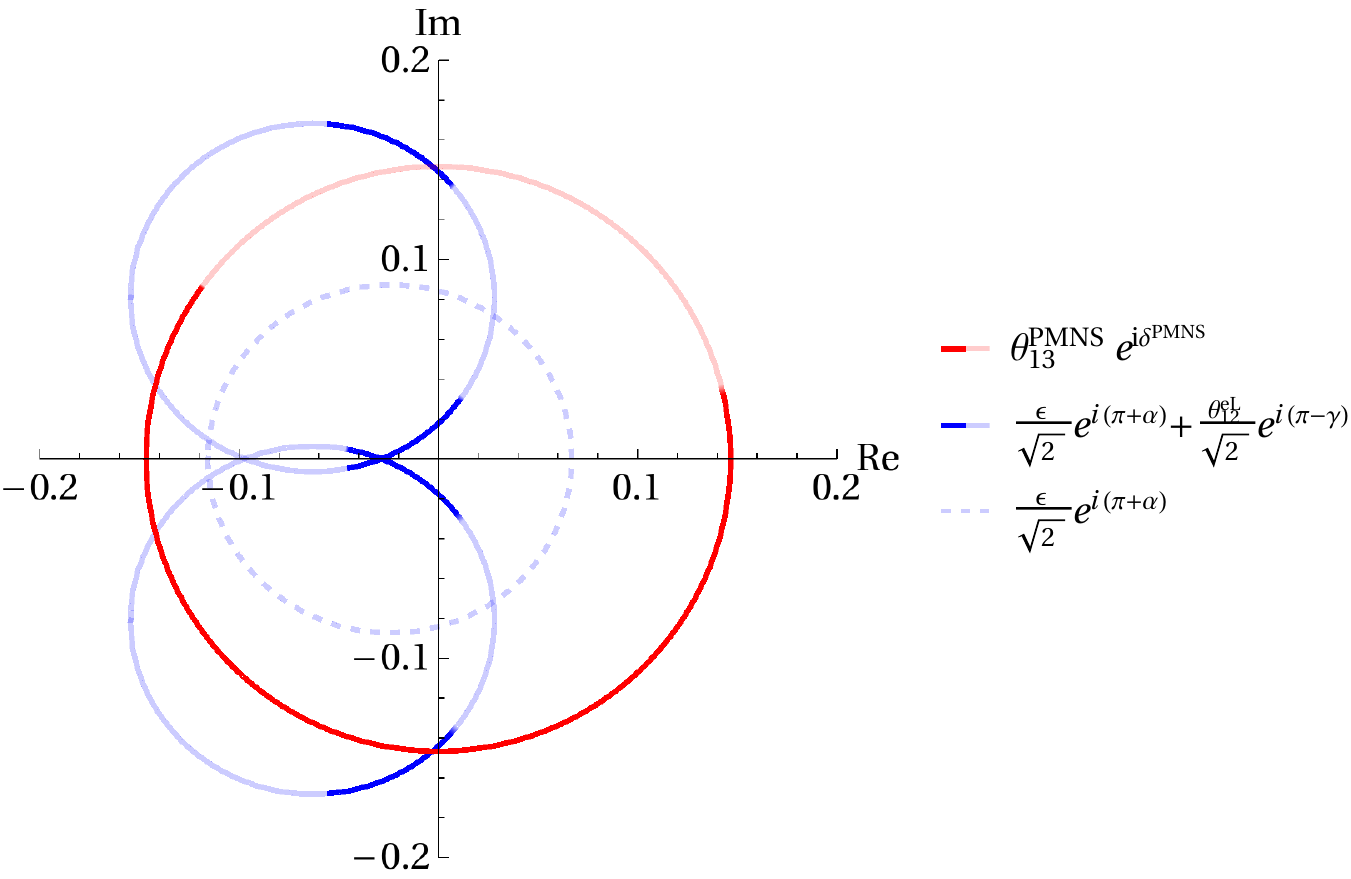}
\caption{The different terms in Eq.~\eqref{eq:graph} are illustrated for $(c_x,c_y,c_z)=(3,\frac{3}{2},\frac{1}{2})$ and the CSD2 scenario $\YNU^{(102)}$, where $\epsilon$ is taken as a function of $\alpha$ induced by neutrino mass fitting. There are two solid blue curves according to the two solutions for $\gamma$ when fitting $\theta_{12}^\text{PMNS}$. The dark blue lines represent the experimental $1\sigma$ range of $\theta_{23}^\text{PMNS}$ for given $\epsilon$ and $\alpha$ using Eq.~\eqref{appeq:23pmns_102}. The radius of the red circle is given by the experimental value of $\theta_{13}^\text{PMNS}$, where the darker part indicates the experimental $3\sigma$ range of $\delta^\text{PMNS}$~\cite{Esteban:2016qun}. The angles $\delta^\text{PMNS}$ and $\alpha$ run from $0$ to $2\pi$, and $\theta_{12}^{eL} = \big|\frac{c_y}{c_x}\frac{y}{x}\big|$ is fixed by the fitting of the Yukawa couplings.}
\label{fig:pmns}
\end{figure}

When fitting the model for fixed CG coefficients to the experimental data the values for $x$, $y$, $z$, $\theta_{12}^{uL}$ and $\eta_q$ are completely fixed in the quark and charged lepton sector by the observables $y_e$, $y_\mu$, $y_d$, $y_s$, $\theta_{12}^\text{CKM}$, $\theta_{13}^\text{CKM}$ and $\delta^\text{CKM}$. In the neutrino sector the parameters $\gamma$, $m_a$, $\epsilon$ and $\alpha$ are then used to fit $\theta_{12}^\text{PMNS}$, $\theta_{13}^\text{PMNS}$, $\theta_{23}^\text{PMNS}$, $\Delta m^2_\text{21}$ and $\Delta m^2_\text{31}$. Once a local minimum of the $\chi^2$ function in the space of these four parameters is found, we expect further local minima with the same or a similar $\chi^2$ value. From an analytical point of view the different minima can be explained in three steps as follows, where for the sake of simplicity running effects are neglected:
\begin{enumerate}
\item \label{item:step1} Consider the neutrino mass matrix in Eq.~\eqref{mass:neutrino}, which is dependent on $m_a$, $\epsilon$ and $\alpha$. The masses of the light neutrinos are given by the singular values of this matrix. When fitting the two neutrino mass squared differences, $m_a$ and $\epsilon$ thus can be expressed as a function of $\alpha$, where they are not sensitive to the sign of $\alpha$.

\item \label{item:step2} Since the left angle $\theta_{12}^{eL} \approx \big|\frac{c_y}{c_x}\frac{y}{x}\big|$ is fixed by the Yukawa couplings and the CKM parameters, the parameter $\gamma$ is fixed too (up to a minus sign) when $\theta_{12}^\text{PMNS}$ is fitted, using Eq.~\eqref{appeq:12pmns_102}:
\begin{align}
\theta_{12}^\text{PMNS} &\approx 35.3^\circ - \frac{\theta_{12}^{eL}}{\sqrt{2}}\cos{\gamma}\,.
\end{align}
This means the solutions for $\gamma$ always come in pairs.

\item The final step in the analysis of minima depends on the CSD2 variant. We shall explicitly state here the argument for the $\YNU^{(102)}$ variant, for which we make use of the identity in Eq.~\eqref{appeq:identity_delta_pmns_102} given in leading order of $\theta_{12}^{eL}$ and $\epsilon$:\footnote{The analysis for the $\YNU^{(120)}$ variant is completely analogous, except that we use Eq.~\eqref{appeq:identity_delta_pmns_120}.}
\begin{align}
\theta_{13}^\text{PMNS} e^{i\delta^\text{PMNS}} \approx \frac{\epsilon}{\sqrt{2}} e^{i(\pi + \alpha)} + \frac{\theta_{12}^{eL}}{\sqrt{2}} e^{i(\pi - \gamma)}\,. \label{eq:graph}
\end{align}

While the left-hand side of the equation is determined by experiment, the right-hand side involves parameters of our model; in particular, the first term on the right can be viewed as a function of $\alpha$ only due to step~\ref{item:step1} and the second term represents merely a constant shift due to step~\ref{item:step2}. A successful fit of the model thus involves finding a good value for $\alpha$, which is the only remaining degree of freedom in Eq.~\eqref{eq:graph}. 

We illustrate the stated features of Eq.~\eqref{eq:graph} in Figure~\ref{fig:pmns}, which is drawn based on data for the model with $(c_x,c_y,c_z)=(3,\frac{3}{2},\frac{1}{2})$ and the CSD2 variant $\YNU^{(102)}$. The left- and right-hand side of the equation are represented by solid red and blue curves in the complex plane, respectively. The red curve is a circle with a radius equal to the central measured value for $\theta_{13}^\text{PMNS}$; the dark red part represents the $3\sigma$ experimental range for $\delta^\text{PMNS}$. The dashed blue line represents the first term on the right, which is to a good approximation shaped as an off-center circle; its exact shape depends on the function $\epsilon(\alpha)$. The solid blue curves in the figure represent the dashed curve shifted by the second term; there are two such curves due to solutions for $\gamma$ coming in $\pm$ pairs. The dark blue curves represent values of $\alpha$ that predict $\theta_{23}^\text{PMNS}$ in the experimental $1\sigma$ range via Eq.~\eqref{appeq:23pmns_102}.

A low $\chi^2$ for $\theta_{13}^\text{PMNS}$ is obtained only when Eq.~\eqref{eq:graph} is satisfied, i.e.~when the red and blue solid curves intersect. Geometrically each blue (approximate) circle can intersect the red circle in either $0$, $1$ (special case when they touch) or $2$ points. We therefore generically expect that if intersection points between the solid blue curves and the red circle exist, there are $4$ of them. Indeed, a geometrical consideration of Figure~\ref{fig:pmns} indicates that once we have found a point with low $\chi^2$ for some values of $(\gamma,\alpha)$, there are further good points for $(-\gamma,-\alpha)$, $(\gamma,2\gamma-\alpha)$ and $(-\gamma,-2\gamma+\alpha)$, where all other parameters are fixed. Since the first two points differ only by a minus sign in $\gamma$ and $\alpha$, they have the same $\chi^2$ value. The same holds true for the last two points. Note that since the dashed blue circle is not centred at the origin, the form stated for the second pair of points is only approximate.

Including the experimental $1\sigma$ range also for $\theta_{23}^\text{PMNS}$~\cite{Esteban:2016qun} (dark blue lines), which is also part of our $\chi^2$ function, $2$ of the $4$ points do not fit anymore. For certain CG coefficients $\theta_{23}^\text{PMNS}$ cannot be fitted well at all; in these cases there is no point with a low $\chi^2$. Otherwise we expect two best fit points with the same $\chi^2$ when fitting the model (with good values for both $\theta_{13}^\text{PMNS}$ and $\theta_{23}^\text{PMNS}$). Usually only one of them provides $\delta^\text{PMNS}$ within the experimental $3\sigma$ range~\cite{Esteban:2016qun} (dark red line). In Figure~\ref{fig:pmns}, the two best fit points predict $\delta^\text{PMNS}$ at around $90^\circ$ and $270^\circ$, the latter one being consistent with the $3\sigma$ range. 
\end{enumerate}
This analytic consideration for minima holds in general: if in a specific model points with low $\chi^2$ exist, we expect $2$ of them, with possibly only $1$ of the $2$ in the correct $\delta_\text{PMNS}$ range. Our numeric results indeed confirm this, as we shall see in the next section.

\section{Results}
\label{sec:numericalresults}
Having specified the implementation of the model at the GUT scale and how observables are compared to experimental data in Section~\ref{sec:implementationmodel}, we investigate in this section the following two questions: First, which tuples of CG coefficients listed in Table~\ref{tab1}, in combination with one of the two CSD2 neutrino Yukawa couplings, are compatible with the experimental data. Second, what are the predictions for $\theta_{23}^\text{PMNS}$, $\delta^\text{PMNS}$, $\frac{y_d}{y_s}$ and $\langle m_{\beta\beta} \rangle$ in these models.

\subsection{Suitable model candidates}
\label{sec:bestfitclebsches}
In Table~\ref{tab2} a complete list of combinations of CG coefficients $(c_x,c_y,c_z)$ that provide a $\chi^2$ less than $15$ is shown. They are ordered with respect to their best fit value and labelled by an integer number. For each tuple $(c_x,c_y,c_z)$ both types of CSD2 neutrino Yukawa couplings ($\YNU^\text{(102)}$ and $\YNU^\text{(120)}$) are considered. According to the analytical discussion in Section~\ref{sec:analyticalconsideration}, the (local) best fit points always come in pairs with opposite sign in $\gamma$ and $\alpha$. In the table only the minima with $\chi^2 < 15$ are shown, and they are distinguished by the labels $a_1,a_2,...$ in the case of $\YNU^\text{(102)}$ and $b_1,b_2,...$ in the case of $\YNU^\text{(120)}$. For each local minimum listed in Table~\ref{tab2}, the values of certain selected quantities are shown:
\begin{itemize}
\item Beside the total $\chi^2$ ($\chi^2_\text{Tot}$), two partial sums are also listed: $\chi^2_\text{q}$ sums over the contributions from the Yukawa couplings and the CKM angles and phase, while $\chi^2_\nu$ sums over the terms for the neutrino mass squared differences and the PMNS angles. As expected, $\chi^2_\text{Yuk}$ usually gives only a minor contribution to $\chi^2_\text{Tot}$, because of the selection of the CG coefficients which was guided by the Yukawa double ratio in Eq.~\eqref{eq:doubleratio-GUT}. Hence, in the models which do not fit well, the main contribution comes from $\chi^2_\nu$, and in particular in most of the cases from $\theta_{23}^\text{PMNS}$. 
\item We list the values of the observables $\theta_{23}^\text{PMNS}$, $\delta^\text{PMNS}$, which are the predictions of each model; we discuss the results in Section~\ref{sec:predictions}. The values of all observables in Table~\ref{tab2} are given at the $Z$-boson mass scale.
\item The best-fit values of the parameters $\gamma$, $\alpha$ and of the $1$-$2$ left angle $\theta_{12}^{eL}$ of the charged leptons are shown. These parameters allow insight into explicitly constructing new models, as for example discussed in the two points below:
	\begin{itemize}
	\item A full flavour model could predict the value for the phase $\gamma$ by a suitable flavon VEV alignment.  A striking feature is that the most promising models (cf.~also Table \ref{tab3}) feature $\gamma$ close to $270^\circ$. As discussed in section \ref{sec:reasoning-phase}, such phases (or phase differences) can emerge in flavour models in various ways, e.g.\ from ``discrete vacuum alignment'' \cite{Antusch:2011sx} combined with spontaneous CP violation, or from other methods for vacuum alignment with non-Abelian discrete symmetries, e.g.\ from a flavon potential as discussed in \cite{Antusch:2013kna}. We would also like to point out the very interesting possibility that the phase difference of $90^\circ$ for the ``phase sum rule mechanism'' and a phase $\gamma = 270^\circ$ could arise from a single imaginary entry in the 2-2 element of $\YE$/$\YD$. Furthermore, in explicit flavour models also the phase $\alpha$ could emerge from the vacuum alignment, and for a specific model candidate one could try to find a model realisation where its value is close to the one given in Table~\ref{tab2} or \ref{tab3}.\footnote{Alternatively, of course, one could try to construct models where $\gamma$ and/or $\alpha$ are kept as free parameters.}
	\item
The values of $\theta_{12}^{eL}$ also allow to explore model building possibilities within the considered $\mathrm{SU}(5)$ GUT setup beyond the CSD2 setup in the neutrino sector. For example, as already mentioned in section \ref{sec:reasoning-Ynu}, one can check whether a tri-bimaximal mixing pattern in the neutrino sector instead of CSD2 could be a valid option, with $\theta_{13}^\text{PMNS}$ generated solely from the charged lepton mixing contribution. The angle $\theta_{13}^\text{PMNS}$ is then predicted as $\theta_{13}^\text{PMNS} = \theta_{12}^{eL}/\sqrt{2}$, and one finds from Table \ref{tab2} that no model candidate would give an acceptable value for $\theta_{13}^\text{PMNS}$.  Analogously, one can also explore whether other leading order mixing patterns in the neutrino sector could be promising for $\mathrm{SU}(5)$ GUT model building in the considered framework.
	\end{itemize}
\end{itemize}

We see from Table~\ref{tab2} that out of the $37$ tuples of CG factors giving potentially viable models listed in Table~\ref{tab1}, only $20$ have minima with $\chi^2 <15$. There are  $10$ combinations of CG coefficients which have an excellent fit of $\chi^2<4$, which means that all observables of the model do not deviate more than $2\sigma$ in total from the experimentally measured values. 

Before the present study, only two representatives from the considered class of models had been studied; model $18$ with the tuple of Clebsch factors $(\tfrac{9}{2},\tfrac{3}{2},\tfrac{3}{2})$ in Ref.~\cite{Antusch:2013wn}, and model $20$ with CG factors $(3,1,1)$ in \cite{Antusch:2017ano}. We can see that the fits of these two models are not as promising given the latest results from NuFIT 3.2 (2018)~\cite{Esteban:2016qun} with a preference for $\theta_{23}^\text{PMNS}> 45^\circ$. 

In Table~\ref{tab3} the model parameters of the $12$ best fit points with lowest $\chi^2$, namely $1a_1$, $3b_2$, $4b_2$, $3a_2$, $6a_2$, $7b_2$, $4a_2$, $8a_2$, $7a_2$, $8b_2$, $9a_2$, $10b_2$, are listed. Note that the models~$2$ and $5$ are not considered in Table~\ref{tab3}, since the tuple of CG coefficients differs only by an overall factor $2$ and $3/2$ compared to the ones in model~$1$ and $4$, respectively. Thus, the predictions for the observables in each of the two pairs of models are essentially the same. 

Another general observation in comparing models is that neither of the CSD2 variants $\YNU^{(102)}$ and $\YNU^{(120)}$ is strongly preferred overall. There exist models where one of the variants is strongly preferred over the other, such as model 6 with CG factors $(\tfrac{9}{2},3,\tfrac{2}{3})$ preferring the $(102)$ flavon VEV alignment; there are also models where there is minimal difference between the variants, such as model 10 with CG factors $(3,\tfrac{3}{2},\tfrac{2}{3})$. Models $15$ to $20$ have a preference for the $(120)$ variant, with the other having $\chi^2 > 15$, and thus not listed. In the list of $12$ best minima in Table~\ref{tab3}, $7$ of them are of the $(102)$ variant and $5$ are of the $(120)$ variant, again showing no strong preference overall. 

\subsection{Predictions}
\label{sec:predictions}
\subsubsection{$\theta_{23}^\text{PMNS}$ and $\delta^\text{PMNS}$}
\label{sec:predictionspmns}
For the $12$ best fit points listed in Table~\ref{tab3} predictions of $\theta_{23}^\text{PMNS}$ and $\delta^\text{PMNS}$ are shown in Figure~\ref{fig:contours1}. In this figure the minimal $\chi^2$ contours in the $\theta_{23}^\text{PMNS}$-$\delta^\text{PMNS}$ plane are shown around each local minimum. For fixed $\theta_{23}^\text{PMNS}$ and $\delta^\text{PMNS}$ the minimal $\chi^2$ is determined by varying the model parameters, with the condition that $\theta_{23}^\text{PMNS}$ and $\delta^\text{PMNS}$ have the correct fixed values. Up to a certain threshold, these $\chi^2$ values are then shown as contours around the chosen best fit point. In order to be in agreement with the experimental data, only best fit points with $\delta^\text{PMNS}$ within the experimental $3\sigma$ range are taken into account. In this way, we demonstrate how well a specific model can be fitted to the known values of the SM parameters assuming a certain $\theta_{23}^\text{PMNS}$ and $\delta^\text{PMNS}$ prediction, showing in which $\theta_{23}^\text{PMNS}$-$\delta^\text{PMNS}$ regions the models work well.

For a given model the range of $\theta_{23}^\text{PMNS}$ with low $\chi^2$, defined by the corresponding plot in Figure~\ref{fig:contours1}, is in most of the cases much smaller than the experimental $3\sigma$ range, given by the interval $[40.3^\circ,51.1^\circ]$~\cite{Esteban:2016qun}. This implies, although $\theta_{23}^\text{PMNS}$ is used to fit the parameters, that the models make distinct predictions for this observable. More accurate measurements of $\theta_{23}^\text{PMNS}$ in future experiments can distinguish between the different models. Furthermore, all models predict the ranges for $\delta^\text{PMNS}$ within around $230^\circ$ and $290^\circ$, which is quite restricted compared the current experimental $3\sigma$ range, given by $[144^\circ,374^\circ]$~\cite{Esteban:2016qun}. Thus, independent of the choice of the CG coefficients and the CSD2 variant, the class of model under consideration delivers a prediction for the PMNS Dirac phase, which can be tested by future experiments. 

To illustrate the above consideration further, all the plots listed in Figure~\ref{fig:contours1} have been combined into one plot in Figure~\ref{fig:contours2}, where the predictions of all the models can be compared in the $\theta_{23}^\text{PMNS}$-$\delta^\text{PMNS}$ plane, together with the experimental $3\sigma$ ranges of the two quantities. We see that all minimal $\chi^2$ regions of models fall onto an almost horizontal trend line; this implies that a future more precise $\theta_{23}^\text{PMNS}$ measurement can indeed further reduce the set of viable models if not outright discriminate between them\footnote{Future measurements by the DUNE experiment, for example, shall determine $\theta_{23}^\mathrm{PMNS}$ with a precision of less than  $1^\circ$, and $\delta^\mathrm{PMNS}$ with a precision of ${\cal O}(10^\circ)$ \cite{Abi:2018dnh,Abi:2018alz,Abi:2018rgm}, which allows for precision model testing.}, while the rough range of $\delta^\text{PMNS}$ is a prediction of the entire class.  There is a slight positive trend noticeable that models with a higher predicted $\theta_{23}^\text{PMNS}$ also predict a slightly higher $\delta^\text{PMNS}$.

\subsubsection{Ratio of $y_d$ and $y_s$}
\label{sec:predictionsyukawaratio}
In order to compute the $1\sigma$ highest posterior density (HPD) interval of the ratio $\frac{y_d}{y_s}$ the Markov chain Monte Carlo (MCMC) method is used. For the different combinations of CG coefficients listed in Table~\ref{tab2} we perform a Markov chain, where among others the posterior density of $\frac{y_d}{y_s}$ is calculated. The posterior density of this ratio only depends on the choice of the CG coefficients but not on the neutrino Yukawa coupling. In Figure~\ref{fig:ratio} the $1\sigma$ HPD intervals for each of the models are indicated as red lines. In addition, the experimental central value of $\frac{y_d}{y_s}$ is indicated by a dotted line and the regions outside the experimental $1\sigma$ range are represented by grey areas. Note that the values for the ratio $\frac{y_d}{y_s}$ in the Markov chain are computed at the GUT scale. Since this ratio is stable under the RGE running and the SUSY threshold corrections, the calculated values can be compared directly with the experimental value at the $M_Z$ scale.

Figure~\ref{fig:ratio} shows that the predicted range of $\frac{y_d}{y_s}$ for a given model is much smaller than the $1\sigma$ experimental range given by $5.06^{+0.78}_{-0.42} \cdot 10^{-2}$~\cite{Antusch:2013jca}. Since different models predict different ranges, more accurate future measurements of the masses $m_d$ and $m_s$, and consequently also of the Yukawa couplings $y_d$ and $y_s$, have the potential to distinguish between different models. The small ranges of $1\sigma$ HPD intervals of $\frac{y_d}{y_s}$ for each model in Figure~\ref{fig:ratio} can be explained as follows: Once the CG coefficients in the charged lepton Yukawa matrix are fixed, the double ratio $d = \frac{y_\mu y_d}{y_e y_s} \approx \Big|\frac{c_x^2}{c_y c_z}\Big|$ given in Eq.~\eqref{eq:doubleratio-GUT} is fixed too in leading order. Since in addition $y_e$ and $y_\mu$ have very small experimental uncertainties, much smaller than $y_d$ and $y_s$, the ratio $\frac{y_d}{y_s}$ is much more constrained in our models than in the experiment.

\subsubsection{Effective mass in $0\nu\beta\beta$ decay}
\label{sec:predictionsmeff}
Once the CG coefficients and the CSD2 variant is chosen, all parameters in the PMNS matrix and in the left-handed neutrino masses are predicted, including the one Majorana phase (there is only one, since the lightest left-handed neutrinos in our setup was taken massless). Therefore Eq.~\eqref{eq:meff} implies that the effective mass in neutrinoless double-beta decay is predicted too; we shall show results for the experimentally more interesting quantity of effective mass rather than for the Majorana phase. 

The $1\sigma$ HPD interval of $\langle m_{\beta\beta} \rangle$ is determined by calculating the posterior density using the MCMC method.  For the twelve best fit points with lowest $\chi^2$ listed in Table~\ref{tab3}, the $1\sigma$ HPD intervals of the effective mass are shown in Figure~\ref{fig:meff} as red lines.

Different combinations of CG coefficients $(c_x,c_y,c_z)$ and CSD2 scenarios ($\YNU^\text{(102)}$ or $\YNU^\text{(120)}$) predict different ranges for $\langle m_{\beta\beta} \rangle$ as shown in Figure~\ref{fig:meff}. Furthermore, all predictions lie roughly in the interval $[2.5,4.0]\cdot10^{-3}\,\mathrm{eV}$. This means the class of model under consideration predicts a well defined range for the effective mass, independent of the choice of the CG coefficients and of the CSD2 scenario. A precise measurement of $\langle m_{\beta\beta} \rangle$ would have the potential to distinguish different models, but unfortunately this is far beyond the reach of currently planned experiments, which have an upper detectable limit of around $0.1\,\mathrm{eV}$ (e.g.~see Table II in \cite{Agostini:2018tnm}).

\subsubsection{SUSY threshold parameter $\eta_q$}
\label{sec:predictionsetaq}
The SUSY threshold parameter $\eta_q$ is actually one of the input parameters we fit. A complete model involving SUSY breaking and a prediction of the SUSY spectrum would need to reproduce, however, the correct threshold effect in the first $2$ fermion families. For this reason, we can consider the $\eta_q$ value also as one of the predictions, despite it not being directly observable experimentally.

We already stated in Section~\ref{sec:GUT-operators} that the $\eta_q$ value is linked to the Clebsch coefficient $c_x$, which determines the ratio $y_\mu/y_s$ at the GUT scale. Using SM and MSSM RGEs with no SUSY threshold corrections, the GUT scale value of $y_\mu/y_s$ is approximately $4.5$, suggesting that any deviation of the Clebsch factor $c_x$ from $4.5$ will need to be compensated by $\eta_q$. This requirement picked only $3$,$\tfrac{9}{2}$ and $6$ as suitable $c_x$ candidates (involving the possibilities of raising/lowering the $y_\mu/y_s$ ratio by $\pm33\%$ ), implying the predicted values of $\eta_q$ to be approximately $-0.33$, $0$ and $+0.33$, respectively. We confirm this expectation with the results in Table~\ref{tab3}.

\section{Summary and Conclusions}
\label{sec:conclusions}
In this paper, we have systematically investigated the predictions of a novel class of supersymmetric $\mathrm{SU}(5)$ GUT flavour models with Constrained Sequential Dominance 2 (CSD2) in the neutrino sector. CSD2 is an attractive building block for flavour model building because it predicts a non-zero leptonic mixing angle $\theta_{13}^\text{PMNS}$, a deviation of $\theta_{23}^\text{PMNS}$ from $\pi /4$, as well as a leptonic Dirac CP phase $\delta^\text{PMNS}$, which is directly linked to the CP violation relevant for generating the baryon asymmetry via the leptogenesis mechanism.

When embedded into a predictive $\mathrm{SU}(5)$ GUT setup, the CSD2 predictions in the neutrino sector are modified in a calculable way by a charged lepton mixing contribution, which is determined by the $\mathrm{SU}(5)$ relations between the charged lepton and down quark Yukawa matrices $\YE$ and $\YD$, respectively. 
The $\mathrm{SU}(5)$ quark-lepton relations in turn depend on GUT operators responsible for generating the entries of fermion Yukawa matrices. Under the assumption of single operator dominance, the choice of GUT operators and consequently the associated Clebsch-Gordan coefficients directly govern the ratios between the entries of $\YE$ and $\YD$ \cite{Antusch:2009gu,Antusch:2013rxa}.

Furthermore, another model building ingredient is the ``phase sum rule mechanism'' \cite{Antusch:2009hq}, used to obtain a valid scheme for CP violation in the quark sector which leads to the prediction of a right-angled unitarity triangle with $\alpha_\mathrm{UT} = 90^\circ$ and thus to a prediction $\delta^\text{CKM}= 1.188\pm 0.016$ in good agreement with the allowed experimental range.

This chosen setup defines the class of models under consideration, with a specific member defined by a $3$-tuple of Clebsch-Gordan factors between $\YD$ and $\YE$ in the $1$-$2$ block and the choice of the CSD2 variant. Once these choices are made, and once concrete values are given to model parameters, all the SM fermion sector parameters are determined: this includes the masses, as well as the mixing angles and CP violating phases of both the CKM and PMNS matrix.

Making use of the approximately invariant double ratio $\tfrac{y_d}{y_s}/\tfrac{y_e}{y_\mu}$, we can narrow down the list of potentially viable Clebsch factors of the model class to $37$; this list is given in Table~\ref{tab1}. For each of the viable $37$ candidates, and for each of the $2$ CSD2 variants, we performed a fit of parameters by minimizing the $\chi^2$ for the observables, thus identifying which models can be viable in at least some part of their parameter space; the minimization results for this are gathered in Table~\ref{tab2}, where all (local) minima with $\chi^2<15$ are listed, with the complete information on the input parameters for the $12$ best minima given in Table~\ref{tab3}. The goal of this study was to systematically explore the predictions of the whole model class to identify the most promising candidates for future model building; up to now only two representatives from this class of models had been studied in Ref.~\cite{Antusch:2013wn,Antusch:2017ano}. 

A general observation from the results in Tables~\ref{tab2} and~\ref{tab3} is that while there may be  a preference for the CSD2 variant $\YNU^{(102)}$ or $\YNU^{(120)}$ for an individual model, there is no strongly preferred overall variant across all models. 

\begin{figure}
\vspace{-1cm}
\centering
\includegraphics[width=0.3\textwidth]{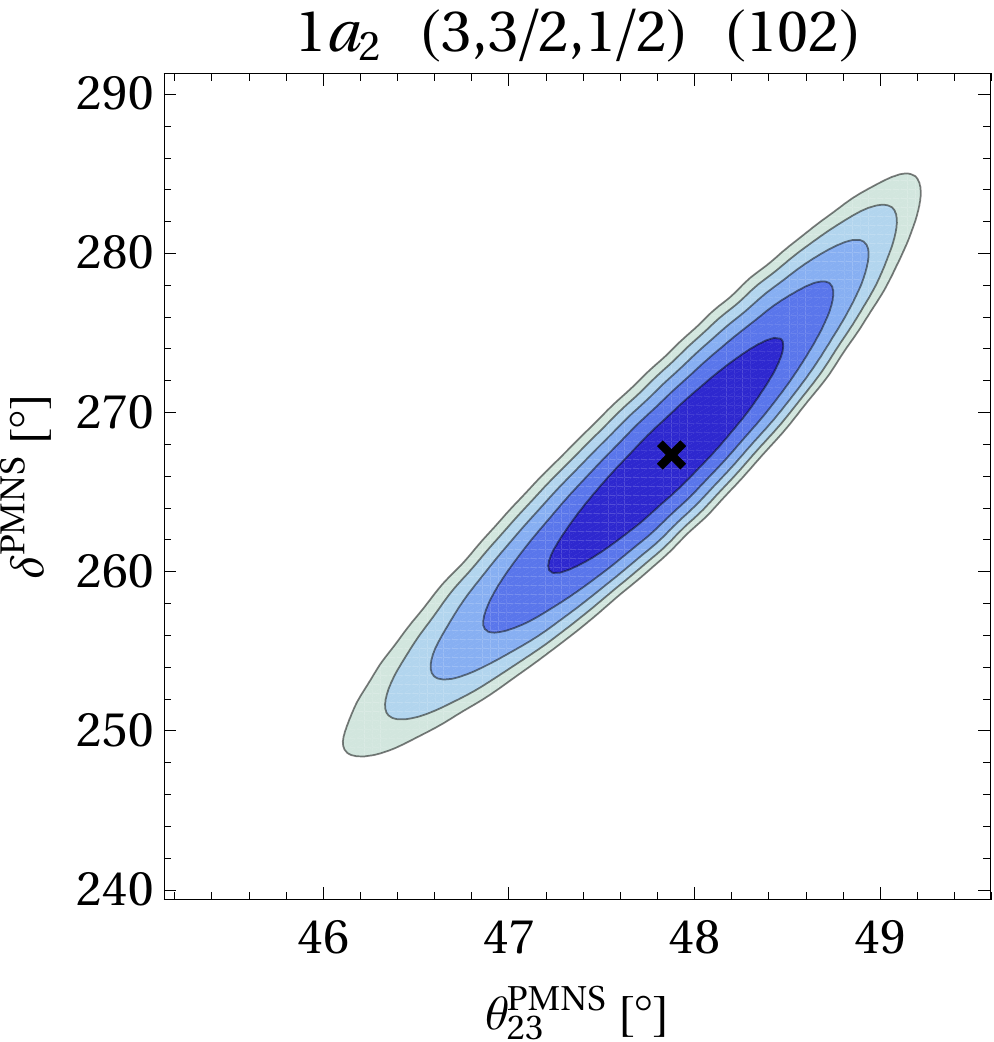} \hspace{0.1cm}
\includegraphics[width=0.3\textwidth]{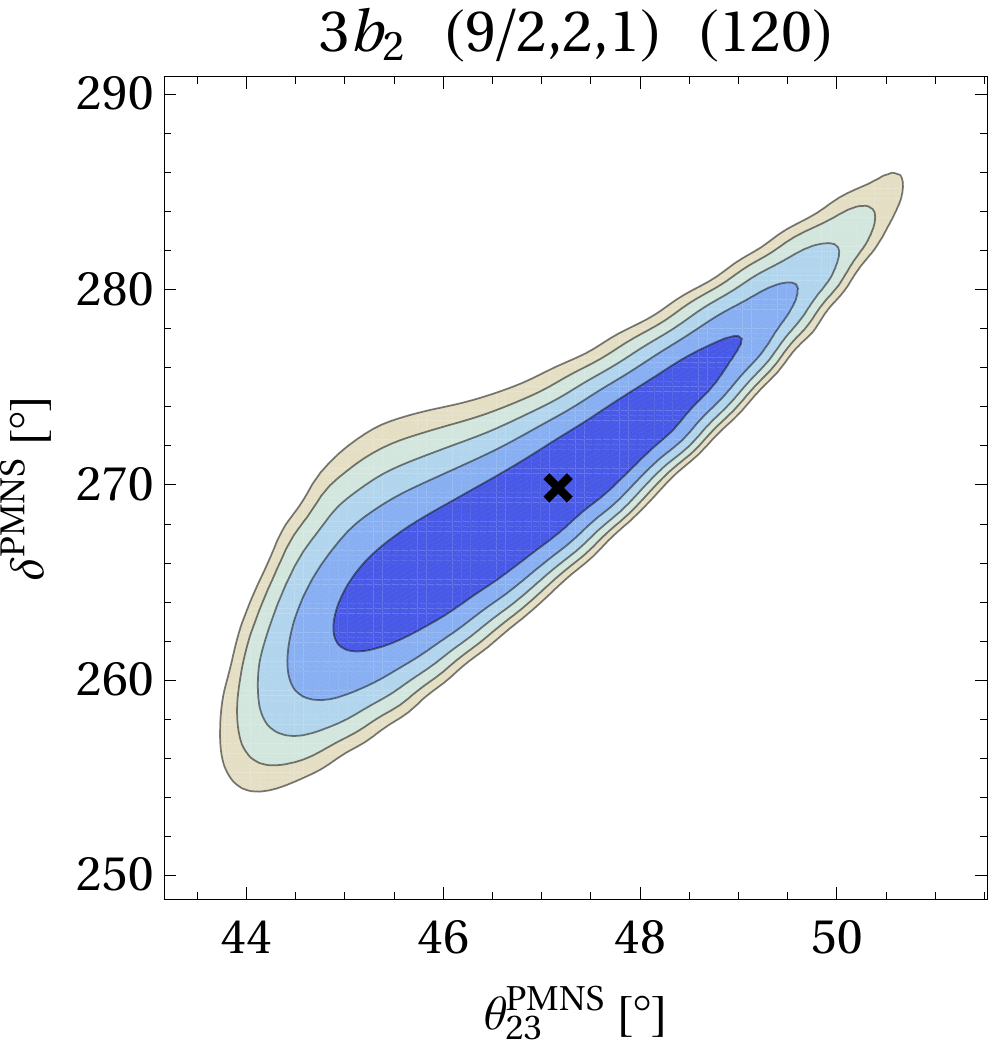} \hspace{0.1cm}
\includegraphics[width=0.3\textwidth]{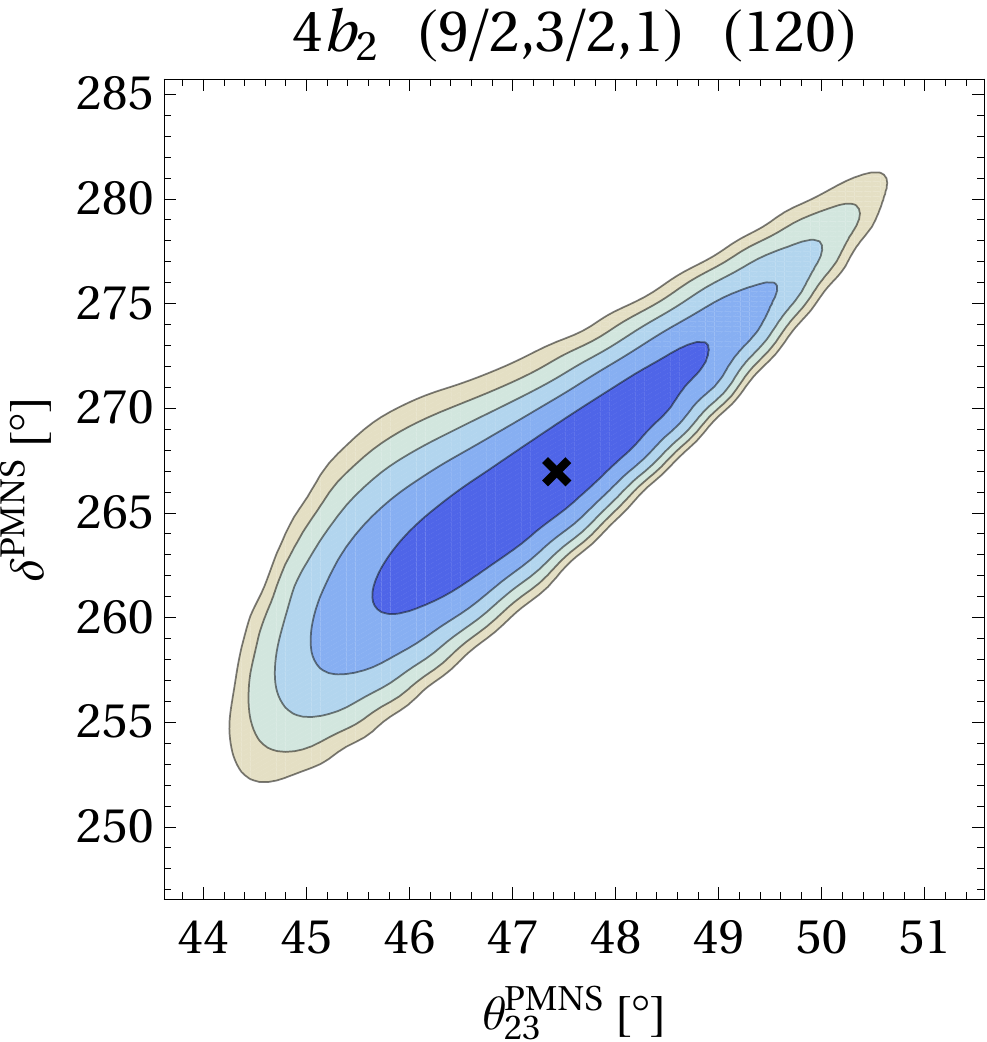}\\ \vspace{0.2cm}
\includegraphics[width=0.3\textwidth]{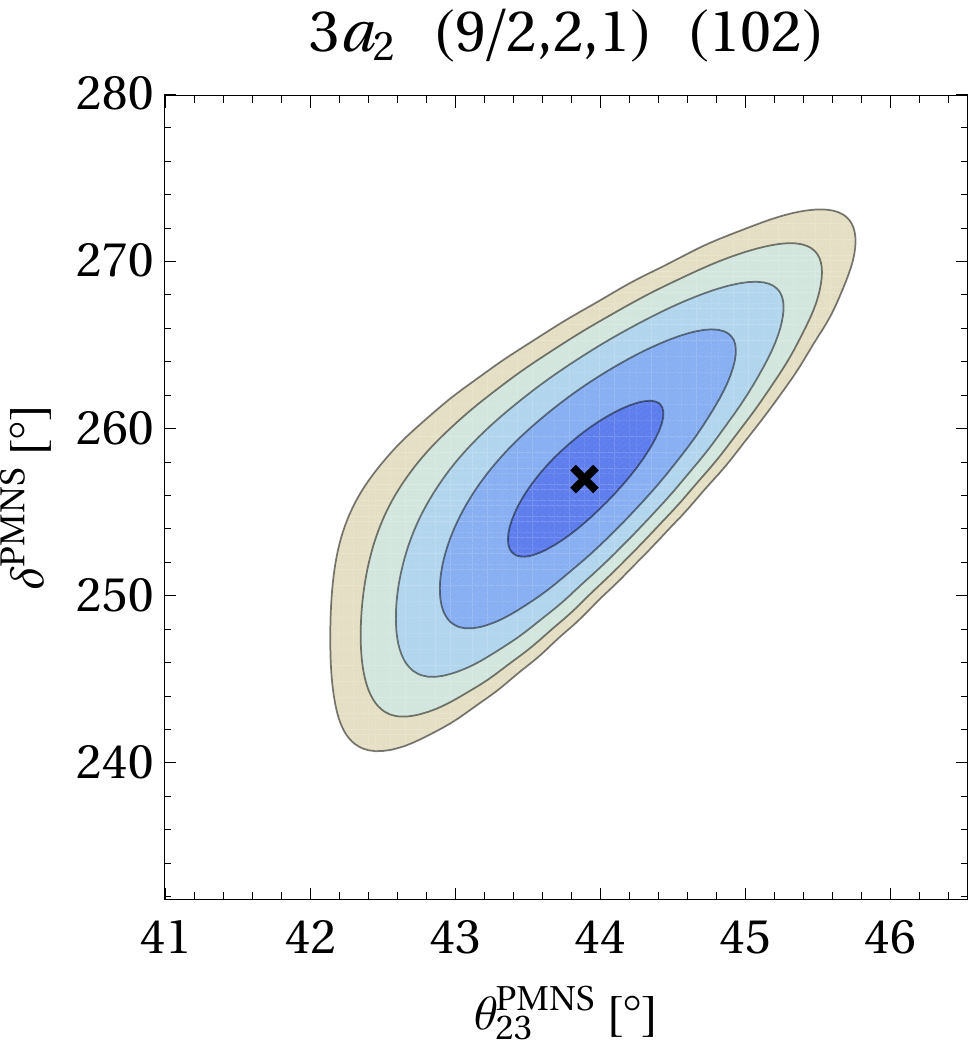} \hspace{0.1cm}
\includegraphics[width=0.3\textwidth]{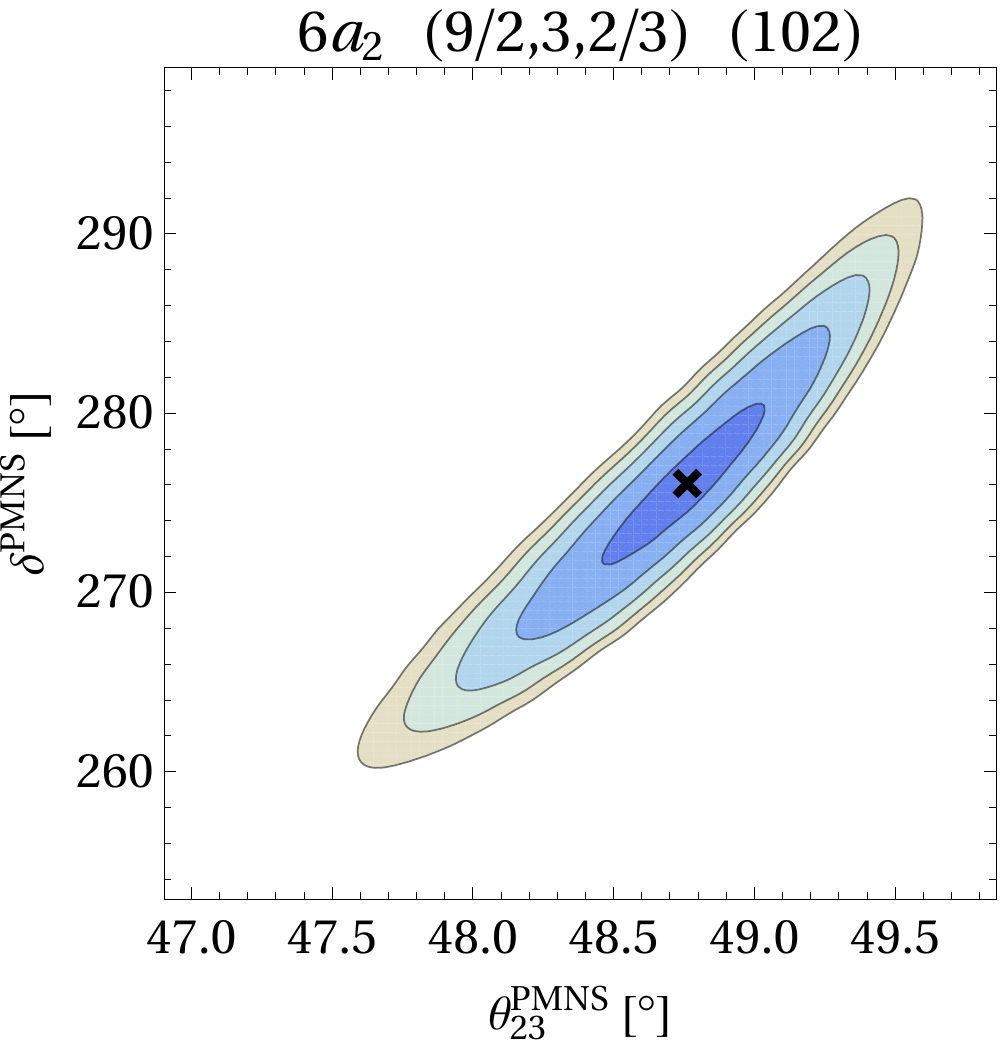} \hspace{0.1cm}
\includegraphics[width=0.3\textwidth]{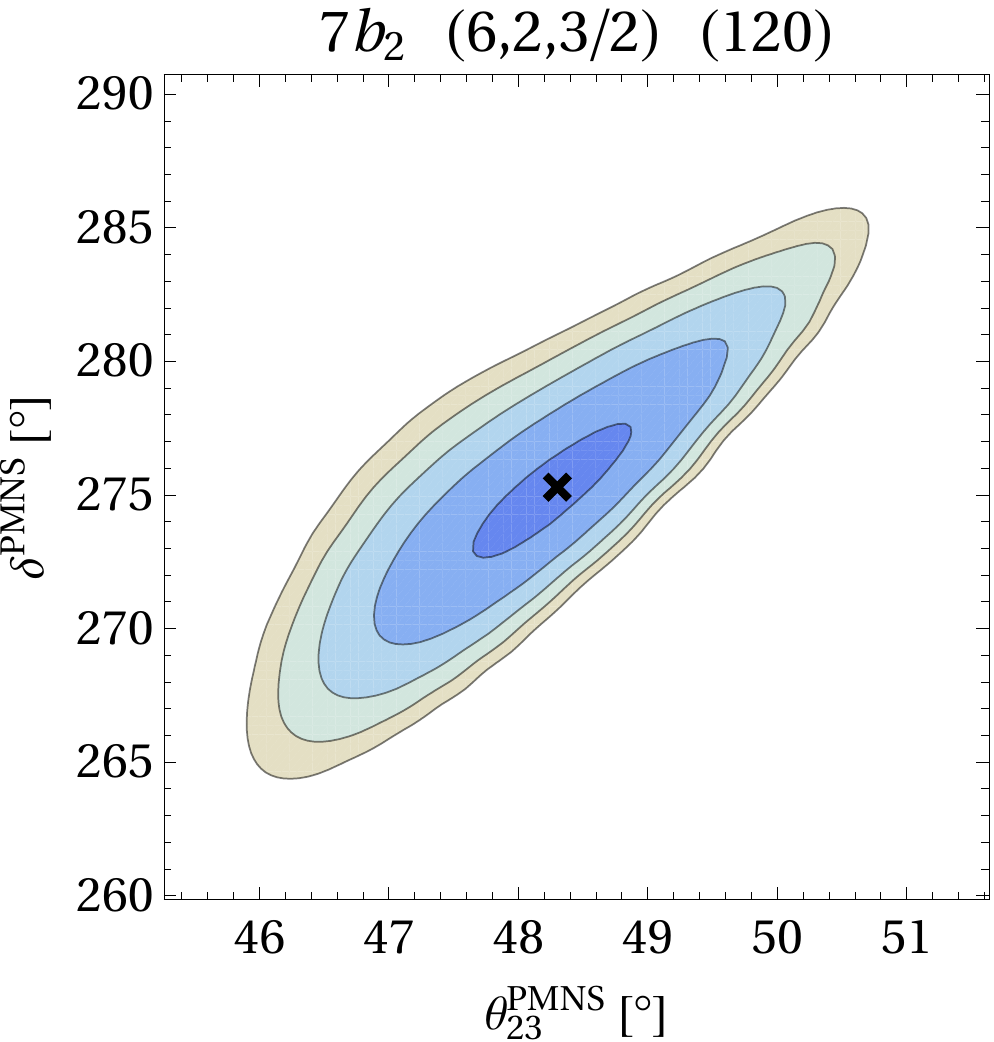}\\ \vspace{0.2cm}
\includegraphics[width=0.3\textwidth]{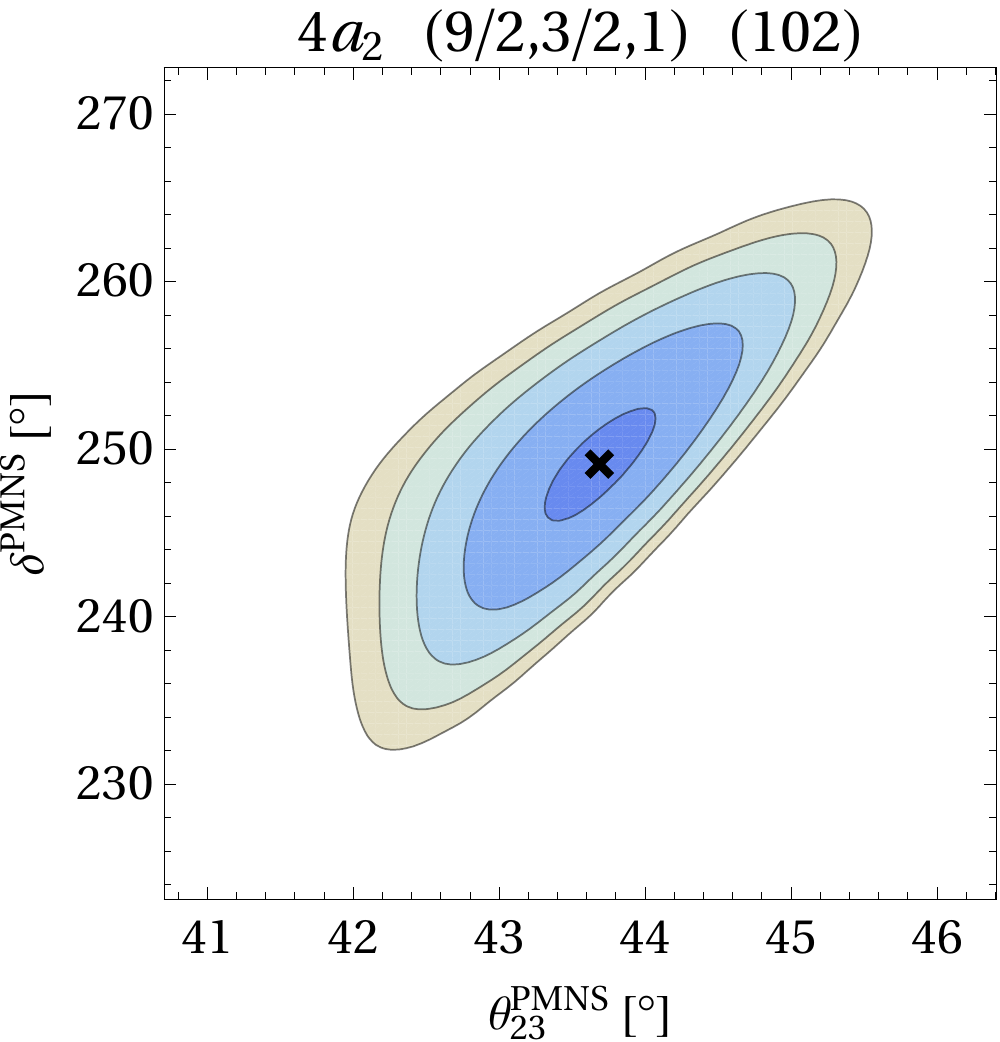} \hspace{0.1cm}
\includegraphics[width=0.3\textwidth]{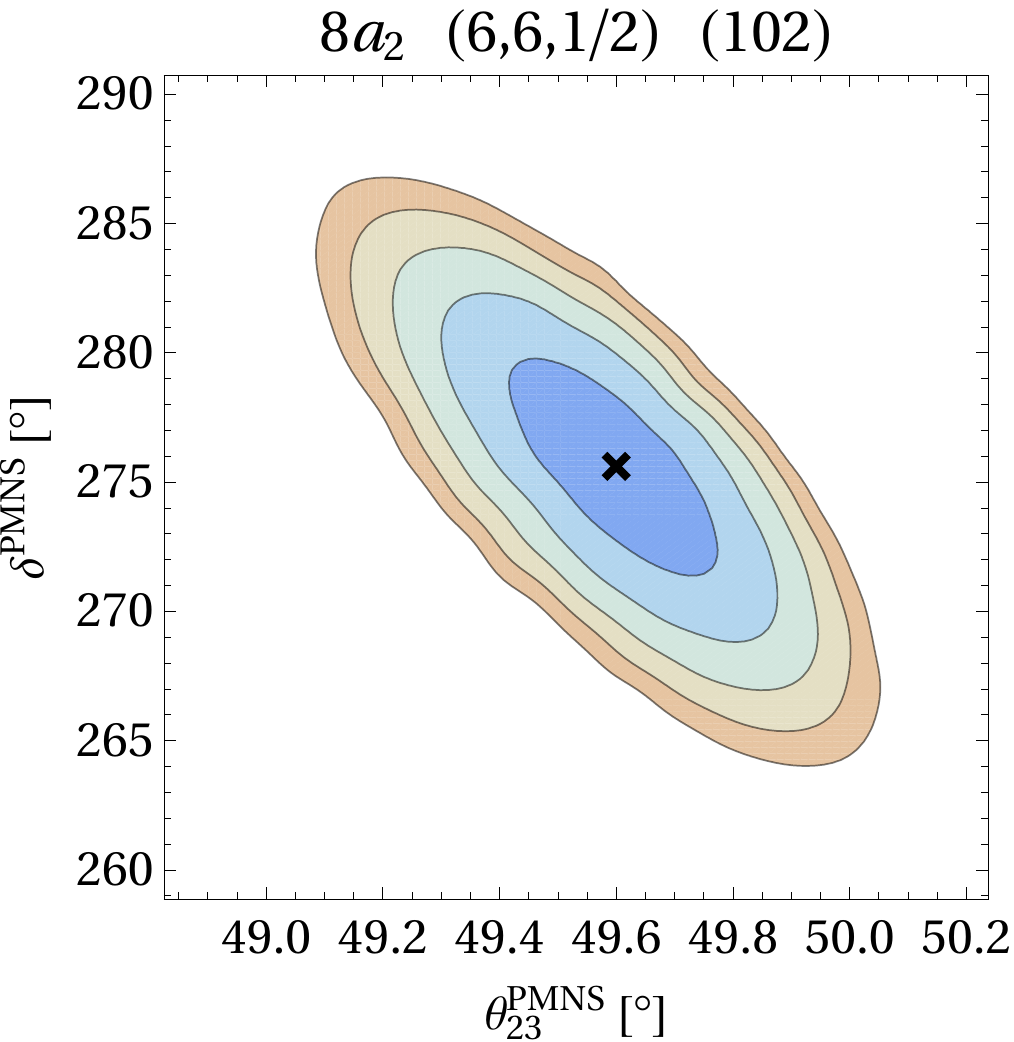} \hspace{0.1cm}
\includegraphics[width=0.3\textwidth]{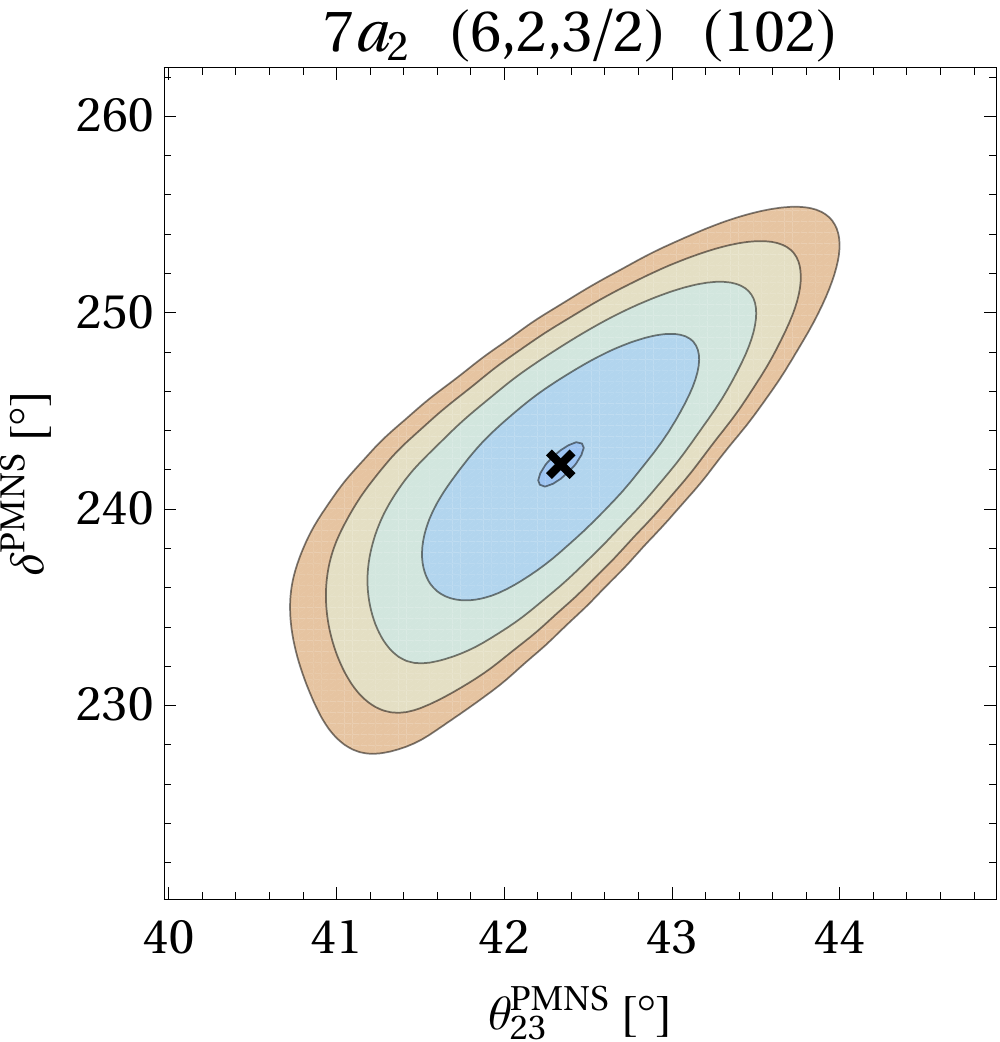}\\ \vspace{0.2cm}
\includegraphics[width=0.3\textwidth]{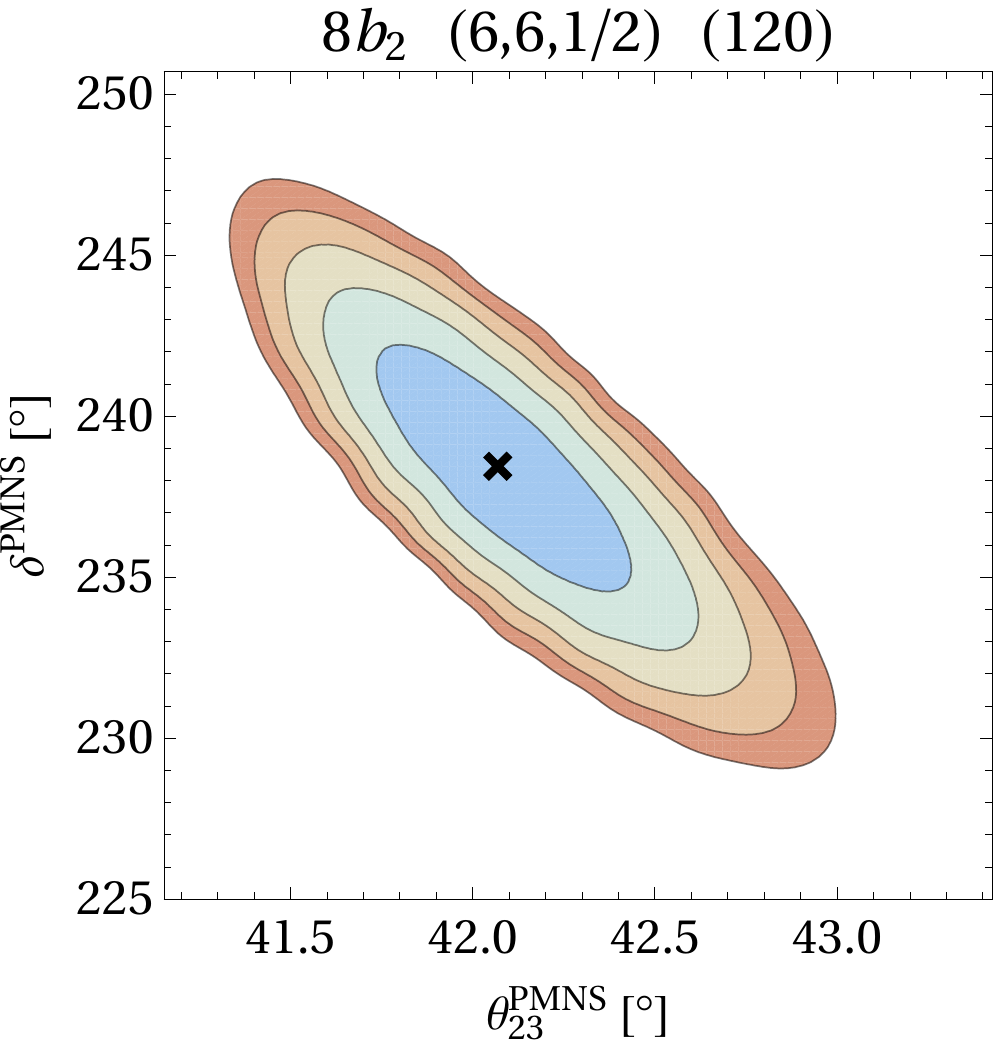} \hspace{0.1cm}
\includegraphics[width=0.3\textwidth]{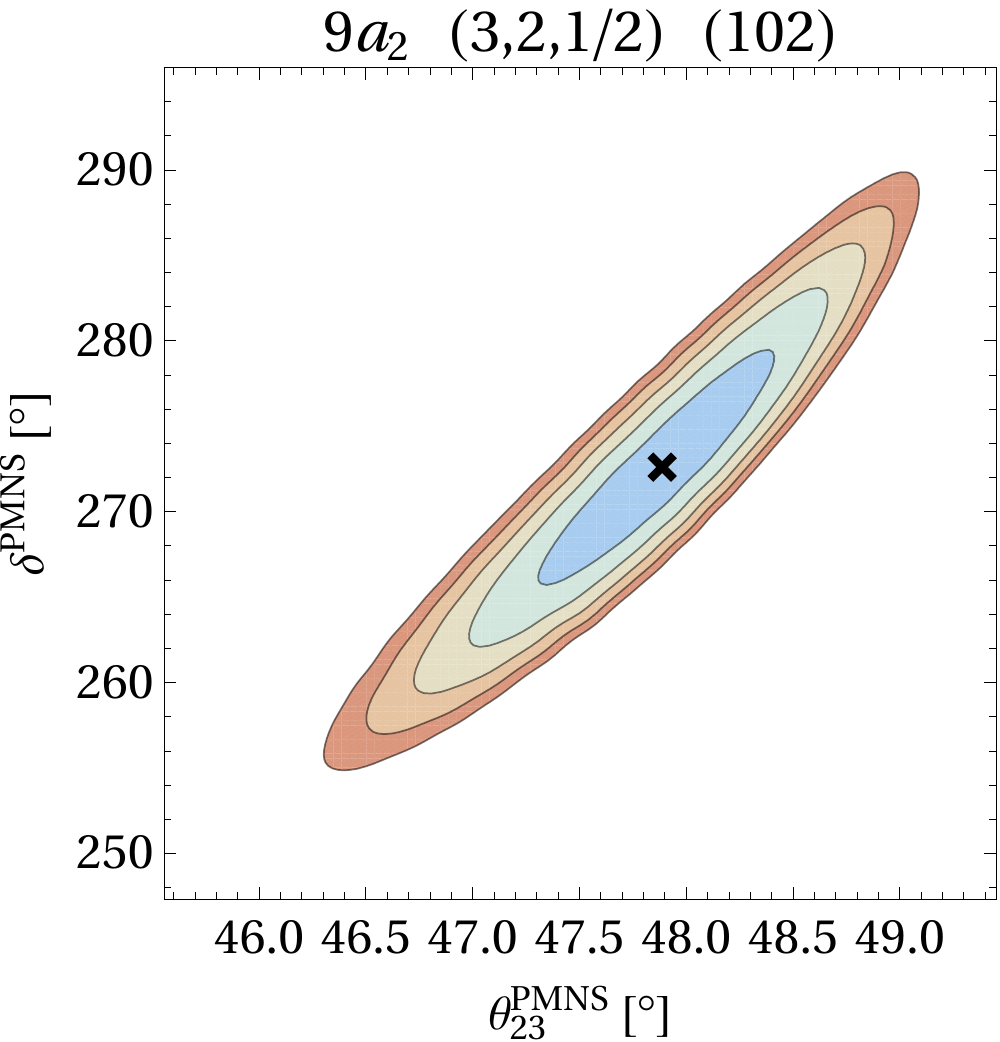} \hspace{0.1cm}
\includegraphics[width=0.3\textwidth]{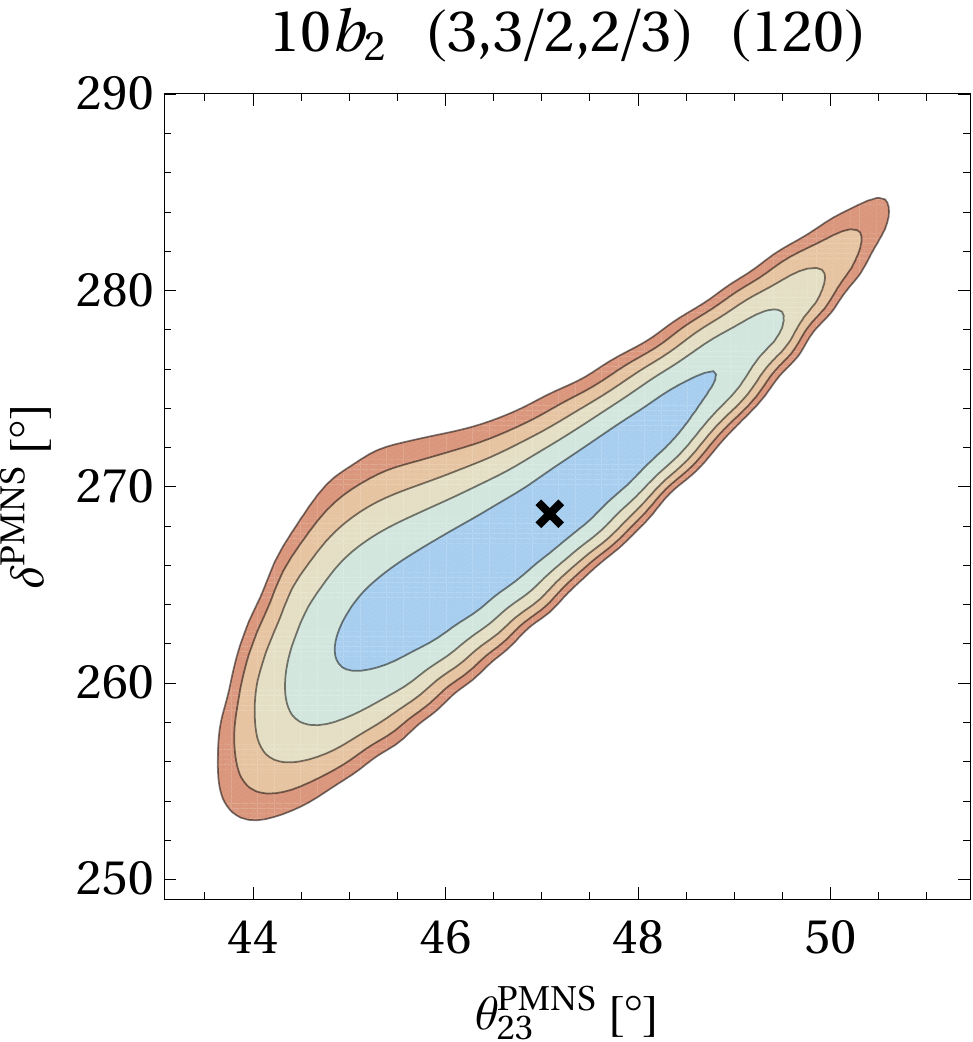}\\ \vspace{0.2cm}
\includegraphics[width=0.25\textwidth]{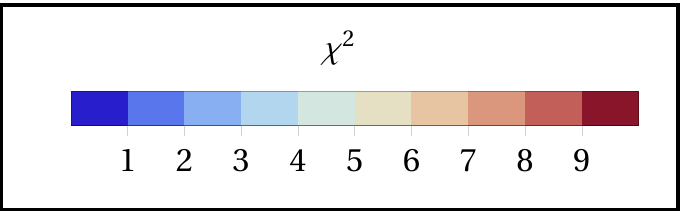}
\caption{Minimal $\chi^2$ contours of the best fit models in the $\theta_{23}^\text{PMNS}$-$\delta^\text{PMNS}$ plane. From top left to bottom right the $12$ best fit points with lowest $\chi^2$ from Table~\ref{tab3} are presented. In each plot the minimal $\chi^2$ for fixed $\theta_{23}^\text{PMNS}$ and $\delta^\text{PMNS}$ is plotted as contours around the local minimum, indicated by a black cross. Beside the chosen CG coefficients $(c_x,c_y,c_z)$ and the CSD2 variant ($\YNU^{(102)}$ or $\YNU^{(120)}$), the title of each plot contains a label specifying the best fit point according to Table~\ref{tab3}. Only local minima with $\delta^\text{PMNS}$ within the experimental $3\sigma$ range are chosen.}
\label{fig:contours1}
\end{figure}

\begin{figure}
\centering
\includegraphics[width=0.7\textwidth]{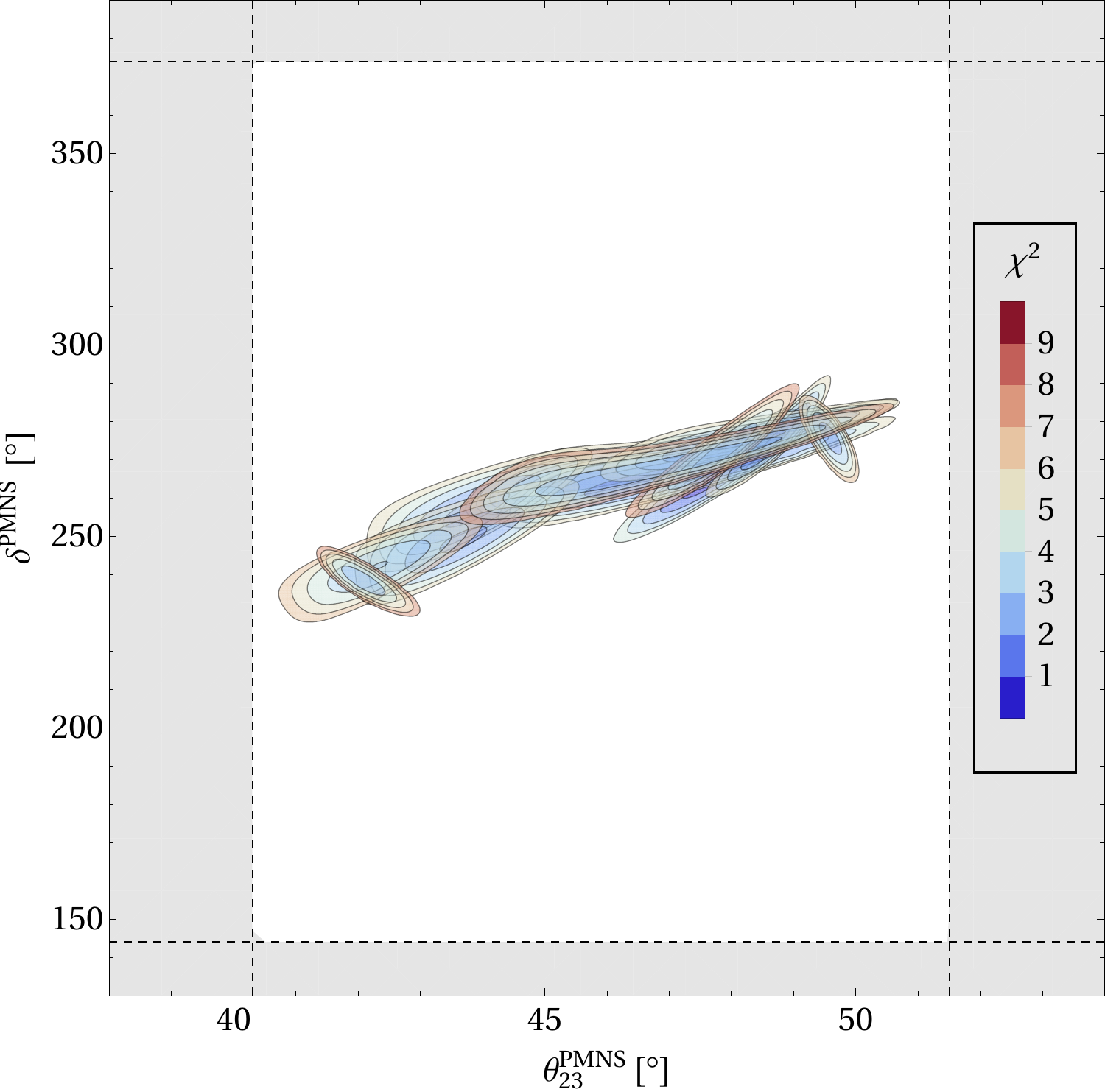}
\caption{Summary of the minimal $\chi^2$ contours of the best fit models in the $\theta_{23}^\text{PMNS}$-$\delta^\text{PMNS}$ plane. The figure shows the combined $\chi^2$ contours of all the plots in Figure~\ref{fig:contours1}. The grey areas represent the regions outside the experimental $3\sigma$ ranges of $\theta_{23}^\text{PMNS}$ and $\delta^\text{PMNS}$ which are given by the intervals $[40.3^\circ,51.5^\circ]$ and $[144^\circ,374^\circ]$, respectively~\cite{Esteban:2016qun}.}
\label{fig:contours2}
\end{figure}

\begin{figure}
\centering
\includegraphics[width=0.8\textwidth]{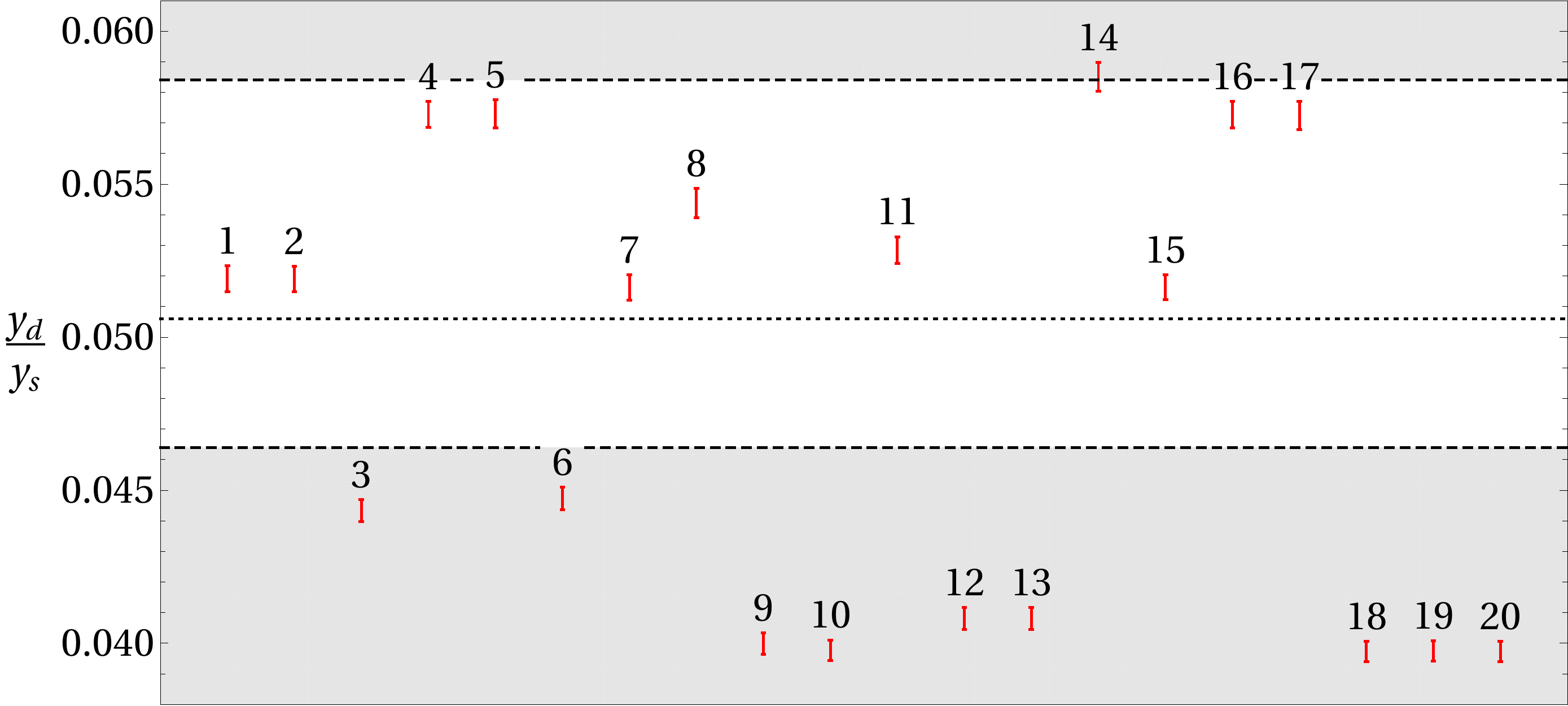}
\caption{Predictions for the Yukawa ratio $y_d/y_s$. For each combination of CG coefficients listed in Table~\ref{tab2} the $1\sigma$ HPD intervals for $y_d/y_s$ are shown as red lines. The HPD intervals do not depend on the choice of the CSD2 scenario. The dotted line indicates the experimental central value of the Yukawa ratio and the grey areas represent the regions outside the experimental $1\sigma$ range, given by $y_d/y_s=5.06^{+0.78}_{-0.42} \cdot 10^{-2}$~\cite{Antusch:2013jca}.}
\label{fig:ratio}
\end{figure}

\begin{figure}
\centering
\includegraphics[width=0.8\textwidth]{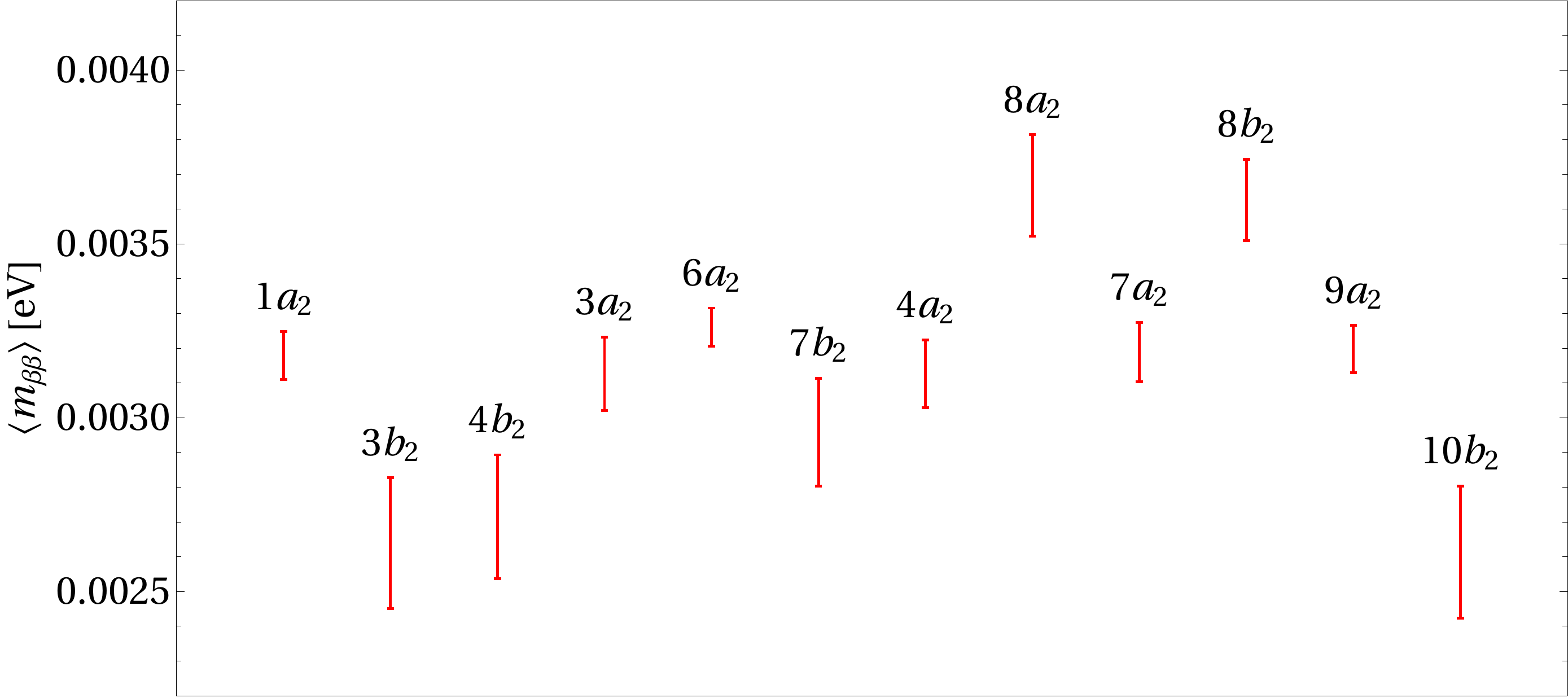}
\caption{Predictions for the effective mass $\langle m_{\beta\beta}\rangle$ in neutrinoless double-beta decay in the best fit models. For the $12$ best fit points with lowest $\chi^2$ listed in Table~\ref{tab3} the $1\sigma$ HPD intervals for $\langle m_{\beta\beta}\rangle$ are shown as red lines.}
\label{fig:meff}
\end{figure}

In the fitting procedure there are $11$ input parameters, which includes $\tan\beta$ and $\eta_b$ with only indirect and minor effects on observables; the $\chi^2$ function we minimize has $12$ terms associated to observables.\footnote{The parameter and observable counting excludes direct pairs of parameter-observable, where a fit of the pair can be performed independently from other quantities.} Our model class is thus predictive with the following results:

\begin{enumerate}
\item The predicted PMNS quantities are $\theta_{23}^\text{PMNS}$ and $\delta^\text{PMNS}$, with results shown in Figures~\ref{fig:contours1} and \ref{fig:contours2}. It shows that the predictions of $\theta_{23}^\text{PMNS}$ vary from model to model, while the entire model class predicts $\delta^\text{PMNS}$ roughly between $230^\circ$ and $290^\circ$. Future measurements planned for example by DUNE~\cite{Abi:2018dnh,Abi:2018alz,Abi:2018rgm} will determine $\theta_{23}^\mathrm{PMNS}$ and $\delta^\mathrm{PMNS}$ with a precision of less than  $1^\circ$ and ${\cal O}(10^\circ)$, respectively, allowing for precision model testing and discrimination between them.
\item Each set of GUT operators predicts the ratio $m_d/m_s$ from the induced quark-lepton mass relations using the precise existing measurements for $m_\mu$ and $m_e$, with very small errors. The predictions for $m_d/m_s$  are summarised in Figure \ref{fig:ratio}.
\item With CSD2 predicting one neutrino mass to be negligible, there is only $1$ Majorana phase in the neutrino sector. We use instead the effective mass $\langle m_{\beta\beta}\rangle$ for neutrinoless double-beta decay as a proxy; the $1\sigma$ HPD interval predictions are given in Figure~\ref{fig:meff}.
\item While the SUSY threshold parameter $\eta_q$ is one of the fit parameters, its value would need to be reproduced by the SUSY particle spectrum in any complete model. The $\eta_q$ value is determined already by the $c_x$ Clebsch factor choice (cf.~Section~\ref{sec:predictionsetaq}).
\end{enumerate}

Beyond the present study, the results of the fits provide useful insight for explicitly constructing new models, especially when better experimental precision for the PMNS parameters will guide the direction. In building a complete flavour GUT model, our results provide the following guidance:
\begin{itemize}
\item A complete theory of flavour would be guided by the Yukawa textures used in the fermion sector, and potentially also by the results for viable values of the phases $\alpha$ and $\gamma$, which would be predicted by a suitable flavon VEV alignment. Interestingly, all of the most promising models in Table \ref{tab3} feature $\gamma$ close to $270^\circ$. We point out the intriguing possibility that such a phase $\gamma$, together with a $\alpha_\text{UT}=90^\circ$ for the ``phase sum rule mechanism'', could arise from a single imaginary entry in the $2$-$2$ element of $\YE$ and $\YD$. Furthermore, the provided values of $\theta_{12}^{eL}$ could allow one to explore model building possibilities within the considered $\mathrm{SU}(5)$ GUT setup even with a neutrino sector texture other than CSD2 (cf.~Section~\ref{sec:bestfitclebsches}).
\item CG coefficients are crucial building blocks of GUT flavour models, since they link predictions for $\theta_{23}^\text{PMNS}$ and $\delta^\text{PMNS}$ to the quark-lepton mass relation from $\mathrm{SU}(5)$ unification. The choice of CG factors actually reveals the choice of the underlying GUT operators in the Yukawa sector, thus suggesting the GUT matter content of the Higgs sector and perhaps guiding even towards a complete Higgs potential, from which the spontaneous breaking of GUT symmetry $\mathrm{SU}(5)\to\text{SM}$ arises. 

\end{itemize}

Finally, an offshoot of the presented work are the extensive RGE data tables for the changes in neutrino observables when run from the GUT scale to the $Z$ scale (cf.~Appendix~\ref{app:rg_running}). Raw data is provided under the link stated in~\cite{running_data}. Interpolating that data, together with the existing data tables from \cite{Antusch:2013jca} for the quark and charged lepton sectors allows to greatly speed up numerical fits of supersymmetric GUT flavour models to the experimental data.

In summary, we provide a systematic study for a novel class of CSD2 neutrino mixing models within a predictive $\mathrm{SU}(5)$ GUT setup. The candidate models have the potential to be highly predictive, and can therefore be tested in future experiments. Our study thus provides a roadmap for future work in constructing new flavour SUSY GUT models of this novel type.

\begin{longtable}{c}
$
\arraycolsep=8pt
\begin{array}{rrd{2}d{2}d{2}d{1}d{1}d{1}d{1}d{2}}
\toprule
 \text{Label} & (c_x,c_y,c_z) & \aligncell{\chi_\text{Tot}^2} & \aligncell{\chi_\text{q}^2} & \aligncell{\chi_\nu^2} & \aligncell{\theta_{23}[{}^\circ]} & \aligncell{\delta[{}^\circ]} & \aligncell{\gamma[{}^\circ]} & \aligncell{\alpha[{}^\circ]} & \aligncell{\theta_{12}^{eL}[{}^\circ]} \\
\midrule
 \spaceclebsches
 1\kern1.3em & \left(3,\frac{3}{2},\frac{1}{2}\right) &  &  &  &  &  &  &  &  \\
 \spacelines
 a_1 & (102) & 0.17 & 0.06 & 0.11 & 47.9 & 92.7 & 68.7 & 233.1 & 7.23 \\
 a_2 &  & 0.17 & 0.06 & 0.11 & 47.9 & 267.3 & 291.3 & 126.9 & 7.23 \\
 b_1 & (120) & 4.05 & 0.06 & 3.99 & 41.6 & 120.1 & 71.6 & 148.2 & 7.22 \\
 b_2 &  & 4.05 & 0.06 & 3.99 & 41.6 & 239.9 & 288.4 & 211.8 & 7.22 \\
 \spaceclebsches
 2\kern1.3em & (6,3,1) &  &  &  &  &  &  &  &  \\
 \spacelines
 a_1 & (102) & 0.19 & 0.06 & 0.14 & 47.9 & 93.7 & 67.7 & 233.9 & 7.23 \\
 a_2 &  & 0.19 & 0.06 & 0.14 & 47.9 & 266.3 & 292.3 & 126.1 & 7.23 \\
 b_1 & (120) & 4.19 & 0.06 & 4.13 & 41.5 & 118.9 & 72.6 & 147.0 & 7.22 \\
 b_2 &  & 4.19 & 0.06 & 4.13 & 41.5 & 241.1 & 287.4 & 213.0 & 7.22 \\
 \spaceclebsches
 3\kern1.3em & \left(\frac{9}{2},2,1\right) &  &  &  &  &  &  &  &  \\
 \spacelines
 a_1 & (102) & 1.62 & 1.06 & 0.56 & 43.9 & 103.0 & 72.9 & 263.2 & 5.49 \\
 a_2 &  & 1.62 & 1.06 & 0.56 & 43.9 & 257.0 & 287.1 & 96.8 & 5.49 \\
 b_1 & (120) & 1.06 & 1.06 & 0.00 & 47.2 & 90.2 & 71.0 & 90.5 & 5.49 \\
 b_2 &  & 1.06 & 1.06 & 0.00 & 47.2 & 269.8 & 289.0 & 269.5 & 5.49 \\
 \spaceclebsches
 4\kern1.3em & \left(\frac{9}{2},\frac{3}{2},1\right) &  &  &  &  &  &  &  &  \\
 \spacelines
 a_1 & (102) & 1.81 & 1.12 & 0.69 & 43.7 & 110.9 & 66.2 & 272.3 & 5.33 \\
 a_2 &  & 1.81 & 1.12 & 0.69 & 43.7 & 249.1 & 293.8 & 87.7 & 5.33 \\
 b_1 & (120) & 1.24 & 1.12 & 0.12 & 47.4 & 93.0 & 68.8 & 94.0 & 5.33 \\
 b_2 &  & 1.24 & 1.12 & 0.12 & 47.4 & 267.0 & 291.2 & 266.0 & 5.33 \\
 \spaceclebsches
 5\kern1.3em & \left(3,1,\frac{2}{3}\right) &  &  &  &  &  &  &  &  \\
 \spacelines
 a_1 & (102) & 1.82 & 1.12 & 0.70 & 43.7 & 110.7 & 66.5 & 272.1 & 5.32 \\
 a_2 &  & 1.82 & 1.12 & 0.70 & 43.7 & 249.3 & 293.5 & 87.9 & 5.32 \\
 b_1 & (120) & 1.24 & 1.12 & 0.12 & 47.4 & 93.1 & 68.7 & 94.1 & 5.32 \\
 b_2 &  & 1.24 & 1.12 & 0.12 & 47.4 & 266.9 & 291.3 & 265.9 & 5.32 \\
 \spaceclebsches
 6\kern1.3em & \left(\frac{9}{2},3,\frac{2}{3}\right) &  &  &  &  &  &  &  &  \\
 \spacelines
 a_1 & (102) & 1.64 & 0.92 & 0.72 & 48.8 & 83.9 & 72.5 & 215.3 & 8.29 \\
 a_2 &  & 1.64 & 0.92 & 0.72 & 48.8 & 276.1 & 287.5 & 144.7 & 8.29 \\
 b_1 & (120) & 9.68 & 0.93 & 8.75 & 40.4 & 117.1 & 78.4 & 155.0 & 8.28 \\
 b_2 &  & 9.68 & 0.93 & 8.75 & 40.4 & 242.9 & 281.6 & 205.0 & 8.28 \\
 \spaceclebsches
 7\kern1.3em & \left(6,2,\frac{3}{2}\right) &  &  &  &  &  &  &  &  \\
 \spacelines
 a_1 & (102) & 2.97 & 0.06 & 2.91 & 42.3 & 117.7 & 64.3 & 283.7 & 4.80 \\
 a_2 &  & 2.97 & 0.06 & 2.91 & 42.3 & 242.3 & 295.7 & 76.3 & 4.80 \\
 b_1 & (120) & 1.77 & 0.05 & 1.72 & 48.3 & 84.7 & 77.8 & 87.7 & 4.80 \\
 b_2 &  & 1.77 & 0.05 & 1.72 & 48.3 & 275.3 & 282.2 & 272.3 & 4.80 \\
\bottomrule
\end{array}$\\
$
\arraycolsep=8pt
\begin{array}{rrd{2}d{2}d{2}d{1}d{1}d{1}d{1}d{2}}
\toprule
 \text{Label} & (c_x,c_y,c_z) & \aligncell{\chi_\text{Tot}^2} & \aligncell{\chi_\text{q}^2} & \aligncell{\chi_\nu^2} & \aligncell{\theta_{23}[{}^\circ]} & \aligncell{\delta[{}^\circ]} & \aligncell{\gamma[{}^\circ]} & \aligncell{\alpha[{}^\circ]} & \aligncell{\theta_{12}^{eL}[{}^\circ]} \\
\midrule
 \spaceclebsches
 8\kern1.3em & \left(6,6,\frac{1}{2}\right) &  &  &  &  &  &  &  &  \\
 \spacelines
 a_1 & (102) & 2.37 & 0.38 & 1.99 & 49.6 & 84.4 & 71.5 & 147.4 & 14.90 \\
 a_2 &  & 2.37 & 0.38 & 1.99 & 49.6 & 275.6 & 288.5 & 212.6 & 14.90 \\
 a_3 &  & 8.62 & 0.43 & 8.19 & 40.9 & 133.9 & 64.5 & 63.6 & 14.93 \\
 a_4 &  & 8.62 & 0.43 & 8.19 & 40.9 & 226.1 & 295.5 & 296.4 & 14.93 \\
 b_1 & (120) & 3.11 & 0.38 & 2.74 & 42.1 & 121.6 & 68.9 & 218.5 & 14.90 \\
 b_2 &  & 3.11 & 0.38 & 2.74 & 42.1 & 238.4 & 291.1 & 141.5 & 14.90 \\
 \spaceclebsches
 9\kern1.3em & \left(3,2,\frac{1}{2}\right) &  &  &  &  &  &  &  &  \\
 \spacelines
 a_1 & (102) & 3.24 & 3.12 & 0.12 & 47.9 & 87.4 & 72.7 & 226.9 & 7.42 \\
 a_2 &  & 3.24 & 3.12 & 0.12 & 47.9 & 272.6 & 287.3 & 133.1 & 7.42 \\
 b_1 & (120) & 8.59 & 3.15 & 5.45 & 41.2 & 114.2 & 77.5 & 144.0 & 7.41 \\
 b_2 &  & 8.59 & 3.15 & 5.45 & 41.2 & 245.8 & 282.5 & 216.0 & 7.41 \\
 b_3 &  & 11.64 & 3.19 & 8.45 & 49.2 & 97.0 & 49.2 & 75.2 & 7.40 \\
 b_4 &  & 11.64 & 3.19 & 8.45 & 49.2 & 263.0 & 310.8 & 284.8 & 7.40 \\
 \spaceclebsches
 10\kern1.3em & \left(3,\frac{3}{2},\frac{2}{3}\right) &  &  &  &  &  &  &  &  \\
 \spacelines
 a_1 & (102) & 3.76 & 3.27 & 0.49 & 44.0 & 102.7 & 72.6 & 262.3 & 5.54 \\
 a_2 &  & 3.76 & 3.27 & 0.49 & 44.0 & 257.3 & 287.4 & 97.7 & 5.54 \\
 b_1 & (120) & 3.28 & 3.27 & 0.01 & 47.1 & 91.4 & 69.8 & 91.5 & 5.54 \\
 b_2 &  & 3.28 & 3.27 & 0.01 & 47.1 & 268.6 & 290.2 & 268.5 & 5.54 \\
 \spaceclebsches
 11\kern1.3em & \left(6,\frac{9}{2},\frac{2}{3}\right) &  &  &  &  &  &  &  &  \\
 \spacelines
 a_1 & (102) & 4.87 & 0.14 & 4.73 & 50.6 & 82.0 & 70.5 & 187.2 & 10.97 \\
 a_2 &  & 4.87 & 0.14 & 4.73 & 50.6 & 278.0 & 289.5 & 172.8 & 10.97 \\
 b_1 & (120) & 8.16 & 0.14 & 8.01 & 40.5 & 128.9 & 69.9 & 188.8 & 10.98 \\
 b_2 &  & 8.16 & 0.14 & 8.01 & 40.5 & 231.1 & 290.1 & 171.2 & 10.98 \\
 \spaceclebsches
 12\kern1.3em & \left(6,6,\frac{2}{3}\right) &  &  &  &  &  &  &  &  \\
 \spacelines
 a_1 & (102) & 5.98 & 2.65 & 3.32 & 50.2 & 78.0 & 73.6 & 182.1 & 11.28 \\
 a_2 &  & 5.98 & 2.65 & 3.32 & 50.2 & 282.0 & 286.4 & 177.9 & 11.28 \\
 b_1 & (120) & 14.58 & 2.66 & 11.92 & 39.8 & 125.1 & 72.7 & 187.3 & 11.28 \\
 b_2 &  & 14.58 & 2.66 & 11.92 & 39.8 & 234.9 & 287.3 & 172.7 & 11.28 \\
 \spaceclebsches
 13\kern1.3em & \left(\frac{9}{2},\frac{9}{2},\frac{1}{2}\right) &  &  &  &  &  &  &  &  \\
 \spacelines
 a_1 & (102) & 6.13 & 2.65 & 3.48 & 50.2 & 78.4 & 73.2 & 182.3 & 11.28 \\
 a_2 &  & 6.13 & 2.65 & 3.48 & 50.2 & 281.6 & 286.8 & 177.7 & 11.28 \\
 b_1 & (120) & 14.66 & 2.66 & 12.00 & 39.8 & 125.0 & 72.8 & 187.2 & 11.28 \\
 b_2 &  & 14.66 & 2.66 & 12.00 & 39.8 & 235.0 & 287.2 & 172.8 & 11.28 \\
 \spaceclebsches
 14\kern1.3em & \left(\frac{9}{2},3,\frac{1}{2}\right) &  &  &  &  &  &  &  &  \\
 \spacelines
 a_1 & (102) & 6.40 & 1.56 & 4.84 & 50.6 & 82.9 & 69.9 & 189.3 & 10.81 \\
 a_2 &  & 6.40 & 1.56 & 4.84 & 50.6 & 277.1 & 290.1 & 170.7 & 10.81 \\
 b_1 & (120) & 9.66 & 1.56 & 8.09 & 40.5 & 128.9 & 70.0 & 187.4 & 10.81 \\
 b_2 &  & 9.66 & 1.56 & 8.09 & 40.5 & 231.1 & 290.0 & 172.6 & 10.81 \\
\bottomrule
\end{array}$\\
$
\arraycolsep=8pt
\begin{array}{rrd{2}d{2}d{2}d{1}d{1}d{1}d{1}d{2}}
\toprule
 \text{Label} & (c_x,c_y,c_z) & \aligncell{\chi_\text{Tot}^2} & \aligncell{\chi_\text{q}^2} & \aligncell{\chi_\nu^2} & \aligncell{\theta_{23}[{}^\circ]} & \aligncell{\delta[{}^\circ]} & \aligncell{\gamma[{}^\circ]} & \aligncell{\alpha[{}^\circ]} & \aligncell{\theta_{12}^{eL}[{}^\circ]} \\
\midrule
 \spaceclebsches
 15\kern1.3em & \left(6,\frac{3}{2},2\right) &  &  &  &  &  &  &  &  \\
 \spacelines
 b_1 & (120) & 11.62 & 0.07 & 11.55 & 50.3 & 67.1 & 99.7 & 74.5 & 3.60 \\
 b_2 &  & 11.62 & 0.07 & 11.55 & 50.3 & 292.9 & 260.3 & 285.5 & 3.60 \\
 \spaceclebsches
 16\kern1.3em & \left(\frac{9}{2},1,\frac{3}{2}\right) &  &  &  &  &  &  &  &  \\
 \spacelines
 b_1 & (120) & 13.25 & 1.10 & 12.16 & 50.3 & 66.4 & 101.3 & 74.3 & 3.54 \\
 b_2 &  & 13.25 & 1.10 & 12.16 & 50.3 & 293.6 & 258.7 & 285.7 & 3.54 \\
 \spaceclebsches
 17\kern1.3em & \left(3,\frac{2}{3},1\right) &  &  &  &  &  &  &  &  \\
 \spacelines
 b_1 & (120) & 13.28 & 1.09 & 12.19 & 50.3 & 66.5 & 101.2 & 74.5 & 3.54 \\
 b_2 &  & 13.28 & 1.09 & 12.19 & 50.3 & 293.5 & 258.8 & 285.5 & 3.54 \\
 \spaceclebsches
 18\kern1.3em & \left(\frac{9}{2},\frac{3}{2},\frac{3}{2}\right) &  &  &  &  &  &  &  &  \\
 \spacelines
 b_1 & (120) & 13.30 & 3.35 & 9.95 & 50.2 & 62.0 & 103.7 & 68.4 & 3.69 \\
 b_2 &  & 13.30 & 3.35 & 9.95 & 50.2 & 298.0 & 256.3 & 291.6 & 3.69 \\
 \spaceclebsches
 19\kern1.3em & (6,2,2) &  &  &  &  &  &  &  &  \\
 \spacelines
 b_1 & (120) & 13.36 & 3.36 & 10.00 & 50.2 & 62.6 & 103.1 & 69.0 & 3.69 \\
 b_2 &  & 13.36 & 3.36 & 10.00 & 50.2 & 297.4 & 256.9 & 291.0 & 3.69 \\
 \spaceclebsches
 20\kern1.3em & (3,1,1) &  &  &  &  &  &  &  &  \\
 \spacelines
 b_1 & (120) & 13.41 & 3.35 & 10.06 & 50.2 & 63.1 & 102.7 & 69.6 & 3.69 \\
 b_2 &  & 13.41 & 3.35 & 10.06 & 50.2 & 296.9 & 257.3 & 290.4 & 3.69 \\
\bottomrule
\end{array}$\\
\caption{Results of the fit for model candidates specified by the CG coefficients and the CSD2 scenario. The table shows a complete list of CG coefficients $(c_x,c_y,c_z)$ with $\chi^2 < 15$, ordered according to their best $\chi^2$ value. For each combination of $(c_x,c_y,c_z)$ and CSD2 scenario ($\YNU^\text{(102)}$ or $\YNU^\text{(120)}$) all local minima with $\chi^2 < 15$ are listed. The 1st column assigns a unique label to each local minimum. The 2nd column specifies the CG coefficients and the type of neutrino Yukawa coupling. The quantity $\chi^2_\text{Tot}$ indicates the $\chi^2$ of the model, which includes all observables. $\chi^2_\text{q}$ contains the contributions of the $\chi^2$ coming from the quark and charged lepton Yukawa couplings and the CKM parameters, whereas in $\chi^2_\nu$ the remaining contributions to the $\chi^2$ from the neutrino mass squared differences and the PMNS angles are incorporated. In the last five columns the values of the two observables $\theta_{23}\equiv\theta_{23}^\text{PMNS}$, $\delta\equiv\delta^\text{PMNS}$, the two parameters $\gamma$, $\alpha$ and the $1$-$2$ left angle $\theta_{12}^{eL}$ of the charged leptons are shown.}\\
\label{tab2}
\end{longtable}

\begin{table}
\begin{align*}
\begin{array}{r@{\kern0.7em}r@{\kern0.7em}r@{\kern0.7em}r@{\kern0.7em}r@{\kern0.7em}r@{\kern0.7em}r@{\kern0.7em}r@{\kern0.7em}r@{\kern0.7em}r@{\kern0.7em}r@{\kern0.7em}r}
\toprule
 \text{Label} & \tan\beta & \eta _b & \eta _q & x & y & z & \gamma[{}^{\circ}] & \theta_{12}^{uL} & m_a[\mathrm{eV}] & \epsilon  & \alpha[{}^{\circ}] \\
\midrule
 1a_2 & 46.9 & 0.449 & -0.344 & 0.00722 & 0.001833 & 0.001642 & 291.3 & 0.0871 & 0.0283 & 0.103 & 126.9 \\
 3b_2 & 33.4 & -0.170 & 0.017 & 0.00347 & 0.000752 & 0.000777 & 289.0 & 0.0871 & 0.0261 & 0.119 & 269.5 \\
 4b_2 & 48.5 & 0.599 & -0.048 & 0.00498 & 0.001396 & 0.001147 & 291.2 & 0.0871 & 0.0266 & 0.117 & 266.0 \\
 3a_2 & 31.1 & -0.147 & 0.016 & 0.00317 & 0.000688 & 0.000710 & 287.1 & 0.0872 & 0.0263 & 0.116 & 96.8 \\
 6a_2 & 31.0 & -0.141 & 0.021 & 0.00314 & 0.000687 & 0.000704 & 287.5 & 0.0870 & 0.0285 & 0.099 & 144.7 \\
 7b_2 & 48.0 & 0.395 & 0.310 & 0.00374 & 0.000945 & 0.000850 & 282.2 & 0.0872 & 0.0263 & 0.121 & 272.3 \\
 4a_2 & 49.3 & 0.600 & -0.048 & 0.00507 & 0.001422 & 0.001169 & 293.8 & 0.0871 & 0.0264 & 0.119 & 87.7 \\
 8a_2 & 48.7 & 0.568 & 0.328 & 0.00365 & 0.000970 & 0.000834 & 288.5 & 0.0872 & 0.0290 & 0.098 & 212.6 \\
 7a_2 & 49.1 & 0.494 & 0.309 & 0.00381 & 0.000964 & 0.000866 & 295.7 & 0.0873 & 0.0258 & 0.125 & 76.3 \\
 8b_2 & 49.6 & 0.590 & 0.328 & 0.00372 & 0.000991 & 0.000852 & 291.1 & 0.0872 & 0.0292 & 0.097 & 141.5 \\
 9a_2 & 32.5 & -0.167 & -0.308 & 0.00502 & 0.000982 & 0.001116 & 287.3 & 0.0871 & 0.0282 & 0.102 & 133.1 \\
 10b_2 & 35.0 & -0.078 & -0.310 & 0.00542 & 0.001053 & 0.001203 & 290.2 & 0.0871 & 0.0262 & 0.119 & 268.5 \\
\bottomrule
\end{array}
\end{align*}
\caption{List of best-fit models. The table shows the model parameters of the $12$ best fit points with lowest $\chi^2$ from Table~\ref{tab2}. Note that the models~$2$ and $5$ are not considered. The corresponding local minima are essentially the same as the ones in model~$1$ and $4$ respectively, since the tuple of CG coefficients in model~$1$ and $2$, and in model~$4$ and $5$, differ only by an overall factor.}
\label{tab3}
\end{table}

\section*{Acknowledgements}
The work of S.A., C.H.~and V.S.~has been supported by the Swiss National Science Foundation. C.K.K. wishes to acknowledge support from the Swiss Government Excellence Scholarship (2017.0527) and the Royal Society-SERB Newton International Fellowship (NF171488).

\appendix

\section{Conventions}
\label{app:conventions}
\subsection{Parametrizations for unitary matrices}
\label{app:conventions_unitary_matrix}
Complex rotation matrices in $3$ dimensions with rotation angles $\theta_{ij}$ and phases $\delta_{ij}$ are written in the form
\begin{align}
\label{eq:complex_rotations}
\begin{split}
\mathbf{U}_{12} &= \begin{pmatrix}
c_{12} & s_{12}e^{-i\delta_{12}} & 0 \\
-s_{12}e^{i\delta_{12}} & c_{12} & 0 \\
0 & 0 & 1
\end{pmatrix},\quad
\mathbf{U}_{13} = \begin{pmatrix}
c_{13} & 0 & s_{13}e^{-i\delta_{13}} \\
0 & 1 & 0 \\
-s_{13}e^{i\delta_{13}} & 0 & c_{13}
\end{pmatrix},\\
\mathbf{U}_{23} &= \begin{pmatrix}
1 & 0 & 0 \\
0 & c_{23} & s_{23}e^{-i\delta_{23}} \\
0 & -s_{23}e^{i\delta_{23}} & c_{23}
\end{pmatrix},
\end{split}
\end{align}
where the abbreviations $c_{ij}\equiv\cos(\theta_{ij})$ and $s_{ij}\equiv\sin(\theta_{ij})$ are used. Real rotations $\mathbf{R}_{ij}$ are defined in the same way, but with vanishing phases, i.e. $\delta_{ij} = 0$. A general $3\times3$ unitary matrix contains $9$ degrees of freedom. A standard way to parametrize such a matrix $\mathbf{U}$ is the following:
\begin{align}
\label{eq:general_matrix_1}
\mathbf{U} &= \mathbf{Q}_1\mathbf{R}_{23}\mathbf{U}_{13}\mathbf{R}_{12}\mathbf{Q}_2\,,
\end{align}
with the diagonal phase matrices $\mathbf{Q}_1=\text{diag}(e^{i\delta_1},e^{i\delta_2},e^{i\delta_3})$ and $\mathbf{Q}_2=\text{diag}(e^{-i\varphi_1/2},e^{-i\varphi_2/2},1)$. The $3$ rotation angles are restricted to the interval $[0,\frac{\pi}{2}]$, while the $6$ phases take values in the full range $[0,2\pi]$. The middle term on the right-hand side in Eq.~\eqref{eq:general_matrix_1} has the explicit form
\begin{align}
\label{eq:matrix_explicit}
\mathbf{R}_{23}\mathbf{U}_{13}\mathbf{R}_{12} &= \begin{pmatrix}
c_{12}c_{13} & s_{12}c_{13} & s_{13}e^{-i\delta} \\
 -s_{12}c_{23} - c_{12}s_{23}s_{13}e^{i\delta} & c_{12}c_{23} - s_{12}s_{23}s_{13}e^{i\delta} & s_{23}c_{13} \\
 s_{12}s_{23} - c_{12}c_{23}s_{13}e^{i\delta} & -c_{12} s_{23} - s_{12}c_{23}s_{13}e^{i\delta} & c_{23}c_{13} \\
\end{pmatrix},
\end{align}
with the notation $\delta\equiv\delta_{13}$. This parametrization is usually used in specifying the CKM and PMNS matrix (see Appendix~\ref{app:ckm_pmns}). Further parametrizations of $3$-dimensional unitary matrices are the following:
\begin{align}
\label{eq:general_matrix_2_1}
\mathbf{U} &= \mathbf{P}\mathbf{U}_{23}\mathbf{U}_{13}\mathbf{U}_{12}\,, \\
\label{eq:general_matrix_2_2}
\mathbf{U} &= \mathbf{U}_{23}\mathbf{U}_{13}\mathbf{U}_{12}\mathbf{P}\,,
\end{align}
with the diagonal phase matrix $\mathbf{P}=\text{diag}(e^{i\eta_1},e^{i\eta_2},e^{i\eta_3})$ and
\begin{align}
\label{eq:general_matrix_3}
\mathbf{U} &= \mathbf{P}_2 \mathbf{R}_{23} \mathbf{R}_{13} \mathbf{P}_1 \mathbf{R}_{12} \mathbf{P}_3\,,
\end{align}
with the diagonal phase matrices $\mathbf{P}_1=\text{diag}(1,e^{i\chi},1)$, $\mathbf{P}_2=\text{diag}(1,e^{i\phi_2},e^{i\phi_3})$ and $\mathbf{P}_3=\text{diag}(e^{i\omega_1},e^{i\omega_2},e^{i\omega_3})$. The relations between the parameters of the four conventions stated in Eqs.~\eqref{eq:general_matrix_1} and \eqref{eq:general_matrix_2_1}--\eqref{eq:general_matrix_3} are listed in Table~\ref{tab:matrix_para}.
\begin{table}
\center
\begin{tabular}{rrrr}
\toprule
(i) & (ii) & (ii') & (iii) \\
\midrule
$\theta_{12}$ & $\theta_{12}$ & $\theta_{12}$ & $\theta_{12}$ \\
$\theta_{13}$ & $\theta_{13}$ & $\theta_{13}$ & $\theta_{13}$ \\
$\theta_{23}$ & $\theta_{23}$ & $\theta_{23}$ & $\theta_{23}$ \\
$\delta$ & $\delta_{13}-\delta_{12}-\delta_{23}$ & $\delta_{13}-\delta_{12}-\delta_{23}$ & $-\chi$ \\
$\varphi_1$ & $-2(\delta_{12}+\delta_{23})$ & $-2(\delta_{12}+\delta_{23}+\eta_1-\eta_3)$ & $2(\omega_3-\omega_1-\chi)$ \\
$\varphi_2$ & $-2\delta_{23}$ & $-2(\delta_{23}+\eta_2-\eta_3)$ & $2(\omega_3-\omega_2-\chi)$ \\
$\delta_1$ & $\eta_1-\delta_{12}-\delta_{23}$ & $\eta_3-\delta_{12}-\delta_{23}$ & $\omega_3-\chi$ \\
$\delta_2$ & $\eta_2-\delta_{23}$ & $\eta_3-\delta_{23}$ & $\omega_3+\phi_2$ \\
$\delta_3$ & $\eta_3$ & $\eta_3$ & $\omega_3+\phi_3$ \\
\bottomrule
\end{tabular}
\caption{The parametrization stated in Eq.~\eqref{eq:general_matrix_1} (i) written in terms of the parametrizations given in Eq.~\eqref{eq:general_matrix_2_1} (ii), \eqref{eq:general_matrix_2_2} (ii') and \eqref{eq:general_matrix_3} (iii).}
\label{tab:matrix_para}
\end{table}

\subsection{CKM and PMNS matrix}
\label{app:ckm_pmns}
A mass matrix $\mathbf{M}_f$ $(f\in\{u,d,e\})$, written in left-right convention, is diagonalized via a singular-value decomposition
\begin{align}
\label{eq:singular_value_decomp}
\mathbf{M}_f^\text{diag} &= \mathbf{U}_f^L \mathbf{M}_f \mathbf{U}_f^{R\dagger}\,,
\end{align}
where $\mathbf{U}_f^L,\mathbf{U}_f^R$ are (unitary) rotation matrices of the left- and right-handed fields respectively, and $\mathbf{M}_f^\text{diag}$ is diagonal with non-negative entries. The CKM matrix $\mathbf{U}^\text{CKM}$ is defined to be equal to $\mathbf{U}_d^{L\dagger}$ in the mass eigenbasis of the up-type quarks. In general the CKM matrix has the following form~\cite{Patrignani:2016xqp}:
\begin{align}
\mathbf{U}^\text{CKM} &= \mathbf{U}_u^L \mathbf{U}_d^{L\dagger}\,.
\end{align}
Using the definitions from Eq.~\eqref{eq:general_matrix_1}, the CKM matrix in the standard form reads
\begin{align}
\mathbf{U}^\text{CKM} &= \mathbf{R}_{23}^\text{CKM}\mathbf{U}_{13}^\text{CKM}\mathbf{R}_{12}^\text{CKM}\,,
\end{align}
which contains three rotation angles $\theta^\text{CKM}_{12},\theta^\text{CKM}_{13},\theta^\text{CKM}_{23}$ and the Dirac CP violating phase $\delta^\text{CKM}$. All other phases are absorbed by redefinitions of the fields. Assuming Majorana neutrinos, the diagonalization of the (symmetric) mass matrix $\mathbf{M}_\nu$ of the light neutrinos is performed by the use of a Takagi decomposition
\begin{align}
\label{eq:takagi_decomp}
\mathbf{M}_\nu^\text{diag} &= \mathbf{U}_\nu^{\textsf{T}} \mathbf{M}_\nu \mathbf{U}_\nu\,,
\end{align}
where $\mathbf{U}_\nu$ is the conjugate transpose of the rotation matrix of the left-handed neutrinos. The PMNS matrix $\mathbf{U}^\text{PMNS}$ is given by $\mathbf{U}_\nu$ in the mass eigenbasis of the charged leptons~\cite{Patrignani:2016xqp}. In general the PMNS matrix reads
\begin{align}
\label{eq:pmns_definition}
\mathbf{U}^\text{PMNS} &= \mathbf{U}_e^L \mathbf{U}_\nu\,,
\end{align}
and, using Eq.~\eqref{eq:general_matrix_1}, the standard parametrization is the following:
\begin{align}
\label{eq:pmns_parametrization}
\mathbf{U}^\text{PMNS} &= \mathbf{R}_{23}^\text{PMNS}\mathbf{U}_{13}^\text{PMNS}\mathbf{R}_{12}^\text{PMNS}\mathbf{Q}_2^\text{PMNS}\,,
\end{align}
where $\theta^\text{PMNS}_{12},\theta^\text{PMNS}_{13},\theta^\text{PMNS}_{23}$ are rotation angles, $\delta^\text{PMNS}$ is the Dirac CP phase and $\varphi_1^\text{PMNS},\varphi_2^\text{PMNS}$ are Majorana phases. The remaining three phases are absorbed into the fields.

\section{Approximate identities for the PMNS parameters}
In the following we provide identities for the lepton mixing angles and phases including charged lepton corrections. We use the definition from Eq.~\eqref{eq:pmns_definition} for the PMNS matrix, with the rotation matrices defined in Eq.~\eqref{eq:singular_value_decomp} and \eqref{eq:takagi_decomp}. The unitary rotation matrices $\mathbf{U}_e^L$ and $\mathbf{U}_\nu$ are parametrized according to Eq.~\eqref{eq:general_matrix_2_1} and \eqref{eq:general_matrix_2_2} respectively:
\begin{align}
\mathbf{U}_e^L &= \mathbf{P}^{eL}\mathbf{U}_{23}^{eL}\mathbf{U}_{13}^{eL}\mathbf{U}_{12}^{eL}\,, \\
\mathbf{U}_\nu &= \mathbf{U}_{23}^\nu\mathbf{U}_{13}^\nu\mathbf{U}_{12}^\nu\mathbf{P}^\nu\,.
\end{align}
The Yukawa matrix $\YE$ is taken from Eq.~\eqref{eq:Yukawa-texture-ude}. Since it is block diagonal, we have $\theta_{13}^{eL}=0$, $\theta_{23}^{eL}=0$, and the phases in $\mathbf{P}^{eL}$ are not fixed. In the following these phases are chosen zero, i.e. $\eta_i^{eL}=0$. Up to $1$st order in $y/x$ (and $z/x$) the parameters in $\mathbf{U}_{12}^{eL}$ are given by 
\begin{align}
\label{appeq:approx_leptons}
\theta_{12}^{eL} \approx \frac{c_y}{c_x}\frac{y}{x}\,,\quad \delta_{12}^{eL} \approx \pi-\gamma\,.
\end{align}
The light neutrino mass matrix $\mathbf{M}_\nu$ is taken from Eq.~\eqref{eq:CSD2-2variants}. Up to $1$st order in $\epsilon$ the parameters in $\mathbf{U}_\nu$ read (cf. \cite{Antusch:2011ic})
\begin{align}
\label{eq:approx_neutrino_102}
\renewcommand{\arraystretch}{1.3}
\begin{array}{l@{\kern3em}l@{\kern3em}l}
\theta_{12}^\nu \approx \arcsin{\Big(\frac{1}{\sqrt{3}}\Big)}\,, & \delta_{12}^\nu \approx \epsilon\sin\alpha\,, & \eta_1^\nu \approx -\frac{1}{2}\epsilon\sin\alpha\,, \\
\theta_{13}^\nu \approx \frac{\epsilon}{\sqrt{2}}\,, & \delta_{13}^\nu \approx \alpha - \frac{7}{2}\epsilon\sin\alpha\,, & \eta_2^\nu \approx -\frac{\alpha}{2} + \frac{3}{2}\epsilon\sin{\alpha}\,, \\
\theta_{23}^\nu \approx \frac{\pi}{4}-\epsilon \cos{\alpha}\,, & \delta_{23}^\nu \approx \pi - 2\epsilon\sin\alpha\,, & \eta_3^\nu \approx \pi - \frac{3}{2}\epsilon\sin\alpha\,,
\end{array}
\renewcommand{\arraystretch}{1}
\end{align}
in the CSD2 scenario $\YNU^{(102)}$, and
\begin{align}
\label{eq:approx_neutrino_120}
\renewcommand{\arraystretch}{1.3}
\begin{array}{l@{\kern3em}l@{\kern3em}l}
\theta_{12}^\nu \approx \arcsin{\Big(\frac{1}{\sqrt{3}}\Big)}\,, & \delta_{12}^\nu \approx -\epsilon\sin\alpha\,, & \eta_1^\nu \approx -\frac{1}{2}\epsilon\sin\alpha\,, \\
\theta_{13}^\nu \approx \frac{\epsilon}{\sqrt{2}}\,, & \delta_{13}^\nu \approx \pi + \alpha - \frac{3}{2}\epsilon\sin\alpha\,, & \eta_2^\nu \approx -\frac{\alpha}{2} - \frac{1}{2}\epsilon\sin{\alpha}\,, \\
\theta_{23}^\nu \approx \frac{\pi}{4}+\epsilon \cos{\alpha}\,, & \delta_{23}^\nu \approx \pi + 2\epsilon\sin\alpha\,, & \eta_3^\nu \approx \pi + \frac{1}{2}\epsilon\sin\alpha\,,
\end{array}
\renewcommand{\arraystretch}{1}
\end{align}
in the CSD2 scenario $\YNU^{(120)}$. In order to write down general formulas for the PMNS parameters we make the assumptions $\theta^{eL}_{13}=\theta^{eL}_{23}=0$ and $\theta^{eL}_{12},\theta_{13}^\nu\ll 1$, which are motivated by the above calculations. Furthermore we use the parametrization from Eq.~\eqref{eq:general_matrix_2_2} for the PMNS matrix, i.e.
\begin{align}
\mathbf{U}^\text{PMNS} &= \mathbf{U}_{23}^\text{PMNS}\mathbf{U}_{13}^\text{PMNS}\mathbf{U}_{12}^\text{PMNS}\mathbf{P}^\text{PMNS}\,.
\end{align}
Up to $1$st order in $\theta^{eL}_{12}$ and $\theta_{13}^\nu$ the identities are the following (cf. \cite{Antusch:2005kw}):
\begin{align}
s_{12}^\text{PMNS} e^{-i\delta_{12}^\text{PMNS}} &\approx s_{12}^{\nu} e^{-i (\delta_{12}^{\nu}+\theta_{12}^{eL} t_{12}^\nu c_{23}^\nu \sin(\delta_{12}^\nu-\delta_{12}^{eL}))} + \theta_{12}^{eL} c_{12}^{\nu} c_{23}^{\nu} e^{-i\delta_{12}^{eL}}\,, \label{appeq:eqs12} \\
s_{13}^\text{PMNS} e^{-i \delta_{13}^\text{PMNS}} &\approx \theta_{13}^\nu e^{-i\delta_{13}^\nu} + \theta_{12}^{eL} s_{23}^\nu e^{-i (\delta_{23}^{\nu}+\delta_{12}^{eL})}\,,  \label{appeq:eqs13} \\
s_{23}^\text{PMNS} e^{-i \delta_{23}^\text{PMNS}} &\approx s_{23}^{\nu} e^{-i\delta_{23}^{\nu}} \,, \label{appeq:eqs23}
\end{align}
with the notation $c^\nu_{ij}\equiv\cos\theta^\nu_{ij}$, $s^\nu_{ij}\equiv\sin\theta^\nu_{ij}$ and $t^\nu_{ij}\equiv\tan\theta^\nu_{ij}$. For completeness we also list the identities for the phases in $\mathbf{P}^\text{PMNS}$:
\begin{align}
\eta_1^\text{PMNS} &\approx \eta_1^\nu-\theta_{12}^{eL} t_{12}^\nu c_{23}^\nu \sin(\delta_{12}^\nu-\delta_{12}^{eL})\,, \\
\eta_2^\text{PMNS} &\approx \eta_2^\nu+\theta_{12}^{eL} t_{12}^\nu c_{23}^\nu \sin(\delta_{12}^\nu-\delta_{12}^{eL}) \,, \\
\eta_3^\text{PMNS} &\approx \eta_3^\nu \,.
\end{align}
Combining Eqs.~\eqref{appeq:approx_leptons}--\eqref{eq:approx_neutrino_120} with Eqs.~\eqref{appeq:eqs12}--\eqref{appeq:eqs23} we get for the lepton mixing angles and phases (cf. \cite{Antusch:2013wn,King:2005bj})\footnote{To calculate the Dirac and the Majorana phase we use the identities from Table~\ref{tab:matrix_para}: $\delta^\text{PMNS}=\delta_{13}^\text{PMNS}-\delta_{12}^\text{PMNS}-\delta_{23}^\text{PMNS}$ and $\varphi_2^\text{PMNS}=-2(\delta_{23}^\text{PMNS}+\eta_2^\text{PMNS}-\eta_3^\text{PMNS})$. The Dirac phase $\delta^\text{PMNS}$ is calculated only in leading order, since it always appears in combination with $\theta_{13}^\text{PMNS}$ in the PMNS matrix.}
\begin{align}
\theta_{12}^\text{PMNS} &\approx 35.3^\circ - \frac{\theta_{12}^{eL}}{\sqrt{2}}\cos{\gamma}\,, \label{appeq:12pmns_102}\\
\theta_{13}^\text{PMNS} &\approx \frac{1}{\sqrt{2}} \big( \epsilon^2 + {\theta_{12}^{eL}}^2 + 2\epsilon\theta_{12}^{eL} \cos{(\alpha+\gamma)} \big)^{1/2}\,, \label{appeq:13pmns_102}\\
\theta_{23}^\text{PMNS} &\approx 45^\circ - \epsilon \cos{\alpha}\,, \label{appeq:23pmns_102}\\
\delta^\text{PMNS} &\approx \text{arg}\big( \epsilon e^{i(\pi+\alpha)} + \theta_{12}^{eL} e^{i(\pi-\gamma)} \big)\,, \label{appeq:deltapmns_102}\\
\varphi_2^\text{PMNS} &\approx \alpha-2\epsilon\sin\alpha + \theta_{12}^{eL}\sin\gamma\,, \label{appeq:majoranapmns_102}
\end{align}
in the CSD2 scenario $\YNU^{(102)}$, and
\begin{align}
\theta_{12}^\text{PMNS} &\approx 35.3^\circ - \frac{\theta_{12}^{eL}}{\sqrt{2}}\cos{\gamma}\,, \label{appeq:12pmns_120}\\
\theta_{13}^\text{PMNS} &\approx \frac{1}{\sqrt{2}} \big( \epsilon^2 + {\theta_{12}^{eL}}^2 - 2\epsilon\theta_{12}^{eL} \cos{(\alpha+\gamma)} \big)^{1/2}\,, \label{appeq:13pmns_120}\\
\theta_{23}^\text{PMNS} &\approx 45^\circ + \epsilon \cos{\alpha}\,, \label{appeq:23pmns_120}\\
\delta^\text{PMNS} &\approx \text{arg}\big( \epsilon e^{i\alpha} + \theta_{12}^{eL} e^{i(\pi-\gamma)} \big)\,, \label{appeq:deltapmns_120}\\
\varphi_2^\text{PMNS} &\approx \alpha-2\epsilon\sin\alpha + \theta_{12}^{eL}\sin\gamma\,, \label{appeq:majoranapmns_120}
\end{align}
in the CSD2 scenario $\YNU^{(120)}$, as an expansion of $\theta_{12}^{eL}$ and $\epsilon$. Since the lightest left-handed neutrino is massless, the Majorana phase $\varphi_1^\text{PMNS}$ is unphysical, and therefore not listed. In particular, we get the following leading order identities for the respective CSD2 scenarios $\YNU^{(102)}$ and $\YNU^{(120)}$:
\begin{align}
s_{13}^\text{PMNS} e^{i \delta^\text{PMNS}} &\approx \frac{\epsilon}{\sqrt{2}} e^{i(\pi+\alpha)} + \frac{\theta_{12}^{eL}}{\sqrt{2}} e^{i(\pi-\gamma)}\,, \label{appeq:identity_delta_pmns_102} \\
s_{13}^\text{PMNS} e^{i \delta^\text{PMNS}} &\approx \frac{\epsilon}{\sqrt{2}} e^{i\alpha} + \frac{\theta_{12}^{eL}}{\sqrt{2}} e^{i(\pi-\gamma)}\,. \label{appeq:identity_delta_pmns_120}
\end{align}

\section{RGE running of neutrino data}
\label{app:rg_running}
The values of the observables of our models are predicted at the GUT scale. The predictions of the Yukawa couplings and the CKM parameters can be compared with the experimental data in a very efficient way by using the data tables of \cite{Antusch:2013jca}. These tables provide the experimental best fit values and the errors at the GUT scale, including SUSY threshold corrections. In order that the PMNS parameters and the neutrino mass squared differences can be fitted to the experimental values in an efficient way too, without performing the running explicitly, we have prepared a data table which contains the running effects of these quantities between the GUT scale $M_\text{GUT}$ and the $Z$-boson mass scale $M_Z$. In contrast to \cite{Antusch:2013jca} the experimental data is not run to $M_\text{GUT}$, but the data table is used to determine the values of the observables at $M_Z$, where they can be compared with the experimental data. In particular the table contains the $7$ quantities
\begin{align}
\label{eq:table_delta_quantities}
\Delta\theta_{12}^\text{PMNS},\,\Delta\theta_{13}^\text{PMNS},\,\Delta\theta_{23}^\text{PMNS},\,\Delta\delta^\text{PMNS},\,\Delta\varphi_2^\text{PMNS},\,\Delta m_{\nu_2},\,\Delta m_{\nu_3},
\end{align}
as functions of the $5$ parameters
\begin{align}
\label{eq:table_parameters}
\tan\beta,\,\eta_b,\,\theta_{23}^\text{PMNS},\,\delta^\text{PMNS},\,\varphi_2^\text{PMNS},
\end{align}
where $\eta_b$ is the threshold parameter defined in Eqs.~\eqref{thresh:yu}--\eqref{thresh:yl}, and the remaining three PMNS parameters are specified at the GUT scale. From left to right, the values of the parameters in Eq.~\eqref{eq:table_parameters} are contained in the ranges $[5,75]$, $[-0.6,0.6]$, $[0,\frac{\pi}{2}]$, $[0,2\pi]$ and $[0,2\pi]$. The $\Delta$-quantities in Eq.~\eqref{eq:table_delta_quantities} describe the difference of the corresponding observables at $M_Z$ compared to $M_\text{GUT}$, e.g.
\begin{align}
\theta_{12}^\text{PMNS}|_{M_Z} = \theta_{12}^\text{PMNS}|_{M_\text{GUT}} + \Delta\theta_{12}^\text{PMNS}.
\end{align}
In the calculation of the data table we used the $2$-loop RGEs of the MSSM and SM\footnote{In the SM the running of the coefficient of the neutrino mass operator, $\boldsymbol{\kappa}\equiv-\frac{4}{v^2_\text{EW}}\mathbf{M}_\nu$, is only known at $1$-loop.} in the $\overline{\text{DR}}$ and $\overline{\text{MS}}$ scheme respectively~\cite{Antusch:2001ck,Martin:1993zk,Antusch:2005gp}. These two models are matched a the SUSY scale $M_\text{SUSY}$, considering $\tan\beta$-enhanced SUSY threshold corrections as described in Section~\ref{sec:GUT-operators}. The mass scales are chosen as follows:
\begin{align}
M_Z=91.2\,\mathrm{GeV}\,,\quad M_\text{SUSY}=3\cdot10^3\,\mathrm{GeV}\,,\quad M_\text{GUT}=2\cdot10^{16}\,\mathrm{GeV}\,.
\end{align}
At the SUSY scale, when we are passing from the $\overline{\text{DR}}$ to the $\overline{\text{MS}}$ scheme (or vice versa) the matching of the gauge couplings $g_i$ is given by 
\begin{align}
(\alpha^{-1}_1)_{\overline{\text{MS}}} &= (\alpha^{-1}_1)_{\overline{\text{DR}}}\,, \\
(\alpha^{-1}_2)_{\overline{\text{MS}}} &= (\alpha^{-1}_2)_{\overline{\text{DR}}} + \frac{1}{6\pi}\,, \\
(\alpha^{-1}_3)_{\overline{\text{MS}}} &= (\alpha^{-1}_3)_{\overline{\text{DR}}} + \frac{1}{4\pi}\,,
\end{align}
where $\alpha_i\equiv \frac{g_i^2}{4\pi}$. Although the threshold corrections of the Yukawa couplings in Eqs.~\eqref{thresh:yu}--\eqref{thresh:yl} contain the parameter $\eta_q$, the $\Delta$-quantities in Eq.~\eqref{eq:table_delta_quantities} do not depend on it, and thus we fix $\eta_q=0$. We neglect the contribution of the neutrino Yukawa coupling $Y_\nu$ to the running. This is justified because $Y_\nu$ would come from effective operators. Furthermore, only two out of three left-handed neutrinos are massive, having normal hierarchy, i.e. $0=m_{\nu_1}<m_{\nu_2}<m_{\nu_3}$. For all parameters which are not listed in Eq.~\eqref{eq:table_parameters} but are used in the RGEs, like Yukawa couplings, CKM and PMNS parameters and gauge couplings, the boundary conditions are fixed at $M_Z$, where they coincide with the experimental values listed in NuFIT 3.2 (2018)~\cite{Esteban:2016qun} and in \cite{Antusch:2013jca}. In particular we used $m_{\nu_2}=\sqrt{\Delta m^2_\text{21}}$ and $m_{\nu_3}=\sqrt{\Delta m^2_\text{31}}$ at $M_Z$.

The data table, together with an example Mathematica notebook, is provided under the link stated in \cite{running_data}.

\end{document}